\documentclass{article}

\usepackage{arxiv}

\usepackage[utf8]{inputenc} 
\usepackage[T1]{fontenc}    
\usepackage{hyperref}       
\usepackage{url}            
\usepackage{booktabs}       
\usepackage{amsfonts}       
\usepackage{nicefrac}       
\usepackage{microtype}      
\usepackage[normalem]{ulem}
\usepackage{lineno}

\usepackage[numbers]{natbib}
\usepackage{xcolor}
\usepackage{wrapfig}
\usepackage{graphicx,caption}
\usepackage{amsmath,amsfonts,amssymb,amsthm}
\usepackage[toc,page,titletoc]{appendix}
\usepackage{bm}

\newenvironment{keypoints}
{
  \centerline
  {\large \bfseries \scshape Key Points}
  \begin{quote}
}
{
  \end{quote}
}

\newcommand{\beginsupplement}{%
        \setcounter{table}{0}
        \renewcommand{\thetable}{S\arabic{table}}%
        \setcounter{figure}{0}
        \renewcommand{\thefigure}{S\arabic{figure}}%
     }

\title{Use Internet Search Data to Accurately Track State-Level Influenza Epidemics}

\author{
  Shihao Yang \thanks{Joint first author. Department of Biomedical Informatics, Harvard Medical School, Boston, MA 02115, USA}
   \and
 Shaoyang Ning \thanks{Joint first author. Department of Mathematics and Statistics, Williams College, Williamstown, MA 01267, USA}
   \and
   S. C. Kou \thanks{To whom correspondence should be addressed. Department of Statistics, Harvard University, Cambridge, MA 02138, USA. kou@stat.harvard.edu} 
}

\begin{document}

\maketitle

\begin{keypoints}
Big data from the Internet has great potential to track social and economic events at multiple geographical levels. Here for real-time estimation of state-level influenza activities in the United States, we propose a statistical model that efficiently combines publicly available Internet search data at multiple resolutions (national, regional, and state-level) with traditional influenza surveillance data from the Centers for Disease Control and Prevention. Our method, across all states, outperforms all existing time-series-based influenza tracking methods. Our model is robust and easy to implement, with the flexibility to incorporate additional information from other sources and resolutions, making it generally applicable to tracking other social, economic or public health events at the state or local level.
\end{keypoints}


\newpage

\begin{abstract}
For epidemics control and prevention, timely insights of potential hot spots are invaluable. Alternative to traditional epidemic surveillance, which often lags behind real time by weeks, big data from the Internet provide important information of the current epidemic trends. Here we present a methodology, ARGOX (Augmented Regression with GOogle data CROSS space), for accurate real-time tracking of state-level influenza epidemics in the United States. ARGOX combines Internet search data at the national, regional and state levels with traditional influenza surveillance data from the Centers for Disease Control and Prevention, and accounts for both the spatial correlation structure of state-level influenza activities and the evolution of people's Internet search pattern. ARGOX achieves on average 28\% error reduction over the best alternative for real-time state-level influenza estimation for 2014 to 2020. ARGOX is robust and reliable and can be potentially applied to track county- and city-level influenza activity and other infectious diseases.
\end{abstract}

\section*{Introduction}

Each year in the United States (US) alone, the seasonal influenza (flu) epidemics may claim up to 61,000 deaths \cite{fluburden}. Quick responses and preventive actions to changes in flu epidemics rely on timely and accurate information on the current flu severity. In particular, due to the geographically varying timing and intensity of disease epidemics, most public health decisions and executive orders for disease control and prevention are made at the state or local level. Accurate \emph{real-time} flu tracking at the state/local level is thus indispensable. Traditional flu surveillance, such as those conducted by the US Centers for Disease Control and Prevention (CDC), however, often lags behind real time by up to two weeks. Here we propose a statistically principled, self-coherent framework ARGOX (Augmented Regression with GOogle data CROSS space) for real-time, accurate flu estimation at the state level. ARGOX efficiently combines publicly available Internet search data with traditional flu surveillance data and coherently utilizes the data from multiple geographical resolutions (national, regional, and state levels). 

For the last two decades, 
tracking of flu activities in the US mainly relies on traditional surveillance systems, such as the US Outpatient Influenza-like Illness Surveillance Network (ILINet) by the CDC. Through the ILINet, thousands of healthcare providers across the US report their numbers of outpatients with Influenza-like Illness (ILI) to CDC on a weekly basis. CDC then aggregates the data and publishes the ILI percentages (\%ILI, i.e., the percentages of outpatients with ILI) in its weekly reports at the national and 
regional levels (there are ten Health and Human Services (HHS) regions in the US, each consisting of multiple states). Starting from 2017, the state-level \%ILI reports became available for selected states, and in late 2018 the state-level \%ILI reports became available for all states except Florida. Owing to the time for administrative processing and aggregation, CDC's flu reports typically lag behind real time for up to 2 weeks and are also subject to subsequent revisions. Such delay and inaccuracy are far from optimal for public health decision making, especially in the face of epidemic outbreaks or pandemics.

Big data from the Internet offers the potential of real-time tracking of public health or social events.
In fact, valuable insights have been gained from the Internet data about current social and economical status of a nation, including epidemic outbreaks \cite{Ginsberg_2009, yang2017advances} and macro economic indices \cite{scott2014predicting,scott2015bayesian}. Furthermore, real-time data from the Internet could also offer insights at the regional, state, or local level. Examples include foreshadowing state-wise housing price index in the US \cite{wu2014}, estimating New York City flu activity \cite{Shaman_2012}, estimating real-time county-level unreported COVID-19 severity in the US \cite{mcneil_2020} among others. For epidemic surveillance, such real-time digital data at local level can be potentially used to provide insights for early epidemic hot-spot detection and timely public health resource allocation (e.g. vaccine campaigns) as well as to gather information on the overall disease prevalence.


Various models have been proposed to utilize Internet data, especially Internet search volume data, to provide real-time estimation of the current flu activity at the national level. Google Flu Trends (GFT), as one of the early examples, uses the search frequency of selected query terms from Google to estimate the real-time \%ILI \cite{Ginsberg_2009}. Recent models on combining CDC's surveillance data with Internet-derived data appear to work well at the national level \cite{yang2015accurate,Yang2017athgt}. Other methods, primarily targeting national flu epidemics, were also developed based on traditional epidemiology data and mechanistic models, such as susceptible-infectious-recovered-susceptible model with ensemble adjustment Kalman filter (SIRS-EAKF)  \cite{Shaman_2012,Yang03032015,shaman2013real,yang2014comparison,shaman2015improved}.

Compared to estimation at the national level, \%ILI estimation at the regional or state level is much more challenging, as documented by FluSight, the CDC-sponsored Flu Prediction Initiative \cite{cdcflusight}. Due to factors like geographical proximity, transportation connectivity, and public health communication, the state-wise epidemic spread exhibits strong spatial structure. However, many digital flu estimation methods \cite{shaman2013real, brooks2015flexible,Farrow2017-dj}, including GFT, ignore such spatial structure and apply the same national-level method to regional, and/or state-level flu estimation. 
A few attempts have been made to incorporate the geographical dependence structure. For example, Ref. \citenum{yang2016forecasting} studied the estimation of ILI activity in the boroughs and neighborhoods of New York City using a traditional epidemiological mechanistic SIRS-network model without Internet search data, where the dynamic system is multivariate with explicit parameters to characterize traffic between locales, and concluded that the spatial network is helpful at the borough scale but not at the neighborhood scale; Ref. \citenum{Davidson2015} utilized an ordinary-least-squares-based network model to improve upon the output of GFT, where a weighted average of GFT from all regions is produced as an network-enhanced final estimate for each individual region; Ref. \citenum{zou2018multi} employs a multi-task nonlinear regression method for regional \%ILI estimation, where a Multi-Task Gaussian Process is proposed to regress each region's \%ILI on the corresponding Google search data; Ref. \citenum{lu2019improved} uses a network approach for \%ILI estimation in a few selected states, where they first built a stand-alone state \%ILI prediction based on the ARGO method\cite{yang2015accurate}, and then obtained a multiple linear regression prediction for a given state's \%ILI from other states' \%ILI, and finally a winner-takes-all approach was adopted for each state separately to select one of the two approaches; Ref. \citenum{ning2019accurate} shows that careful spatial structure modeling can lead to much improved accuracy in \%ILI estimation at the regional level. An ensemble approach has also been proposed to utilize the output of a variety of available models to achieve better accuracy \cite{reich2019accuracy}.

Nevertheless, at the state level, no existing methods provide real-time flu tracking with satisfactory accuracy and reliability. (i) 
There are no unified approaches to combine multi-resolution and cross-state information effectively to provide national, regional and state-level estimates within the same framework. (ii) Few existing models can outperform a naive estimation method, which, for each state, without any modeling effort, simply uses CDC's reported \%ILI from the previous week as the \%ILI estimate for the current week (see Fig \ref{fig_heatmap} for an illustration). This would be particularly worrisome for public health officials who rely on accurate flu estimation at the local level to make informed decisions.


In this article we introduce ARGOX, a unified spatial-temporal statistical framework that combines multi-resolution, multi-source information to provide real-time state-level \%ILI estimates while maintaining coherency with \%ILI estimation at the regional and national levels (in a cascading fashion). To illustrate the underlying idea of ARGOX, let us take estimating the \%ILI in California as an example. The real-time Google search volumes for flu-related terms like "flu symptoms" or "flu duration" from California reflect its current state-level flu intensity to some extent. In addition, California's flu epidemics could be highly correlated with flu epidemics of nearby states such as Oregon and Nevada, as well as with geographically distant but transportation-wise well-connected states such as Illinois. California's current flu situation may also depend heavily on the recent trends of flu epidemics, in particular, the overall national and Pacific-west regional flu trends. Taken these considerations together, ARGOX operates in two steps: at the first step, it extracts Google search information of most relevant query terms at three geographical resolutions -- national, regional, and state levels; at the second step, the cross-time, cross-resolution, cross-state information mentioned above, together with Internet-extracted information, are integrated through careful modeling of their temporal-spatial dependence structure, which yields significant enhancement in the estimation accuracy. 

ARGOX was inspired in part by Refs. \citenum{yang2015accurate} and \citenum{ning2019accurate}, which studied the \%ILI estimation at the national and regional levels respectively. Although the methods introduced in Refs. \citenum{yang2015accurate} and \citenum{ning2019accurate} worked well for flu-tracking at the national or regional level, these methods cannot be directly applied to accurately track state-level \%ILI for a number of reasons, which are specifically solved by ARGOX. In particular, ARGOX addresses the following issues: (i) how to \emph{simultaneously} provide accurate, real-time flu tracking at the higher-resolution level for all 51 US states (district/city), as opposed to only at the national or regional level, (ii) how to effectively combine multi-resolution information from the national, regional and state levels for state \%ILI estimation, i.e., how to leverage the information from the national and regional levels, in addition to the information at a particular state, for the \%ILI estimation at a given state; (iii) how to solve the challenge of declining quality of Internet search data at higher geographical resolution, since compared to the Internet search data at the national level, the state-level Internet search data are of much inferior quality; (iv) how to determine when to borrow information from other states for the \%ILI estimation at a given state and when not to borrow, since the states have varying degree of connections --- for a state well connected with others, borrowing information probably would help its \%ILI estimation, but for a state not (geographically or epidemically) well connected with others, using information from other states might hurt (as opposed to help) its \%ILI estimation; and (v) how to model the correlation structure of \%ILI across the ``well-connected'' states to effectively borrow such cross-state information to improve prediction accuracy. ARGOX, therefore, significantly advances accurate flu tracking from the national and regional levels to the state level, which could help public health officials make much more informed decisions.

Through the ARGOX framework, the state-level flu activity estimates are produced in a unified and coherent way with the national and regional estimates. ARGOX achieves on average 28\% mean squared error (MSE) reduction compared to the 
best alternative and shows strong advantages over all benchmark methods, including GFT, time-series-based vector autoregression (VAR), and another recent Internet-search-based method developed in Lu et al. (2019) \cite{lu2019improved}. ARGOX achieves its high estimation accuracy through a few features: (i) it automatically selects the most relevant 
search queries to address the problem of lower-quality Google search information at state or regional level; (ii) it incorporates time-series momentum of flu activity; (iii) it pools the multi-resolution information by combining the national-, regional-, and state-level data; (iv) it explicitly models the spatial correlation structure of state-level flu activities; (v) it adapts to the evolution in people's search pattern, Google's search engine algorithms, epidemic trends, and other time-varying factors \cite{Burkom_etal_2007} with a dynamic two-year rolling window for training; and (vi) it achieves selective pooling of most immediately relevant information for a handful of stand-alone states (details in Methods).

\section*{Results}

We conducted retrospective estimation of the weekly \%ILI at the US state level -- 50 states excluding Florida whose ILI data is not available from CDC, plus Washington DC and New York City -- for the period of Oct 11, 2014 to March 21, 2020. For each week during this period, we only used the data that would have been available -- the historical CDC's ILI reports up to the previous week and Google search data up to the current week -- to estimate state-level \%ILI of the current week. To evaluate the accuracy of our estimation, we compared the estimates with the actual \%ILI released by CDC weeks later in multiple metrics, including the mean squared error (MSE), the mean absolute error (MAE),  and the correlation with the actual \%ILI (detailed in Methods). We also compared the performance of ARGOX with several benchmark methods, including (a) GFT (last estimate available: the week ending on August 15, 2015), (b) estimates by the lag-1 vector autoregressive model (VAR model), (c) the naive estimates, which for each state without any modeling effort simply use CDC’s reported \%ILI of the previous week as the estimate for the current week, and (d) a recent Internet-search-based state-level estimation model developed in Lu et al. (2019) \cite{lu2019improved}. As ARGOX uses a two-year training window, for fair comparison we keep the same two-year training window for VAR as well. Also for fair comparison, the numerical results of the method of Lu et al. (2019) were 
directly quoted from the article \cite{lu2019improved} (which reported results through May 14, 2017). 

Table \ref{tab_overall} summarizes the overall results of ARGOX, VAR, GFT, and the naive method, averaging over the 51 states/district/city for the whole period of 2014 to 2020 (up to March 21, 2020). Table \ref{tab:lu-et-al-short} summarizes the comparison between ARGOX and the method of Lu et al. (2019), averaging over 37 states for the period of 2014 to 2017. We need to compare ARGOX with Lu et al. (2019) in a separate Table \ref{tab:lu-et-al-short} because the results of Lu et al. (2019) are only available for 37 states and only for the period of 2014 to 2017.

\begin{table*}
\centering
\begin{tabular}{crrrrrrr}
  \hline
  & Whole period & '14-'15 & '15-'16 & '16-'17 & '17-'18 & '18-'19 & '19-'20 \\ 
  \hline  \multicolumn{1}{l}{MSE}\\ARGOX & \textbf{0.340} & \textbf{0.488} & \textbf{0.217} & \textbf{0.421} & \textbf{0.445} & \textbf{0.301} & \textbf{0.835} \\ 
  VAR & 1.556 & 1.606 & 0.819 & 1.629 & 2.615 & 1.277 & 3.747 \\ 
  GFT & -- & 2.186 & -- & -- & -- & -- & -- \\ 
  naive & 0.473 & 0.665 & 0.257 & 0.551 & 0.779 & 0.434 & 1.150 \\ 
   \hline  \multicolumn{1}{l}{MAE}\\ARGOX & \textbf{0.340} & \textbf{0.380} & \textbf{0.311} & \textbf{0.407} & \textbf{0.423} & \textbf{0.359} & \textbf{0.580} \\ 
  VAR & 0.597 & 0.633 & 0.516 & 0.693 & 0.825 & 0.668 & 1.058 \\ 
  GFT & -- & 0.944 & -- & -- & -- & -- & -- \\ 
  naive & 0.393 & 0.435 & 0.340 & 0.464 & 0.547 & 0.443 & 0.696 \\ 
   \hline  \multicolumn{1}{l}{Correlation}\\ARGOX & \textbf{0.949} & \textbf{0.914} & \textbf{0.832} & \textbf{0.875} & \textbf{0.937} & \textbf{0.921} & \textbf{0.902} \\ 
  VAR & 0.857 & 0.806 & 0.693 & 0.752 & 0.854 & 0.813 & 0.772 \\ 
  GFT & -- & 0.904 & -- & -- & -- & -- & -- \\ 
  naive & 0.931 & 0.885 & 0.803 & 0.842 & 0.902 & 0.890 & 0.874 \\ 
   \hline
\end{tabular}
\captionsetup{width=.9\linewidth}
\caption{Comparison of different methods for state-level \%ILI estimation. The evaluation is based on the average of 51 US states/district/city in multiple periods and multiple metrics. The MSE, MAE, and correlation are reported. The method with the best performance is highlighted in boldface for each metric in each period. Methods considered here include ARGOX, VAR, GFT, and the naive method. All comparisons are conducted on the original scale of CDC’s \%ILI. The whole period is Oct 11, 2014 to March 21, 2020. Columns 3 to 8 correspond to the regular flu seasons (week 40 to week 20 next year, defined by CDC’s Morbidity and Mortality Weekly Report; 19’-20’ season is up to March 21, 2020). }   \label{tab_overall}
\end{table*}

\begin{table*}
  \begin{center}
    \begin{tabular}{crrrr}
  \hline
  & Overall ('14-'17) & '14-'15 & '15-'16 & '16-'17 \\ 
  \hline  \multicolumn{1}{l}{MSE}\\ARGOX & \textbf{0.269} & \textbf{0.406} & \textbf{0.163} & \textbf{0.339} \\ 
  Lu et al. (2019) \cite{lu2019improved} & 0.418 & 0.467 & 0.528 & 0.544 \\ 
   \hline  \multicolumn{1}{l}{Correlation}\\ARGOX & \textbf{0.919} & \textbf{0.914} & \textbf{0.836} & \textbf{0.890} \\ 
  Lu et al. (2019) \cite{lu2019improved} & 0.912  & 0.912 & 0.808 & 0.858\\ 
   \hline
\end{tabular}
  \end{center}
  \captionsetup{width=.9\linewidth}
  \caption{Comparison of ARGOX to the method of Lu et al. (2019) \cite{lu2019improved} for state-level \%ILI estimation. The numbers of Lu et al. (2019) are directly obtained from \cite{lu2019improved}, which reported its estimation results of 37 states over three flu seasons: 2014 to 2017. For fair comparison, the result of ARGOX is  restricted to the same 37 states and the same time period
  to match \cite{lu2019improved}. The method with best performance for each metric in each period is highlighted in boldface.} \label{tab:lu-et-al-short} 
\end{table*}

Table \ref{tab_overall} shows that ARGOX gives the leading performance uniformly through all flu seasons in all metrics. Particularly, ARGOX achieves up to 28\% error reduction in MSE and about 15 \% error reduction in MAE compared to the best alternative in the whole period. ARGOX also keeps consistent season-by-season performance, with at least 15\% error reduction in MSE compared to the best alternative method in every season from 2014 to 2019. For the current 2019-2020 flu season with the (onset of) COVID-19 pandemic, ARGOX's accuracy still maintains. 
Compared with other benchmarks, ARGOX's advantages in state-level flu tracking are substantial. VAR and GFT fail to outperform the naive method in any of the evaluated flu seasons; both methods have MSE two or three times larger than the naive method. 
Table \ref{tab:lu-et-al-short} shows that ARGOX also uniformly outperforms Lu et al. (2019) in all three seasons when the benchmark is available. More detailed results comparing ARGOX with the benchmarks can be found in the Supplementary Information (Table \ref{tab:lu-et-al-full}). 
The advantage of ARGOX over the method of Lu et al. (2019) \cite{lu2019improved} could be attributed to (i) incorporating multi-resolution information in the modeling that pools national, regional and state-level information together, (ii) capturing the spatio-temporal information using one joint statistically structured variance-covariance matrix as opposed to ad hoc regression of each individual state's \%ILI on other states', and (iii) using a statistically principled and interpretable method to dichotomously select between either joint modeling for statistically ``connected'' states or stand-alone modeling for statistically/geographically ``disconnected'' states.

\begin{figure*}
\centering
\includegraphics[width=15.8cm]{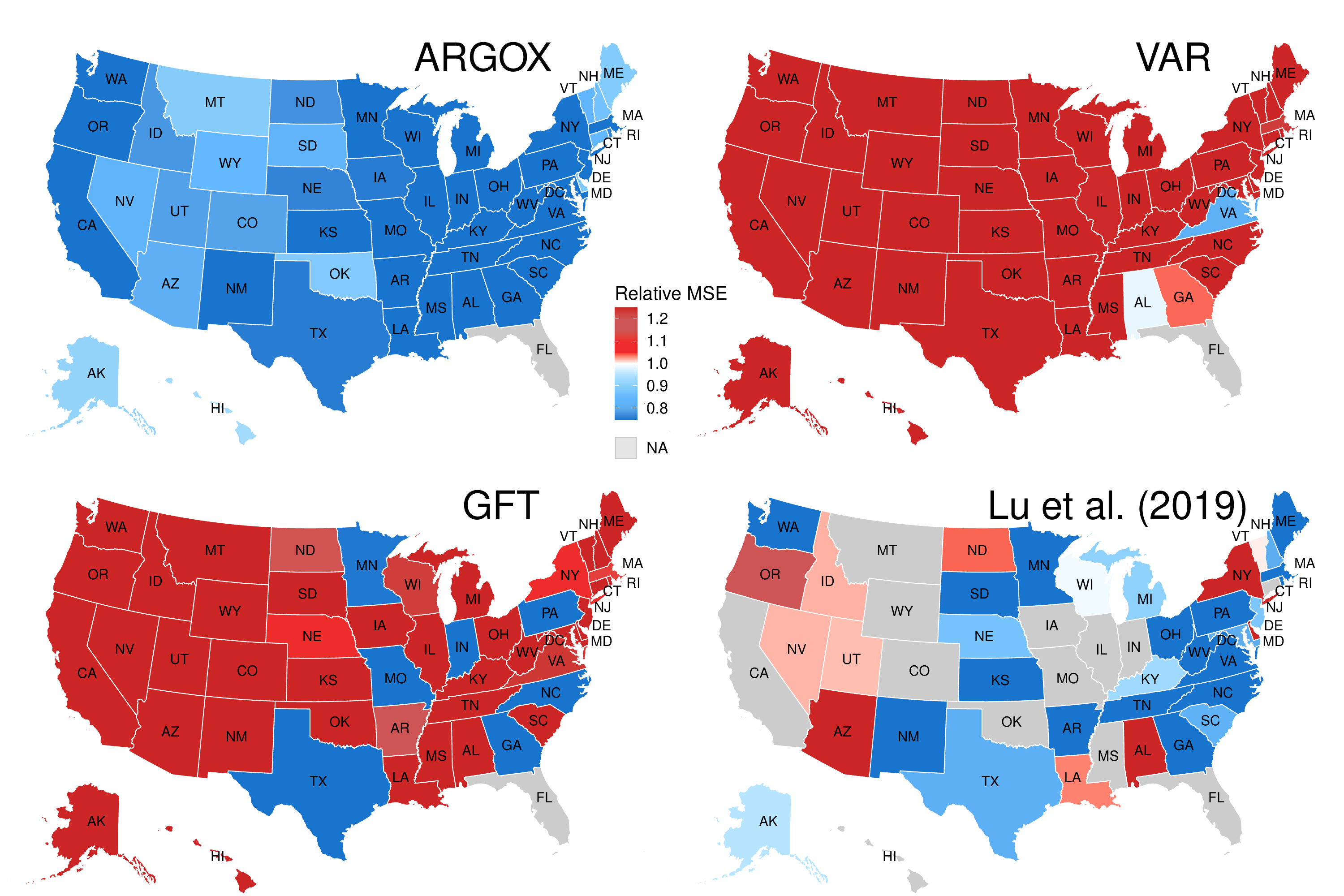}
\captionsetup{width=.9\linewidth}
\caption{State-by-state Heatmap of Relative Mean Squared Error of ARGOX, VAR, GFT, and Lu et al. (2019) \cite{lu2019improved} to the naive method. The relative MSE is the ratio of the MSE of a given method to that of the naive method. Blue color means smaller MSE (i.e., better performance) than the naive method; red color means larger MSE (i.e., worse performance ) than the naive method; grey color means result not available. ARGOX with all blue colors uniformly dominates the naive method, while mixed colors in the rest of the plots show that VAR, GFT, and Lu et al. (2019) were worse than the naive method in a large proportion of states. ARGOX and VAR are evaluated for the whole period of Oct 11, 2014 to March 21, 2020; GFT is evaluated for the period of Oct 11, 2014 to August 15, 2015 due to GFT data availability; Lu et al. (2019) is evaluate from Oct 11, 2014 to May 14, 2017 due to its availability. The figure was generated by the programming language \texttt{R}. The US maps were drawn based on the publicly available \texttt{R} package \texttt{urbnmapr}, which uses map shapefiles from the US Census Bureau (\url{https://www.census.gov/geographies/mapping-files/time-series/geo/tiger-line-file.html}).}
\label{fig_heatmap}
\end{figure*}

Among all the methods that we numerically compared, ARGOX is the only one that uniformly outperforms the naive method in all 51 states/district/city in terms of MSE for the whole period of evaluation. Fig \ref{fig_heatmap} plots the state-by-state estimation results, showing the ratio of the MSE of a given method to the MSE of the naive method. The results of four methods are plotted: ARGOX, VAR, GFT, and Lu et al. (2019). For each state, a blue color means that the MSE of a method is smaller (better) than the MSE of the naive method for that state, and a red color means the MSE of the method is larger (worse) than the MSE of the naive method. Darker blue means more advantage over the naive method, while darker red means more disadvantage than the naive method. It is noteworthy that ARGOX with all blue colors is the only method that gives uniformly better performance than the native method across all states. All other methods in comparison fail to do so for a large portion of the states investigated. Note that the naive method provides a model-free baseline benchmark that solely relies on information from CDC's flu reports. Therefore, ARGOX is the only method that effectively utilizes the Internet data to uniformly improve flu tracking from the traditional surveillance system, indicating ARGOX's reliability and adaptability. With its universally enhanced accuracy over the alternative methods for real-time state-level flu situation estimate, it appears that ARGOX could aid timely, proper public health decision making for the local monitoring and control of the disease. 

Detailed numerical results for each state and for each flu season are reported in Tables \ref{tab_state1}-\ref{tab_state51} and the figures in Supporting Information (SI), 
where ARGOX holds lead over other methods in the vast majority of the cases, 
further revealing its robustness over geographical and seasonal variability in flu epidemics.

In addition to the point estimate, ARGOX also provides 95\% confidence intervals for each week's estimates. For the entire period from 2014 to 2020, over all 51 states/district/city, the intervals provided by ARGOX successfully cover the actual \%ILI in 92.5\% of the cases (Table \ref{tab_CI}), which is close to the nominal 95\%, demonstrating ARGOX's accurate uncertainty quantification. 

\section*{Discussion}

ARGOX effectively combines state-, regional-, and national-level publicly available data from Google searches and CDC's traditional flu surveillance system. It incorporates geographical and temporal correlation of flu activities to provide accurate, reliable real-time flu tracking at the state level. Across all the available states, ARGOX outperforms time-series-based benchmark models, GFT, and the method of Lu et al. (2019). ARGOX's weekly \%ILI estimations are accompanied by reliable interval estimates as a measure for uncertainty. The state-level real-time tracking of flu epidemics by ARGOX could help public health officials and the general public to make more informed decisions to control and prevent the flu epidemics at the state or local levels. In particular, with the real-time estimates of flu activities by ARGOX in their home states and neighboring states, local public health officials could make more proper and timely decisions on the allocation of relevant resources, such as vaccines, hospitalization, medical equipment, personnel, etc. Also, informed with the current local flu situation provided by ARGOX, the general public could take necessary measures accordingly, such as taking the flu shot, social distancing, and mask wearing to reduce the risk of contracting flu; knowing the real-time flu severity at other states could help the general public make travel decisions and plan/arrange care for relatives and friends. More discussion on the usefulness of influenza forecasts to public health decision making can be found in Ref. \citenum{biggerstaff2016results} and Ref. \citenum{reich2019accuracy}.

ARGOX's adaptive pooling of the most-relevant information among the 51 US states/district/city 
plays an important role in its performance. 
To avoid the possibility of overfitting, a structured covariance matrix on the \%ILI increments is utilized. 
Such structured dynamic modeling of the cross-state covariance serves to capture the ever-changing geographic spread pattern of the flu. It aggregates state-to-state, time-varying connectivity factors such as commuting traffic, airline frequency, geographic proximity, and climatic patterns. The utilization of cross-state correlation also helps pool information from different states, regions and the entire nation in addition to the information at a given state. The pooling from national and regional level estimates incorporates the shared seasonality component in flu trends across all the states, which further helps reduce the risk of overfitting.

ARGOX operates in two steps: the first step extracts Internet search information at the state level, and the second step enhances the estimates using cross-state and cross-resolution information (detailed in Methods). Such two-step design of ARGOX has broad applicability. With the general availability of ubiquitous Internet search data, ARGOX's two-step framework could be flexibly adapted to track flu activities at even higher resolutions, such as county or city levels, when such weekly \%ILI data become available. In addition, the first step could be substituted by other models or include other data sources, while the second step remains adaptable for multi-resolution spatial-temporal boosting. A wide spectrum of flu estimation models, including susceptible-infectious-recovered-susceptible model \cite{Shaman_2012}, empirical Bayes method \cite{brooks2015flexible}, Wisdom-of-crowds forecast \cite{Farrow2017-dj}, or ensemble of them \cite{Santillana2015_ensemble} can be fitted into the cross-state boosting step (the second step) of ARGOX.

Like all big-data-based models, our result has certain limitations. ARGOX's accuracy depends on the reliability of its inputs -- Google Trends data and historical \%ILI data from CDC. Google Trends data have increasing amount of missing data and zero counts as the resolution goes from national to regional and state levels (Table  \ref{tab:gt-zeros}). Such degeneracy in data quality is a challenge for high-resolution inference. Google search information could also be sensitive to media coverage  \cite{Lazer_etal_14,butler2013, lampos2020tracking}. Furthermore, Google search data may only be representative of the search interests among Google users rather than the entire population. In states with less Internet penetration, such Google search data may be less predictive of the overall \%ILI. 
The $L_1$ penalty and the dynamic training of ARGOX aims to correct for the sparsity, over-shooting, and representative issues of Google data, where only the most relevant search terms to \%ILI estimation are selected at each state's level. Models to further alleviate or eliminate the bias  in Internet search data (e.g. by incorporating data on media coverage intensity) could be an interesting future direction. In addition, we should be aware that our estimation target, the CDC’s \%ILI, is only a proxy for the true flu incidence in the population, as it's calculated from a sample of outpatient visits with influenza-like symptoms. 
The reported \%ILI at the state level could have (i) high noise due to its limited sample size, (ii) subsequent revision when healthcare providers update their information, and (iii) bias towards those with easy healthcare access. Nevertheless, accurate estimation of  CDC’s \%ILI at the state level is  valuable for optimizing resource allocations. More detailed discussion about the importance of alternative indicators for flu incidence in the population can be found in Ref. \citenum{lipsitch_2011,nsoesie2014systematic,chretien2014influenza}.

ARGOX is accurate, reliable, flexible and generalizable, making it adaptable to other spatial and temporal resolutions for tracking or forecasting other diseases and social/economic events that leave traces on people's Internet activity records. 
The ARGOX framework can be potentially adapted for COVID-19 tracking by incorporating additional coronavirus-related query terms at city, state, regional, and national level \cite{stephens-davidowitz_2020}. With the current development of COVID-19 pandemic, it is likely that the coronavirus would come back in the future winters. In light of this, accurate localized tracking of epidemic activity has become more important than ever before.

\section*{Methods}

\subsection*{CDC's ILINet data}
Every Friday, CDC releases a report of \%ILI for the previous week, which gives the percent of outpatient visits with influenza-like illness for the whole nation, each HHS region, each state (except Florida), Washington DC, and New York City (separated from New York State) (\url{http://www.cdc.gov/flu/weekly/overview.htm}). CDC also revises the initial report numbers in the subsequent weeks when more information become available (\url{gis.cdc.gov/grasp/fluview/fluportaldashboard.html}). Consequently, CDC's \%ILI data lag behind real-time for up to 2 weeks and are less accurate for more recent weeks. CDC's \%ILI data for this study were downloaded on Mar 27, 2020.

\subsection*{Google Data}

The Internet search volume data from Google are publicly available through Google Trends (\url{trends.google.com}). A user can specify the desired query term, geographical location, and time frame on Google Trends; the website then will return a (weekly) time series in integer values from 0 to 100, which corresponds to the normalized search volume of the query term within the specified time frame, where 100 represents the historical maximum, and 0 represents missing data due to inadequate search intensity. This integer-valued time series from Google Trends is based on sampling Google's raw search logs.

The search query terms that we use are based on previous work for national and regional flu estimation \cite{yang2015accurate, ning2019accurate}. We also included several additional queries and topics in this study, which were obtained from ``Related queries'' and ``Related topics'' on the Google Trends website when searching for flu related information. Table \ref{tab:query} in the Supplementary Information lists these search terms.

As one benchmark, we downloaded the discontinued Google Flu Trends (GFT) data (\url{https://www.google.org/flutrends/about/data/flu/us/data.txt}). GFT has national, regional, and state-level prediction for the weekly \%ILI from Jan 1, 2004 to August 9, 2015. 

Google search data may only be representative of the search interests among Google users rather than the entire population. The ARGOX attempts to correct for such potential bias in the modeling.

\subsection*{Regional-Enrichment of state-level Google search data}
Google Trends provides (normalized) search volume data at both national and state levels. However, for the state-level data, there is a high level of sparsity (i.e., zero observations) among the returned integer-valued time series (see Table \ref{tab:gt-zeros}). These zeros, which correspond to missing data due to inadequate search intensity, significantly lower the data quality at the state level (compared to the national level), which in turn severely reduces the prediction accuracy at the state level. To enhance the predictive power of state-level Google data, we use a simple approach to borrow information from the regional level. First, we reconstruct regional-level search frequency for each region in the US by weighting the state-level search frequencies within a given region, where the weights are proportional to the state's population. Second, instead of using the state-level Google Trends time-series, for each search term, we use a weighted average of the state-level search frequency (2/3 weight) and the regional-level search frequency (1/3 weight) as the input for state-level \%ILI estimation. We carry out this regional-enrichment process for all states/district/city, except seven states -- Hawaii (HI), Alaska (AK), Vermont (VT), Montana (MT), North Dakota (ND), Maine (ME), and South Dakota (SD) -- because these seven states are modeled with a separate stand-alone model (as detailed in the following sections). For these seven states, the raw Google Trends state-level times series, not the regional-enriched time series, are used as input.

\subsection*{Evaluation metrics}
We use three metrics to evaluate the accuracy of an estimate against the actual \%ILI released by CDC: the mean squared error (MSE), the mean absolute error (MAE), and the Pearson correlation (Correlation). MSE between an estimate $\hat{p}_t$ and the true value $p_t$ over period $t=1,\ldots, T$ is $\frac{1}{T}\sum_{t=1}^T \left(\hat{p}_t - p_t\right)^2$. MAE between an estimate $\hat{p}_t$ and the true value $p_t$ over period $t=1,\ldots, T$ is $\frac{1}{T}\sum_{t=1}^T \left|\hat{p}_t - p_t\right|$. 
Correlation is the Pearson correlation coefficient between $\hat{\boldsymbol{p}}=(\hat{p}_1, \dots, \hat{p}_T)$ and $\boldsymbol{p}=(p_1,\dots, p_T)$.


\subsection*{Prediction model of ARGOX}
ARGOX operates in two steps: the first step extracts Internet search information at the state level, and the second step enhances the estimates using cross-state and cross-resolution information. 

At the second step, we take a dichotomous approach for the 51 US states/district/city (50 states except Florida, which does not have \%ILI data, plus Washington DC and New York City). We set apart seven states: HI, AK, VT, MT, ND, ME, and SD. The first two (HI and AK) are geographically separated from the contiguous US. The last five (VT, MT, ND, ME, and SD) are the states that have the lowest multiple correlations (a.k.a. the $R$) in \%ILI to the \%ILI of the entire nation, the \%ILI of the other states, and the \%ILI of the other regions (detailed calculation method is given in Supplementary Information). 
A low multiple correlation of a state implies that the state's flu activity is not well correlated with other states' or other regions'. 
For these seven states, due to either the geological discontinuity or the low multiple correlation, it is not clear if using information cross the other states or other regions can help the state-level \%ILI estimation. Therefore, we adopt the dichotomous approach: For the 44 states/district/city (the vast majority), 
we apply a joint estimation approach at the second step to enhance the state-level \%ILI estimation by using all information, including information from other states and other regions; for the above-mentioned seven states, we use a stand-alone estimation approach at the second step to enhance the \%ILI estimation (not using information from other states and regions). The two steps of ARGOX are detailed below.

\subsubsection*{First step: extracting Internet search information at the state level}

This step concerns extracting Google search information at each state. In particular, for a given state/district/city $m$, $m= 1, \dots, 51$, let $X_{i,t,m}$ be the logarithm of 1 plus the state-level Google Trends data of search term $i$ at week $t$ (note: 1 is added to each state-level Google Trends data point to avoid taking logarithm of zero); let $y_{t,m}$ be the logit-transformation of CDC's \%ILI at time $t$ for state $m$. 
To estimate $y_{T,m}$, an $L_1$ regularized linear estimator is used in the first step based on the vector $\bm{X}_{T,m} = (x_{i,T,m})$:
\[
\hat{y}_{T,m} = \hat{\beta}_{0,m} + \bm{X}_{T,m}^{^\intercal}\hat{\bm{\beta}}_m,
\]
where the coefficients $(\hat{\beta}_{0,m}, \hat{\bm{\beta}}_m)$ are obtained via
\begin{equation}\label{eqn:argo_1st}
\underset{\beta_{0,m}, \bm{\beta}_m}{\mathrm{argmin}}\sum_{t=T-N}^{T-1} \left(y_{t,m} -\beta_{0,m} - \bm{X}^{^\intercal}_{t,m}\bm{\beta}_m\right)^2 + \lambda\|\bm{\beta}_m\|_1.
\end{equation}
We set $N=104$, i.e., a two-year window, as recommended in previous studies \cite{yang2015accurate,ning2019accurate,Burkom_etal_2007}. We set $\lambda$ through cross-validation. 

In addition, we obtain an accurate estimate $\hat{p}^{nat}_T$ for the national \%ILI by using the ARGO method \cite{yang2015accurate}, which uses national level Google search data. We also obtain an estimate $(\hat{p}_{T, 1}^{reg}, \dots,\hat{p}_{T,10}^{reg})$ for the ten HHS regional \%ILI by the {first step of} ARGO2 method \cite{ning2019accurate}, which uses aggregated regional level Google search data.

\subsubsection*{Second step: joint model for the 44 states/district/city other than HI, AK, ND, VT, MT, ME, and SD}  

For the 44 states, let $\bm{p}_t=(p_{t,1},\dots, p_{t,44})^\intercal$ denote CDC's \%ILI at the state level; they are related to $y_{t,m}$ through $p_{t,m}=\exp(y_{t,m})/(1+\exp(y_{t,m}))$. Our raw estimate for $\bm{p}_t$ from the first step is $\hat{\bm{p}}^{GT}_{t} = (\hat{p}_{t,1},\dots, \hat{p}_{t,44})^\intercal$, where $\hat{p}_{t,m} = \exp(\hat{y}_{t,m})/(1+\exp(\hat{y}_{t,m}))$. Our estimate of the national \%ILI from the first step is $\hat{p}_t^{nat}$. 
Let the boldface $\hat{\bm{p}}_t^{nat}$ denote the length-44 vector $\hat{\bm{p}}_t^{nat}=(\hat{p}_t^{nat}, \dots,\hat{p}_t^{nat})^\intercal$. We also have the regional \%ILI estimate $(\hat{p}_{t, 1}^{reg}, \dots,\hat{p}_{t,10}^{reg})$ from the first step. Let $\hat{\bm{p}}_t^{reg}$ denote the length-44 vector $\hat{\bm{p}}_t^{reg}=(\hat{p}_{t, r_1}^{reg}, \dots,\hat{p}_{t, r_{44}}^{reg})^\intercal$, where $r_m$ is the region number for state $m$. 

Estimating $\bm{p}_t$ is equivalent to estimating the time series increment $\Delta\bm{p}_{t} = \bm{p}_{t} - \bm{p}_{t-1}$. We denote $\bm{Z}_{t} = \Delta\bm{p}_{t}$ for notational simplicity. For the estimation of $\bm{Z}_{t}$, we want to incorporate the cross-state, cross-source correlations. We have four predictors for $\bm{Z}_{t}$ after the first step: (i) $\bm{Z}_{t-1}=\Delta\bm{p}_{t-1}$, (ii) $\hat{\bm{p}}_{t}^{GT} - \bm{p}_{t-1}$, (iii) $\hat{\bm{p}}_{t}^{reg} - \bm{p}_{t-1}$, and (iv) $\hat{\bm{p}}_{t}^{nat} - \bm{p}_{t-1}$
; they represent time series information, information from the state level Google search, information from the regional level estimation, and information from the national level estimation, respectively. Let $\bm{W}_{t}$ denote the collection of these four vectors $\bm{W}_{t}=(\bm{Z}_{t-1}^\intercal, (\hat{\bm{p}}_{t}^{GT} - \bm{p}_{t-1})^\intercal, (\hat{\bm{p}}_{t}^{reg} - \bm{p}_{t-1})^\intercal, (\hat{\bm{p}}_{t}^{nat} - \bm{p}_{t-1})^\intercal)^\intercal$.

To combine the four predictors, we use the best linear predictor formed by them:
\begin{align}
\hat{\bm{Z}}_{t} = \bm{\mu}_{Z} + \Sigma_{ZW}\Sigma_{WW}^{-1}(\bm{W}_t-\bm{\mu}_{W}),\label{eq: blp}
\end{align}
where $\mu_Z$ and $\mu_W$ are the mean vectors of $\bm{Z}$ and $\bm{W}$ respectively, and $\Sigma_{ZZ}$, $\Sigma_{ZW}$, and $\Sigma_{WW}$ are the covariance matrices of and between $\bm{Z}$ and $\bm{W}$. The best linear predictor gives the optimal way to linearly combine the four predictors to form a new one. The variance of $\hat{\bm{Z}}_{t}$ is
\begin{equation}
\mathrm{Var}(\hat{\bm{Z}}_{t}|\bm{W}_t) = \Sigma_{ZZ} - \Sigma_{ZW}\Sigma_{WW}^{-1}\Sigma_{WZ}.
\label{eq:var}
\end{equation}

Consistent with the first step, we adopt a sliding two-year training window to estimate $\mu_Z$, $\mu_W$, $\Sigma_{ZZ}$, $\Sigma_{ZW}$, and $\Sigma_{WW}$ in Eq. \eqref{eq: blp} and \eqref{eq:var}. For $\mu_Z$ and $\mu_W$, we use the empirical mean of the corresponding variables as the estimates. However, for the covariance matrices, due to their large sizes and the small number of observations, we need to structure the covariance matrices for reliable estimation.

We assume the following structure: 
\begin{enumerate}
\item The covariances between the time series increments satisfy $\mathrm{Var}(\bm{Z}_{t})=\mathrm{Var}(\bm{Z}_{t-1})=\Sigma_{ZZ}$ and $\mathrm{Cov}(\bm{Z}_{t}, \bm{Z}_{t-1})=\rho \Sigma_{ZZ}$, where $0<\rho<1$. This essentially assumes that the time series increments are stationary and have a stable autocorrelation across time and states.
\item Independence among the different sources of information: time series increment, the estimation error of the first-step state-level estimate, the estimation error of the regional estimate, and the estimation error of the national estimate, i.e., $\bm{Z}_{t}, \hat{\bm{p}}_{t}^{GT} - \bm{p}_{t}, \hat{\bm{p}}_{t}^{reg} - \bm{p}_{t}, \hat{\bm{p}}_{t}^{nat} - \bm{p}_{t}$ are all mutually independent. 
\end{enumerate}

The covariance matrices are thereby simplified as:
\begin{equation}\label{eq:sigma_zw}
\Sigma_{ZW}=
\begin{pmatrix}
\rho\Sigma_{ZZ}&
\Sigma_{ZZ}&
\Sigma_{ZZ}&
\Sigma_{ZZ}
\end{pmatrix}
\end{equation}
\begin{equation}\label{eq:sigma_ww}
\Sigma_{WW}=  
 \begin{pmatrix}
\Sigma_{ZZ}&\rho\Sigma_{ZZ}& \rho\Sigma_{ZZ}& \rho\Sigma_{ZZ}\\
\rho\Sigma_{ZZ}& \Sigma_{ZZ} + \Sigma^{GT}& \Sigma_{ZZ}& \Sigma_{ZZ}\\
\rho\Sigma_{ZZ}& \Sigma_{ZZ}& \Sigma_{ZZ} +\Sigma^{reg}& \Sigma_{ZZ}\\
\rho\Sigma_{ZZ}& \Sigma_{ZZ}& \Sigma_{ZZ}& \Sigma_{ZZ} +\Sigma^{nat}\\
\end{pmatrix}
\end{equation}
where $\Sigma^{reg} = \mathrm{Var}(\hat{\bm{p}}_{t}^{reg} - \bm{p}_{t})$, $\Sigma^{nat} = \mathrm{Var}(\hat{\bm{p}}_{t}^{nat} - \bm{p}_{t})$, and $\Sigma^{GT} = \mathrm{Var}(\hat{\bm{p}}_{t}^{GT} - \bm{p}_{t})$. 
To further control the estimation stability, we incorporate a ridge-regression-inspired shrinkage \cite{hoerl1970ridge} to the linear predictor \eqref{eq: blp}, replacing the joint covariance matrix of $(\bm Z_t^\intercal, \bm W_t^\intercal)^\intercal$ by the average of the structured covariance matrix and its empirical diagonal. 
Effectively, in Eq. \eqref{eq: blp}, $\Sigma_{ZW}$ is replaced by $\frac{1}{2} \Sigma_{ZW}$, and $\Sigma_{WW}$ is replaced by $(\frac{1}{2}\Sigma_{WW}+\frac{1}{2}D_{WW})$, where $D_{WW}$ is the diagonal of the empirical covariance of $\bm{W}_t$:

\begin{equation}
\hat{\bm{Z}}_{t} = \bm{\mu}_{Z} + \frac{1}{2}\Sigma_{ZW}(\frac{1}{2}\Sigma_{WW}+\frac{1}{2}D_{WW})^{-1}(\bm{W}_t-\bm{\mu}_{W}).\label{eq:2ndstep}
\end{equation}

$\Sigma_{ZZ}$, $\Sigma^{nat}$, $\Sigma^{reg}$, $\Sigma^{GT}$ and $D_{WW}$ are estimated by the corresponding sample covariance from the data in the most recent 2-year training window; $\rho$ is estimated by minimizing the Frobenius norm ($L_2$ distance) between the empirical correlation and structured correlation. Based on Eq. \eqref{eq:var}, the variance estimate is similarly updated by 
\[
\mathrm{Var}(\hat{\bm{Z}}_{t}|\bm{W}_t) = \Sigma_{ZZ} - \frac{1}{2}\Sigma_{ZW}(\frac{1}{2}\Sigma_{WW}+\frac{1}{2}D_{WW})^{-1}\frac{1}{2}\Sigma_{WZ}.
\]

Our final state-level \%ILI estimate for week $T$ after the second step is:
\begin{equation}\label{eq:2ndstep_est}
{\hat{\bm{p}}}_{T} = {\bm{p}}_{T-1} + \hat{\bm{\mu}}_Z + \hat{\Sigma}_{ZW}(\hat{\Sigma}_{WW}+\hat{D}_{WW})^{-1}(\bm{W}_T-\hat{\bm{\mu}}_{W}),
\end{equation}
with corresponding 95\% interval estimate
\[
\left[{\hat{\bm{p}}}_{T} \pm 1.96 \cdot \sqrt{\mathrm{diagonal} \left(\hat\Sigma_{ZZ} - \frac{1}{2}\hat\Sigma_{ZW}(\hat\Sigma_{WW}+\hat{D}_{WW})^{-1}\hat\Sigma_{WZ} \right)}\right].
\]

\subsubsection*{Second step: stand-alone model for HI, AK, ND, VT, MT, ME and SD}
For $m \in \{\text{HI, AK, ND, VT, MT, ME, SD}\}$, we take a stand-alone modeling approach. For each of these states, which is either non-contiguous or has the lowest multiple correlation with out-of-state \%ILI (detailed in Supplementary Information), we focus on estimating the individual state's \%ILI by integrating the within-state and national information in the second step. Thereby, our target is a scalar $Z_{t}^{(m)} = p_{t, m} - p_{t-1, m}$, the state's \%ILI increment at the current week. The predictor vector in the second step for state $m$ is $\bm{W}_{t}^{(m)} =({Z}_{t-1}^{(m)}, (\hat{{p}}_{t, m}^{GT} - {p}_{t-1, m}), (\hat{{p}}_{t}^{nat} - {p}_{t-1, m}))$, where the regional terms are dropped. The best linear predictor with ridge-regression inspired shrinkage is then used to get the final estimate
\begin{equation}
\hat{{Z}}_{t}^{(m)} = {\mu}_{Z}^{(m)} + \frac{1}{2}\Sigma_{ZW}^{(m)}(\frac{1}{2}\Sigma_{WW}^{(m)}+\frac{1}{2}D_{WW}^{(m)})^{-1}(\bm{W}_t^{(m)}-\bm{\mu}_{W}^{(m)}).\label{eq:2ndstep_ind}
\end{equation}

The corresponding covariance matrices between the components $\Sigma_{ZW}^{(m)} = \mathrm{Cov}(Z^{(m)}, \bm{W}^{(m)})$, $\Sigma_{WW}^{(m)} = \mathrm{Var}(\bm{W}^{(m)})$, and $D_{WW}^{(m)} = \mathrm{diagonal}(\Sigma_{WW}^{(m)})$ are estimated by the corresponding sample covariance from the data in the most recent 2-year training window. 

The final state-level \%ILI estimate for week $T$ after the second step for $m \in \{\text{HI, AK, ND, VT, MT, ME, SD}\}$ is:
\begin{equation}\label{eq:2ndstep_est_ind}
\hat{{p}}_{T, m} = {{p}}_{T-1, m} + \hat{\mu}_{Z}^{(m)} + \hat{\Sigma}_{ZW}^{(m)}(\hat{\Sigma}_{WW}^{(m)}+\hat{D}_{WW}^{(m)})^{-1}(\bm{W}_T^{(m)}-\hat{\bm \mu}_{W}^{(m)}),
\end{equation}
with corresponding 95\% interval estimate
$$
\left[{\hat{{p}}}_{T,m} \pm 1.96 \cdot \sqrt{\hat \Sigma_{ZZ}^{(m)} - \frac{1}{2}\hat\Sigma_{ZW}^{(m)}(\hat\Sigma_{WW}^{(m)}+\hat{D}_{WW}^{(m)})^{-1}\hat\Sigma_{WZ}^{(m)} }\right],
$$
where $\Sigma_{ZZ}^{(m)} = \mathrm{Var}(Z^{(m)})$ is the scalar variance of the univariate time series $Z_{t}^{(m)}$.

\subsection*{Availability of data and material}
All analyses were performed with the R statistical software \cite{rcore}. The R package that implements the ARGOX method is available on CRAN at \url{https://cran.r-project.org/web/packages/argo/}, which uses the \texttt{glmnet} package \cite{glmnet}. All datasets analyzed in the current study are available in the Harvard Dataverse repository, doi:XXX/XXX/XXXX.

\section*{Acknowledgements}
SCK's research was supported in part by National Science Foundation grant DMS-1810914. The authors thank Professor Herman Chernoff for helpful comments. All analyses were performed with the R statistical software \cite{rcore}. The R package that implements the ARGOX method is available on CRAN at \url{https://cran.r-project.org/web/packages/argo/}, which uses the \texttt{glmnet} package \cite{glmnet}. All datasets analyzed in the current study are available in the Harvard Dataverse repository, \url{https://doi.org/10.7910/DVN/2IVDGK}.

\section*{Author contributions statement}
S.Y. and S.N. contributed equally to this work. S.Y., S.N., and S.C.K. designed the research; S.Y., S.N., and S.C.K. performed the research; S.Y. and S.N. analyzed data; and S.Y., S.N., and S.C.K. wrote the paper.

\section*{Additional information}

\subsection*{Competing interests}
The authors declare that they have no conflict of interest.

\clearpage
\renewcommand{\appendixtocname}{Supplementary Material}
\renewcommand\appendixname{Supplementary Material}
\renewcommand\appendixpagename{Supplementary Material}

\begin{appendices}
\beginsupplement
This Supplementary Material is organized as following: (1) the detailed calculation procedure for the multiple correlation behind the stand-alone modeling of HI, AK, ND, VT, MT, ME, and SD is presented; (2) a table for the confidence interval coverage is presented; (3) all the Google query terms used in this study are listed; (4) Google Trends data quality at different geographic area is studied; (5) full comparison to another Google-search-based benchmark method is presented; (6) detailed estimation results for each of 51 studied states/district/city are reported in tables and plotted in figures.

\section*{Multiple correlation}
For each state, the multiple correlation of its flu activity level to the other states', other regions' and the national flu activity levels is calculated as follows. First, the states of HI and AK are excluded because they are not part of the contiguous US; the state of FL is excluded because FL data is not available from CDC. Then for the in-sample time period of 2010-10-09 to 2014-09-27, we regress each state's \%ILI to (i) all other 48 states' \%ILI (including DC and NYC but excluding FL, HI, and AK), (ii) all the other 9 regions' \%ILI (i.e., regions other than the one that the specific state belongs to), and (iii) the national \%ILI. After the regression, we obtain the R-squared, which is the square of multiple correlation. The five states with the lowest multiple correlations are ND, VT, MT, ME, and SD. We, therefore, would not use spatial pooling on HI, AK, ND, VT, MT, ME, and SD. Instead, we only use the state-specific data together with national level data for cross-resolution boosting on those seven aforementioned states in the second step of ARGOX.


\section*{Confidence interval coverage}
We study the goodness of our confidence intervals by examining its actual coverage (coverage of the actual \%ILI released by CDC weeks later). The result is shown in Table \ref{tab_CI}. In general, the coverage of 95\% confidence interval is quite close to the nominal value, suggesting that our model quantifies the uncertainty reasonably well.

\begin{table}[ht]
\small
\centering
\begin{tabular}{llllllllll}

  \hline
 AL &  AK &  AZ &  AR &  CA &  CO &  CT &  DE &  DC &  GA \\
0.919 & 0.916 & 0.909 & 0.930 & 0.944 & 0.905 & 0.909 & 0.930 & 0.923 & 0.926 \\
   \hline
   
      \hline
   HI &  ID &  IL &  IN &  IA &  KS &  KY &  LA &  ME &  MD \\

  0.947 & 0.958 & 0.930 & 0.926 & 0.926 & 0.940 & 0.867 & 0.916 & 0.916 & 0.947 \\
     \hline
     
        \hline
   MA &  MI &  MN &  MS &  MO &  MT &  NE &  NV &  NH &  NJ \\ 
  0.937 & 0.947 & 0.940 & 0.930 & 0.923 & 0.937 & 0.909 & 0.916 & 0.898 & 0.930 \\
     \hline  
     
     \hline
   NM &  NY &  NC &  ND &  OH &  OK &  OR &  PA &  RI &  SC \\
0.933 & 0.902 & 0.951 & 0.944 & 0.926 & 0.937 & 0.926 & 0.930 & 0.909 & 0.926 \\ 
     \hline
     
        \hline
   SD &  TN &  TX &  UT &  VT &  VA &  WA &  WV &  WI &  WY \\
  0.940 & 0.919 & 0.926 & 0.898 & 0.947 & 0.951 & 0.905 & 0.909 & 0.937 & 0.874 \\    \hline
  
   \hline
   NYC &  &  &  &  &  &  &  &  &  \\ 
    0.912 &  &  &  &  &  &  &  &  &  \\ 
   \hline
\end{tabular}
\caption{The actual coverage of the confidence intervals by ARGOX for 51 states/district/city. The coverage is for 95\% nominal confidence level. The average coverage over all the 51 states/district/city is 92.5\%.} \label{tab_CI}
\end{table}

\clearpage
\section*{Query terms for Google Trends}
Table \ref{tab:query} lists all query terms/phrases used in this study. Most of them are taken from previous studies \cite{yang2015accurate, ning2019accurate} with a few additional terms identified through ``Related topics'' and ``Related queries'' from Google Trends when search for flu-related information.

\begin{table}[ht]
\footnotesize
\centering
\caption{All search query terms used in this study. The last 21 terms separated by a horizontal line from the first 140 terms are new ``Related topics'' and ``Related queries'' identified from Google Trends.}\label{tab:query}
\begin{tabular}{llll}
  \hline
flu incubation & flu incubation period & influenza type a & symptoms of the flu \\ 
  flu symptoms & influenza symptoms & flu contagious & influenza a \\ 
  a influenza & symptoms of flu & flu duration & influenza incubation \\ 
  type a influenza & flu treatment & symptoms of influenza & influenza contagious \\ 
  flu in children & cold or flu & symptoms of bronchitis & flu recovery \\ 
  tessalon & influenza incubation period & symptoms of pneumonia & tussionex \\ 
  signs of the flu & flu treatments & remedies for the flu & walking pneumonia \\ 
  flu test & tussin & upper respiratory & respiratory flu \\ 
  acute bronchitis & bronchitis & sinus infections & flu relief \\ 
  painful cough & how long does the flu last & flu cough & sinus \\ 
  expectorant & strep & strep throat & influenza treatment \\ 
  flu reports & flu remedy & robitussin & rapid flu \\ 
  treatment for the flu & chest cold & cough fever & oscillococcinum \\ 
  flu fever & treat the flu & how to treat the flu & over the counter flu \\ 
  how long is the flu & flu medicine & flu or cold & normal body \\ 
  is flu contagious & treat flu & body temperature & reduce fever \\ 
  flu vs cold & how long is the flu contagious & fever reducer & get over the flu \\ 
  treating flu & having the flu & treatment for flu & human temperature \\ 
  dangerous fever & the flu & remedies for flu & influenza a and b \\ 
  contagious flu & fever flu & flu remedies & how long is flu contagious \\ 
  cold vs flu & braun thermoscan & fever cough & signs of flu \\ 
  how long does flu last & normal body temperature & get rid of the flu & i have the flu \\ 
  taking temperature & flu versus cold & how long flu & flu germs \\ 
  flu and cold & thermoscan & flu complications & high fever \\ 
  flu children & the flu virus & how to treat flu & pneumonia \\ 
  flu headache & ear thermometer & how to get rid of the flu & flu how long \\ 
  cold and flu & over the counter flu medicine & treating the flu & flu care \\ 
  how long contagious & fight the flu & reduce a fever & cure the flu \\ 
  medicine for flu & flu length & cure flu & exposed to flu \\ 
  low body & early flu symptoms & flu report & incubation period for flu \\ 
  break a fever & flu contagious period & cold versus flu & what to do if you have the flu \\ 
  medicine for the flu & flu and fever & flu lasts & incubation period for the flu \\ 
  do i have the flu & type a flu symptoms & flu texas & how long am i contagious with the flu \\ 
  how to break a fever & fever breaks & type a flu & how to bring a fever down \\ 
  how to treat the flu at home & flu how long are you contagious & flu a symptoms & flu \\ \hline
  Influenza vaccine & Influenza & Fever & Influenza A virus \\ 
  Influenza B virus & Common cold & Cough & Sore throat \\ 
  Virus & Avian influenza & Spanish flu & Headache \\ 
  Nausea & Flu season & Oseltamivir & Nasal congestion \\ 
  Canine influenza & Rapid influenza diagnostic test & Theraflu & Dextromethorphan \\ 
  Rhinorrhea &  &  &  \\
   \hline
\end{tabular}
\end{table}

\clearpage
\section*{Google Trends data quality}
As stated at \url{trends.google.com}, the numbers in Google Trends ``represent search interest relative to the highest point on the chart for the given region and time; a value of 100 is the peak popularity for the term; a value of 50 means that the term is half as popular; a score of 0 means there was not enough data for this term.'' As such, the proportion of zeros in the Google Trends data reflects the data quality: higher proportion of zeros indicates lower quality of Google Trends data. Table \ref{tab:gt-zeros} summarizes the average proportion of zeros for the query terms listed in Table \ref{tab:query} in each of the geographic areas. As we can see, Google Trends data at the US national level have far fewer zeros than any of the states, implying a significant drop in quality from national-level data to state-level data.

\begin{table}[ht]
\small
\caption{Average proportion of zeros in Google Trends data for the query terms in Table \ref{tab:query}. Higher proportion of zeros indicates lower quality of Google Trends data since ``a score of 0 means there was not enough data for this term'' (\url{trends.google.com}). The proportion of zeros in Google Trends at the US national level is in the upper sub-table, while the proportions of zeros at state/district/city level are in the lower sub-table. 
}
\label{tab:gt-zeros}
\centering
\begin{tabular}{rl}
  \hline
 & US National \\ 
 
proportion of zeros &  1.37\% \\ 
\hline
\end{tabular}
\vskip0.1in
   \begin{tabular}{llllllllll}
  \hline
AK & AL & AR & AZ & CA & CO & CT & DC & DE & FL \\ 
  
78.33\% & 46.13\% & 56.16\% & 39.53\% & 13.52\% & 42.77\% & 50.99\% & 63.00\% & 73.34\% & 20.93\% \\ 
   \hline

  \hline
GA & HI & IA & ID & IL & IN & KS & KY & LA & MA \\ 
  
29.86\% & 67.14\% & 54.90\% & 64.58\% & 26.39\% & 40.78\% & 55.50\% & 47.06\% & 48.70\% & 37.68\% \\ 
   \hline

  \hline
MD & ME & MI & MN & MO & MS & MT & NC & ND & NE \\ 
  
41.26\% & 67.47\% & 31.78\% & 41.99\% & 40.67\% & 56.99\% & 73.07\% & 30.69\% & 75.61\% & 59.34\% \\ 
   \hline

  \hline
NH & NJ & NM & NV & NY & OH & OK & OR & PA & RI \\ 
  
67.21\% & 34.86\% & 63.73\% & 55.14\% & 18.92\% & 30.64\% & 50.78\% & 48.99\% & 30.17\% & 69.28\% \\ 
   \hline

  \hline
SC & SD & TN & TX & UT & VA & VT & WA & WI & WV \\ 
  
45.97\% & 74.43\% & 38.48\% & 16.77\% & 52.78\% & 32.98\% & 77.42\% & 37.85\% & 42.78\% & 64.07\% \\ 
   \hline

  \hline
WY & NYC &  &  &  &  &  &  &  &  \\ 
  
79.82\% & 18.25\% &  &  &  &  &  &  &  &  \\ 
   \hline

\end{tabular}

\end{table}

\clearpage
\section*{More comparison with the result of Lu et al. (2019)}

Lu et al. (2019) \cite{lu2019improved} proposed another Google-search-based method for state-level influenza tracking, utilizing a network approach. We compare ARGOX with Lu et al. (2019) together with other methods here. The retrospective results of Lu et al. (2019) are available for seasons 2014-15, 2015-16, and 2016-17, and it only studied 37 selected states, which are: AK, AL, AR, AZ, DE, GA, ID, KS, KY, LA, MA, MD, ME, MI, MN, NC, ND, NE, NH, NJ, NM, NV, NY, OH, OR, PA, RI, SC, SD, TN, TX, UT, VA, VT, WA, WI, WV. For completeness, ARGOX, VAR, GFT, and the naive method are compared here for the same period and for the same 37 states. Overall ARGOX takes the lead in this subset of 37 states for all three seasons of 2014-15, 2015-16 and 2016-17.

\begin{table*}[hb]
\centering
\begin{tabular}{crrrr}
  \hline
  & Overall '14-'17 & '14-'15 & '15-'16 & '16-'17 \\ 
  \hline  \multicolumn{1}{l}{MSE}\\ARGOX & \textbf{0.269} & \textbf{0.406} & \textbf{0.163} & \textbf{0.339} \\ 
  VAR & 0.873 & 1.234 & 0.503 & 1.214 \\ 
  GFT & -- & 1.464 & -- & -- \\ 
  Lu et al. (2019) & 0.418 & 0.467 & 0.528 & 0.544 \\ 
  naive & 0.383 & 0.618 & 0.201 & 0.471 \\ 
   \hline  \multicolumn{1}{l}{Correlation}\\ARGOX & \textbf{0.919} & 0.914 & \textbf{0.836} & \textbf{0.890} \\ 
  VAR & 0.808 & 0.809 & 0.684 & 0.754 \\ 
  GFT & -- & \textbf{0.915} & -- & -- \\ 
  Lu et al. (2019) & 0.912 & 0.912 & 0.808 & 0.858 \\ 
  naive & 0.894 & 0.880 & 0.806 & 0.860 \\ 
   \hline
\end{tabular}
\caption{Comparison of different methods for state-level \%ILI estimation, averaging over the 37 states, for the period of 2014 to 2017, due to the availability of Lu et al. (2019). The MSE and Correlation are reported. 
The method with the best performance is highlighted in boldface for each metric in each period. Methods considered here include ARGOX, VAR, GFT, Lu et al. (2019), and the naive method. }\label{tab:lu-et-al-full}
\end{table*}

\clearpage
\section*{Detailed estimation results for each state/district/city}


\begin{table}[ht]
\centering
\begin{tabular}{crrrrrrrr}
  \hline
  & Whole period '14-'20 & Overall '14-'17 & '14-'15 & '15-'16 & '16-'17 & '17-'18 & '18-'19 & '19-'20 \\ 
  \hline  \multicolumn{1}{l}{MSE}\\ARGOX & \textbf{ 0.726} & \textbf{ 1.121} &  2.819 &  0.369 & \textbf{ 0.321} & \textbf{ 0.773} & \textbf{ 0.267} & \textbf{ 0.692} \\ 
  VAR &  1.214 &  1.378 &  2.617 &  0.551 &  1.226 &  1.611 &  1.650 &  1.553 \\ 
  GFT & -- & -- &  7.441 & -- & -- & -- & -- & -- \\ 
  Lu et al. (2019) & -- &  4.875 & \textbf{ 2.509} & 12.152 &  3.437 & -- & -- & -- \\ 
  naive &  1.233 &  1.792 &  4.594 & \textbf{ 0.349} &  0.691 &  1.678 &  0.616 &  1.156 \\ 
   \hline  \multicolumn{1}{l}{MAE}\\ARGOX & \textbf{0.469} & \textbf{0.608} & \textbf{1.022} & 0.472 & \textbf{0.444} & \textbf{0.440} & \textbf{0.328} & \textbf{0.635} \\ 
  VAR & 0.707 & 0.801 & 1.055 & 0.629 & 0.861 & 0.780 & 0.788 & 0.912 \\ 
  GFT & -- & -- & 1.750 & -- & -- & -- & -- & -- \\ 
  Lu et al. (2019) & -- & -- & -- & -- & -- & -- & -- & -- \\ 
  naive & 0.593 & 0.709 & 1.200 & \textbf{0.444} & 0.639 & 0.738 & 0.573 & 0.812 \\ 
   \hline  \multicolumn{1}{l}{Correlation}\\ARGOX & \textbf{0.954} & \textbf{0.928} & 0.917 & 0.826 & 0.955 & \textbf{0.974} & \textbf{0.969} & \textbf{0.929} \\ 
  VAR & 0.929 & 0.912 & 0.924 & 0.844 & 0.883 & 0.959 & 0.906 & 0.846 \\ 
  GFT & -- & -- & \textbf{0.963} & -- & -- & -- & -- & -- \\ 
  Lu et al. (2019) & -- & 0.885 & 0.924 & 0.683 & \textbf{0.969} & -- & -- & -- \\ 
  naive & 0.923 & 0.888 & 0.867 & \textbf{0.851} & 0.906 & 0.942 & 0.925 & 0.883 \\ 
   \hline
\end{tabular}
\caption{Comparison of different methods for state-level \%ILI estimation in Alabama (AL).  The MSE, MAE, and correlation are reported. The method with the best performance is highlighted in boldface for each metric in each period. \label{tab_state1}} 
\end{table}

\begin{figure}[!h] 
  \centering 
\includegraphics[width=\linewidth, page=1]{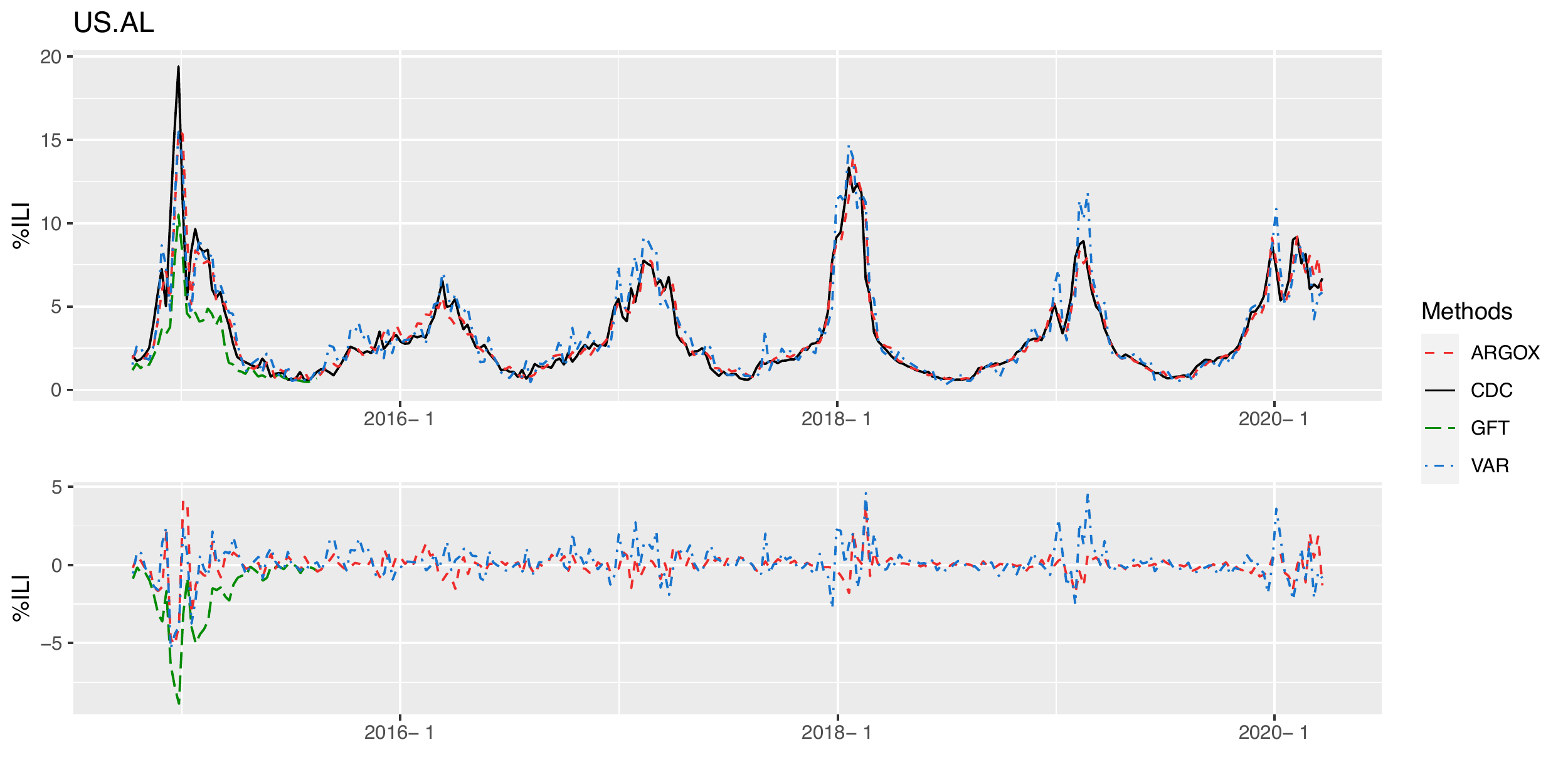} 
\caption{Plots of the \%ILI estimates (top) and the estimation errors (bottom) for Alabama (AL).}
\end{figure}
\newpage      
\begin{table}[ht]
\centering
\begin{tabular}{crrrrrrrr}
  \hline
  & Whole period '14-'20 & Overall '14-'17 & '14-'15 & '15-'16 & '16-'17 & '17-'18 & '18-'19 & '19-'20 \\ 
  \hline  \multicolumn{1}{l}{MSE}\\ARGOX & \textbf{ 0.911} & \textbf{ 0.475} & \textbf{ 0.381} &  0.695 & \textbf{ 0.490} & \textbf{ 0.371} & \textbf{ 1.217} & \textbf{ 4.487} \\ 
  VAR &  4.902 &  4.659 &  6.417 &  4.153 &  5.154 & 10.475 &  2.379 &  8.727 \\ 
  GFT & -- & -- &  0.918 & -- & -- & -- & -- & -- \\ 
  Lu et al. (2019) & -- &  0.510 &  0.450 & \textbf{ 0.692} &  0.602 & -- & -- & -- \\ 
  naive &  0.996 &  0.534 &  0.395 &  0.710 &  0.612 &  0.404 &  1.269 &  4.898 \\ 
   \hline  \multicolumn{1}{l}{MAE}\\ARGOX & \textbf{0.591} & \textbf{0.477} & \textbf{0.459} & \textbf{0.565} & \textbf{0.520} & \textbf{0.493} & \textbf{0.768} & \textbf{1.310} \\ 
  VAR & 1.338 & 1.299 & 1.332 & 1.331 & 1.546 & 2.094 & 1.128 & 1.768 \\ 
  GFT & -- & -- & 0.780 & -- & -- & -- & -- & -- \\ 
  Lu et al. (2019) & -- & -- & -- & -- & -- & -- & -- & -- \\ 
  naive & 0.619 & 0.510 & 0.461 & 0.596 & 0.604 & 0.504 & 0.821 & 1.326 \\ 
   \hline  \multicolumn{1}{l}{Correlation}\\ARGOX & \textbf{0.872} & \textbf{0.805} & 0.746 & 0.815 & \textbf{0.800} & \textbf{0.923} & 0.879 & \textbf{0.564} \\ 
  VAR & 0.593 & 0.291 & 0.166 & 0.363 & 0.227 & 0.380 & 0.754 & 0.457 \\ 
  GFT & -- & -- & 0.638 & -- & -- & -- & -- & -- \\ 
  Lu et al. (2019) & -- & 0.781 & 0.723 & 0.815 & 0.750 & -- & -- & -- \\ 
  naive & 0.865 & 0.797 & \textbf{0.762} & \textbf{0.820} & 0.769 & 0.912 & \textbf{0.880} & 0.555 \\ 
   \hline
\end{tabular}
\caption{Comparison of different methods for state-level \%ILI estimation in Alaska (AK).  The MSE, MAE, and correlation are reported. The method with the best performance is highlighted in boldface for each metric in each period. \label{tab_state2}} 
\end{table}

\begin{figure}[!h] 
  \centering 
\includegraphics[width=\linewidth, page=2]{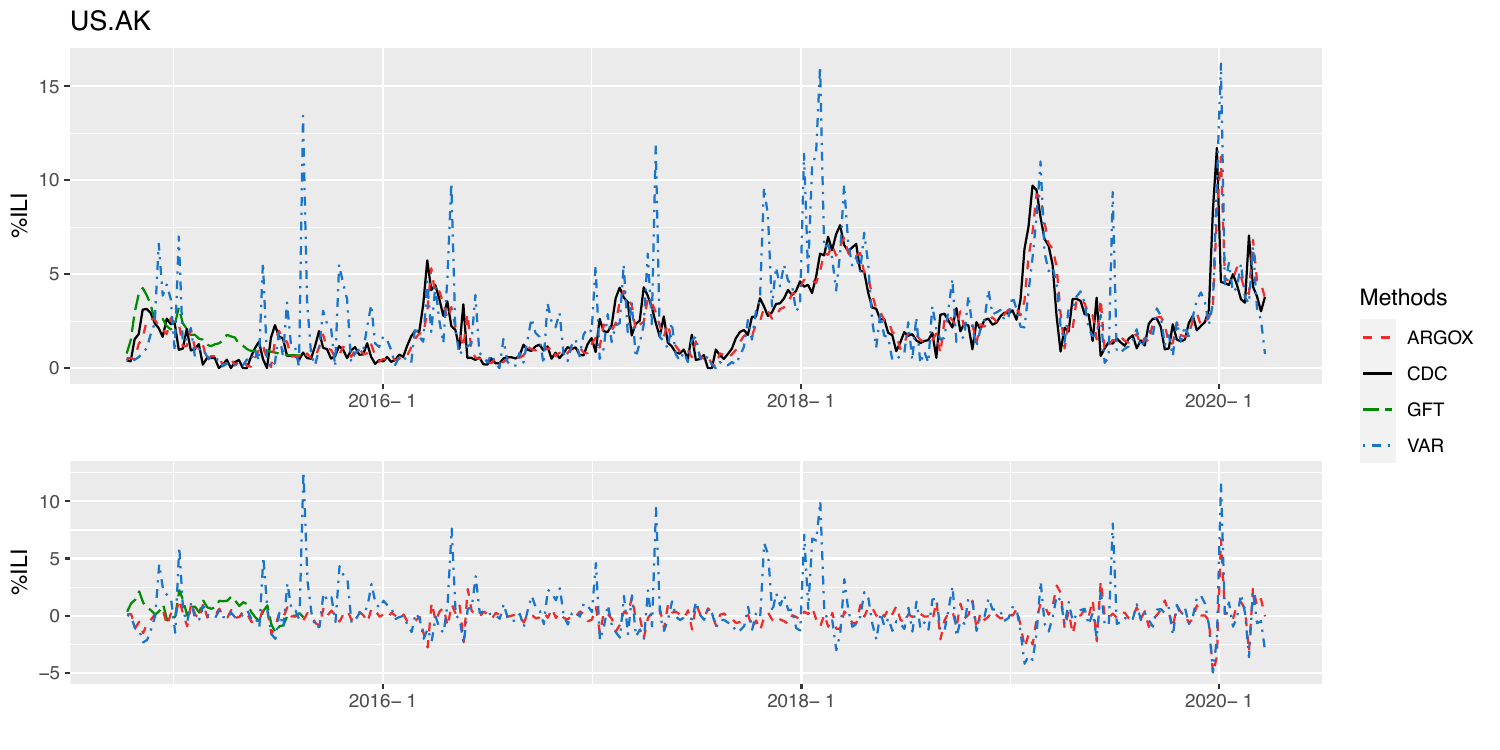} 
\caption{Plots of the \%ILI estimates (top) and the estimation errors (bottom) for Alaska (AK).}
\end{figure}
\newpage      
\begin{table}[ht]
\centering
\begin{tabular}{crrrrrrrr}
  \hline
  & Whole period '14-'20 & Overall '14-'17 & '14-'15 & '15-'16 & '16-'17 & '17-'18 & '18-'19 & '19-'20 \\ 
  \hline  \multicolumn{1}{l}{MSE}\\ARGOX & \textbf{0.151} & \textbf{0.093} & 0.064 & \textbf{0.157} & 0.098 & \textbf{0.418} & \textbf{0.094} & \textbf{0.208} \\ 
  VAR & 0.663 & 0.188 & 0.062 & 0.270 & 0.362 & 3.360 & 0.262 & 0.781 \\ 
  GFT & -- & -- & 1.213 & -- & -- & -- & -- & -- \\ 
  Lu et al. (2019) & -- & 0.126 & \textbf{0.052} & 0.237 & 0.139 & -- & -- & -- \\ 
  naive & 0.185 & 0.095 & 0.070 & 0.161 & \textbf{0.092} & 0.602 & 0.132 & 0.276 \\ 
   \hline  \multicolumn{1}{l}{MAE}\\ARGOX & \textbf{0.282} & \textbf{0.218} & \textbf{0.174} & \textbf{0.291} & 0.254 & \textbf{0.480} & \textbf{0.257} & \textbf{0.385} \\ 
  VAR & 0.459 & 0.319 & 0.199 & 0.376 & 0.497 & 0.999 & 0.403 & 0.543 \\ 
  GFT & -- & -- & 0.940 & -- & -- & -- & -- & -- \\ 
  Lu et al. (2019) & -- & -- & -- & -- & -- & -- & -- & -- \\ 
  naive & 0.297 & 0.230 & 0.198 & 0.293 & \textbf{0.254} & 0.529 & 0.269 & 0.387 \\ 
   \hline  \multicolumn{1}{l}{Correlation}\\ARGOX & \textbf{0.951} & \textbf{0.957} & 0.957 & \textbf{0.944} & 0.861 & \textbf{0.927} & \textbf{0.925} & \textbf{0.867} \\ 
  VAR & 0.835 & 0.923 & \textbf{0.963} & 0.921 & 0.555 & 0.788 & 0.812 & 0.699 \\ 
  GFT & -- & -- & 0.934 & -- & -- & -- & -- & -- \\ 
  Lu et al. (2019) & -- & 0.949 & 0.960 & 0.930 & \textbf{0.896} & -- & -- & -- \\ 
  naive & 0.941 & 0.956 & 0.949 & 0.939 & 0.865 & 0.886 & 0.901 & 0.810 \\ 
   \hline
\end{tabular}
\caption{Comparison of different methods for state-level \%ILI estimation in Arizona (AZ).  The MSE, MAE, and correlation are reported. The method with the best performance is highlighted in boldface for each metric in each period. \label{tab_state3}} 
\end{table}

\begin{figure}[!h] 
  \centering 
\includegraphics[width=\linewidth, page=3]{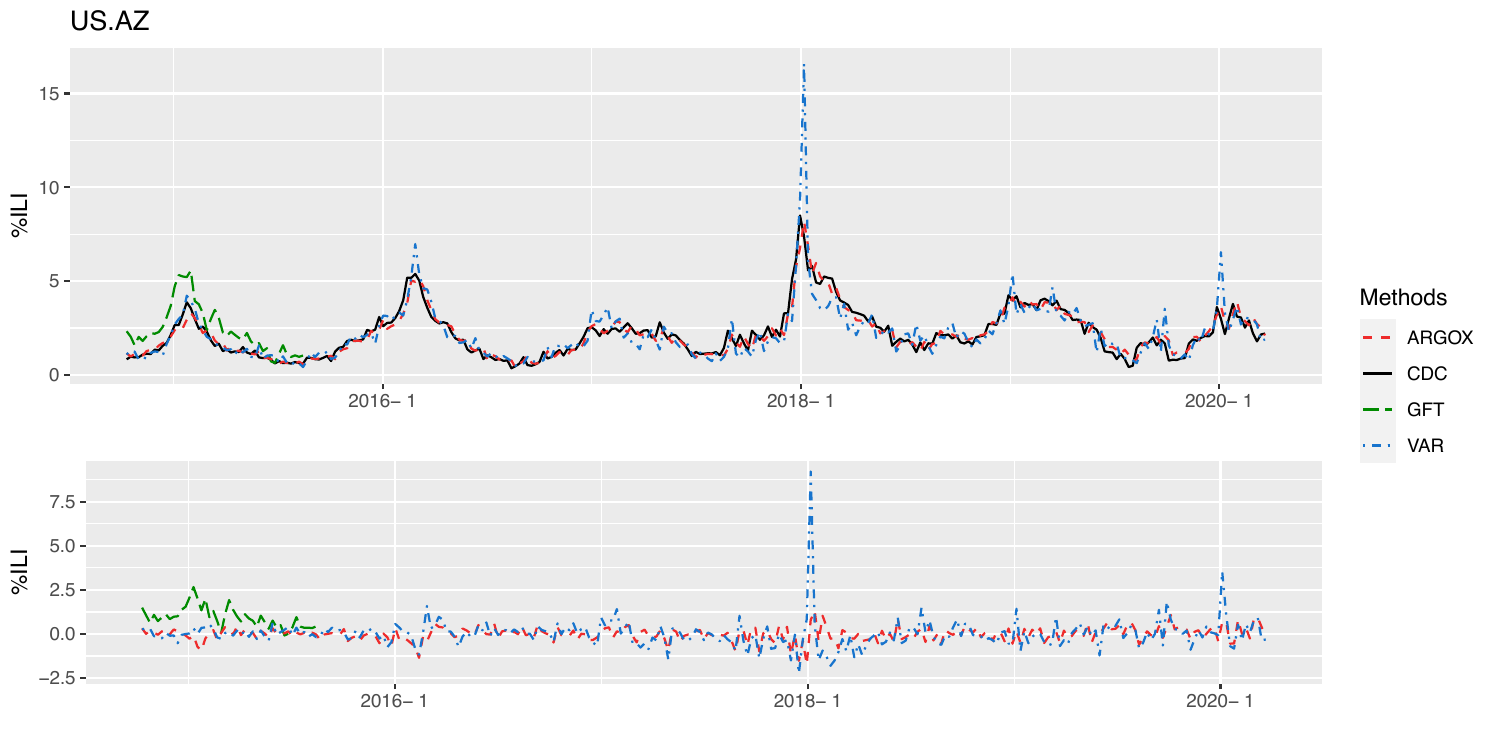} 
\caption{Plots of the \%ILI estimates (top) and the estimation errors (bottom) for Arizona (AZ).}
\end{figure}
\newpage      
\begin{table}[ht]
\centering
\begin{tabular}{crrrrrrrr}
  \hline
  & Whole period '14-'20 & Overall '14-'17 & '14-'15 & '15-'16 & '16-'17 & '17-'18 & '18-'19 & '19-'20 \\ 
  \hline  \multicolumn{1}{l}{MSE}\\ARGOX & \textbf{0.786} & \textbf{0.585} & \textbf{1.024} & \textbf{0.286} & \textbf{0.677} & \textbf{0.849} & \textbf{0.551} & \textbf{3.791} \\ 
  VAR & 2.487 & 1.529 & 2.202 & 0.879 & 2.109 & 9.018 & 2.619 & 3.827 \\ 
  GFT & -- & -- & 1.820 & -- & -- & -- & -- & -- \\ 
  Lu et al. (2019) & -- & 0.679 & 1.325 & 0.341 & -- & -- & -- & -- \\ 
  naive & 1.201 & 0.922 & 1.570 & 0.418 & 1.166 & 2.024 & 0.974 & 4.553 \\ 
   \hline  \multicolumn{1}{l}{MAE}\\ARGOX & \textbf{0.532} & \textbf{0.457} & \textbf{0.586} & \textbf{0.389} & \textbf{0.549} & \textbf{0.673} & \textbf{0.550} & 1.424 \\ 
  VAR & 0.894 & 0.806 & 0.964 & 0.686 & 1.075 & 1.584 & 1.125 & \textbf{1.275} \\ 
  GFT & -- & -- & 1.153 & -- & -- & -- & -- & -- \\ 
  Lu et al. (2019) & -- & -- & -- & -- & -- & -- & -- & -- \\ 
  naive & 0.647 & 0.548 & 0.677 & 0.472 & 0.708 & 1.016 & 0.729 & 1.523 \\ 
   \hline  \multicolumn{1}{l}{Correlation}\\ARGOX & \textbf{0.937} & \textbf{0.940} & 0.927 & \textbf{0.904} & \textbf{0.940} & \textbf{0.972} & \textbf{0.951} & \textbf{0.667} \\ 
  VAR & 0.856 & 0.860 & 0.907 & 0.685 & 0.783 & 0.842 & 0.884 & 0.633 \\ 
  GFT & -- & -- & \textbf{0.974} & -- & -- & -- & -- & -- \\ 
  Lu et al. (2019) & -- & 0.928 & 0.909 & 0.875 & -- & -- & -- & -- \\ 
  naive & 0.906 & 0.906 & 0.889 & 0.866 & 0.893 & 0.920 & 0.903 & 0.620 \\ 
   \hline
\end{tabular}
\caption{Comparison of different methods for state-level \%ILI estimation in Arkansas (AR).  The MSE, MAE, and correlation are reported. The method with the best performance is highlighted in boldface for each metric in each period. \label{tab_state4}} 
\end{table}

\begin{figure}[!h] 
  \centering 
\includegraphics[width=\linewidth, page=4]{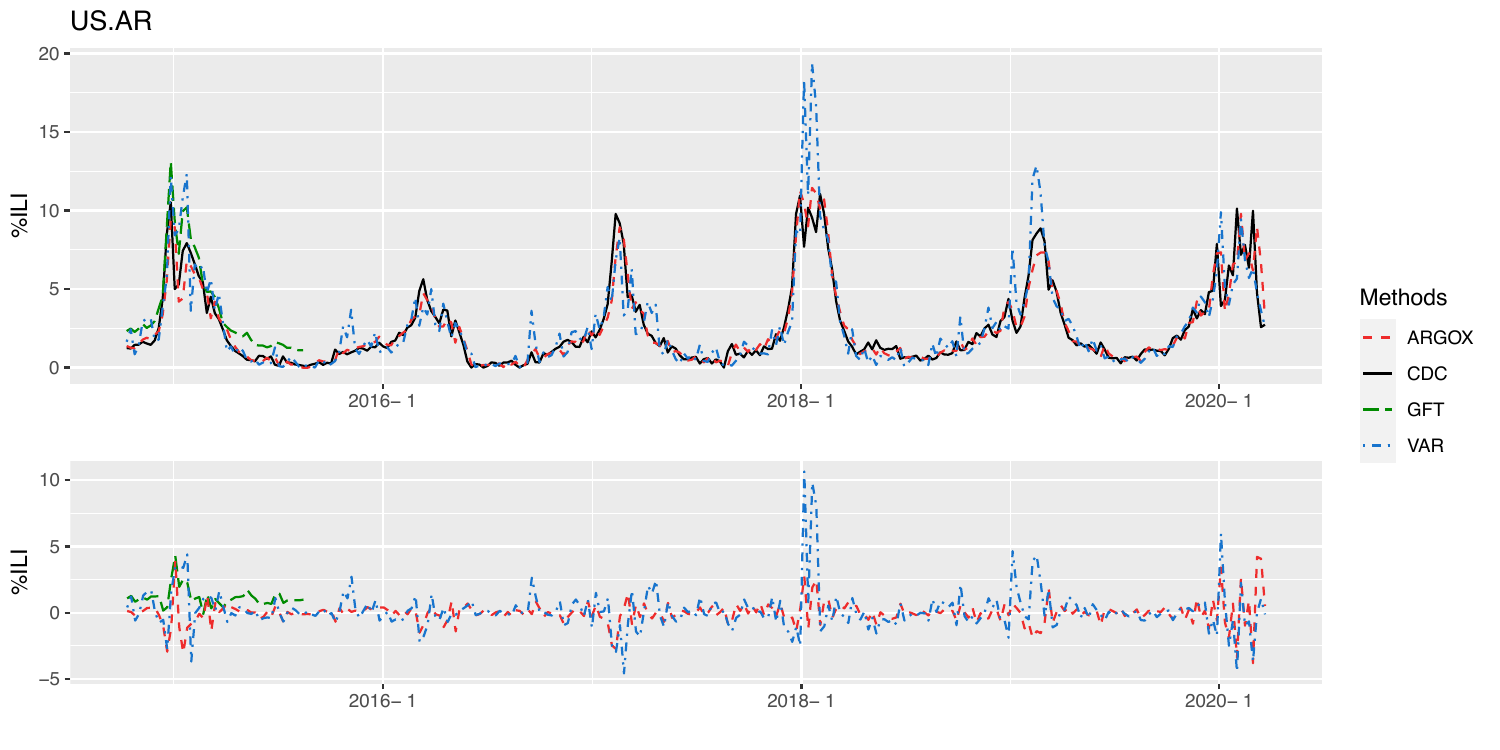} 
\caption{Plots of the \%ILI estimates (top) and the estimation errors (bottom) for Arkansas (AR).}
\end{figure}
\newpage      
\begin{table}[ht]
\centering
\begin{tabular}{crrrrrrrr}
  \hline
  & Whole period '14-'20 & Overall '14-'17 & '14-'15 & '15-'16 & '16-'17 & '17-'18 & '18-'19 & '19-'20 \\ 
  \hline  \multicolumn{1}{l}{MSE}\\ARGOX & \textbf{0.076} & \textbf{0.067} & \textbf{0.081} & \textbf{0.078} & 0.068 & \textbf{0.155} & \textbf{0.047} & \textbf{0.195} \\ 
  VAR & 0.169 & 0.162 & 0.172 & 0.190 & 0.190 & 0.409 & 0.093 & 0.257 \\ 
  GFT & -- & -- & 0.266 & -- & -- & -- & -- & -- \\ 
  Lu et al. (2019) & -- & -- & -- & -- & -- & -- & -- & -- \\ 
  naive & 0.123 & 0.092 & 0.124 & 0.125 & \textbf{0.064} & 0.398 & 0.069 & 0.222 \\ 
   \hline  \multicolumn{1}{l}{MAE}\\ARGOX & \textbf{0.187} & \textbf{0.186} & \textbf{0.199} & \textbf{0.218} & \textbf{0.192} & \textbf{0.250} & \textbf{0.172} & \textbf{0.314} \\ 
  VAR & 0.288 & 0.284 & 0.272 & 0.320 & 0.310 & 0.457 & 0.255 & 0.379 \\ 
  GFT & -- & -- & 0.388 & -- & -- & -- & -- & -- \\ 
  Lu et al. (2019) & -- & -- & -- & -- & -- & -- & -- & -- \\ 
  naive & 0.224 & 0.214 & 0.233 & 0.277 & 0.194 & 0.381 & 0.220 & 0.330 \\ 
   \hline  \multicolumn{1}{l}{Correlation}\\ARGOX & \textbf{0.971} & \textbf{0.962} & \textbf{0.966} & \textbf{0.940} & \textbf{0.926} & \textbf{0.958} & \textbf{0.958} & \textbf{0.950} \\ 
  VAR & 0.941 & 0.938 & 0.949 & 0.908 & 0.872 & 0.926 & 0.924 & 0.931 \\ 
  GFT & -- & -- & 0.927 & -- & -- & -- & -- & -- \\ 
  Lu et al. (2019) & -- & -- & -- & -- & -- & -- & -- & -- \\ 
  naive & 0.952 & 0.947 & 0.944 & 0.907 & 0.917 & 0.892 & 0.940 & 0.943 \\ 
   \hline
\end{tabular}
\caption{Comparison of different methods for state-level \%ILI estimation in California (CA).  The MSE, MAE, and correlation are reported. The method with the best performance is highlighted in boldface for each metric in each period. \label{tab_state5}} 
\end{table}

\begin{figure}[!h] 
  \centering 
\includegraphics[width=\linewidth, page=5]{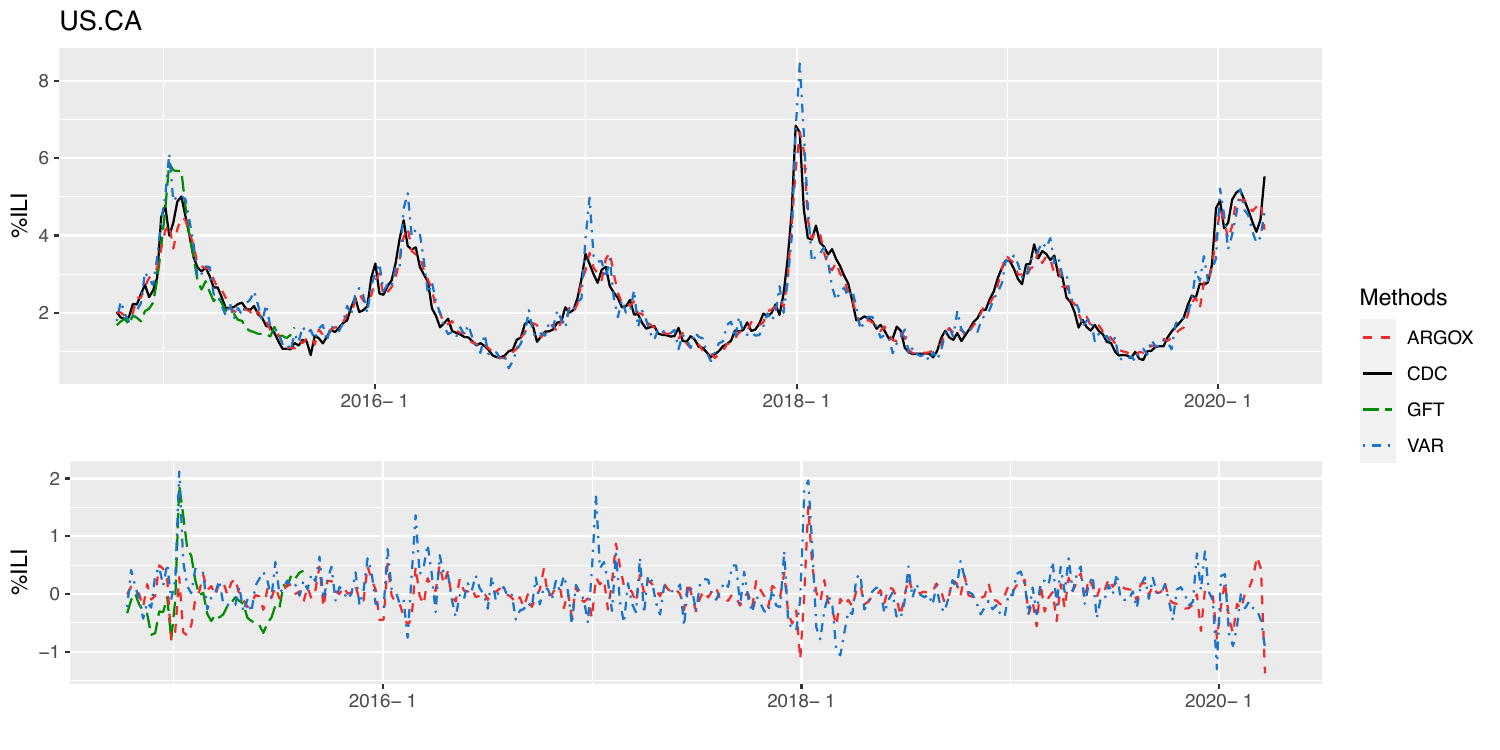} 
\caption{Plots of the \%ILI estimates (top) and the estimation errors (bottom) for California (CA).}
\end{figure}
\newpage      
\begin{table}[ht]
\centering
\begin{tabular}{crrrrrrrr}
  \hline
  & Whole period '14-'20 & Overall '14-'17 & '14-'15 & '15-'16 & '16-'17 & '17-'18 & '18-'19 & '19-'20 \\ 
  \hline  \multicolumn{1}{l}{MSE}\\ARGOX & \textbf{0.178} & \textbf{0.153} & \textbf{0.166} & \textbf{0.028} & \textbf{0.343} & \textbf{0.047} & 0.624 & \textbf{0.196} \\ 
  VAR & 0.770 & 0.456 & 0.514 & 0.070 & 1.042 & 0.814 & 1.971 & 2.211 \\ 
  GFT & -- & -- & 0.585 & -- & -- & -- & -- & -- \\ 
  Lu et al. (2019) & -- & -- & -- & -- & -- & -- & -- & -- \\ 
  naive & 0.225 & 0.201 & 0.204 & 0.043 & 0.470 & 0.071 & \textbf{0.525} & 0.533 \\ 
   \hline  \multicolumn{1}{l}{MAE}\\ARGOX & \textbf{0.272} & \textbf{0.241} & \textbf{0.236} & \textbf{0.119} & \textbf{0.430} & \textbf{0.161} & 0.635 & \textbf{0.332} \\ 
  VAR & 0.531 & 0.374 & 0.363 & 0.204 & 0.676 & 0.578 & 1.062 & 1.085 \\ 
  GFT & -- & -- & 0.420 & -- & -- & -- & -- & -- \\ 
  Lu et al. (2019) & -- & -- & -- & -- & -- & -- & -- & -- \\ 
  naive & 0.310 & 0.287 & 0.279 & 0.166 & 0.506 & 0.213 & \textbf{0.562} & 0.534 \\ 
   \hline  \multicolumn{1}{l}{Correlation}\\ARGOX & \textbf{0.972} & \textbf{0.864} & \textbf{0.927} & \textbf{0.926} & \textbf{0.484} & \textbf{0.979} & \textbf{0.910} & \textbf{0.975} \\ 
  VAR & 0.888 & 0.665 & 0.754 & 0.824 & 0.461 & 0.811 & 0.661 & 0.784 \\ 
  GFT & -- & -- & 0.850 & -- & -- & -- & -- & -- \\ 
  Lu et al. (2019) & -- & -- & -- & -- & -- & -- & -- & -- \\ 
  naive & 0.964 & 0.832 & 0.908 & 0.885 & 0.374 & 0.961 & 0.904 & 0.930 \\ 
   \hline
\end{tabular}
\caption{Comparison of different methods for state-level \%ILI estimation in Colorado (CO).  The MSE, MAE, and correlation are reported. The method with the best performance is highlighted in boldface for each metric in each period. \label{tab_state6}} 
\end{table}

\begin{figure}[!h] 
  \centering 
\includegraphics[width=\linewidth, page=6]{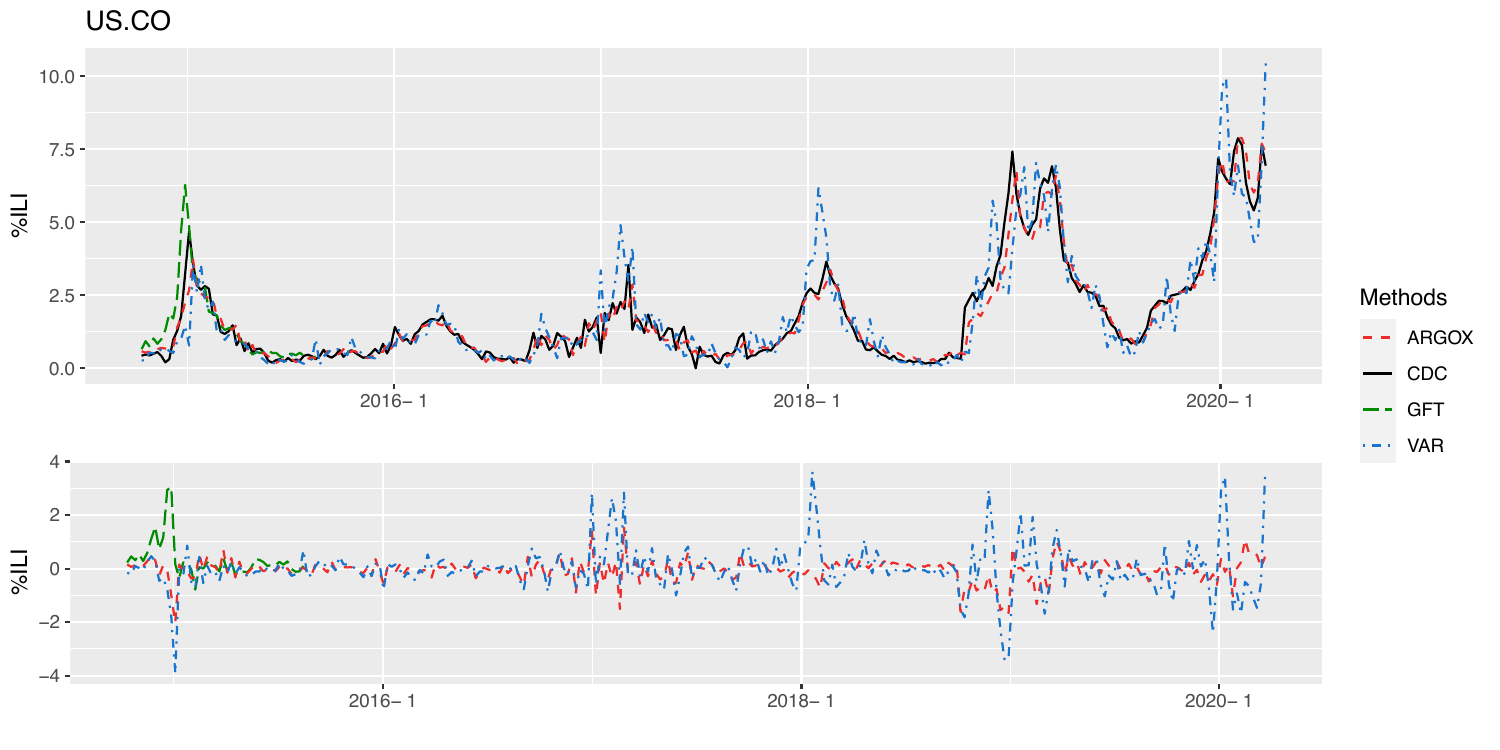} 
\caption{Plots of the \%ILI estimates (top) and the estimation errors (bottom) for Colorado (CO).}
\end{figure}
\newpage      
\begin{table}[ht]
\centering
\begin{tabular}{crrrrrrrr}
  \hline
  & Whole period '14-'20 & Overall '14-'17 & '14-'15 & '15-'16 & '16-'17 & '17-'18 & '18-'19 & '19-'20 \\ 
  \hline  \multicolumn{1}{l}{MSE}\\ARGOX & \textbf{ 0.281} & \textbf{ 0.228} &  0.411 &  0.222 & \textbf{ 0.145} & \textbf{ 0.283} & \textbf{ 0.175} & \textbf{ 0.462} \\ 
  VAR &  3.626 &  5.482 & 15.079 &  1.046 &  0.512 &  3.095 &  1.515 &  3.967 \\ 
  GFT & -- & -- &  1.944 & -- & -- & -- & -- & -- \\ 
  Lu et al. (2019) & -- & -- & -- & -- & -- & -- & -- & -- \\ 
  naive &  0.330 &  0.230 & \textbf{ 0.349} & \textbf{ 0.213} &  0.242 &  0.399 &  0.243 &  0.636 \\ 
   \hline  \multicolumn{1}{l}{MAE}\\ARGOX & \textbf{0.359} & 0.314 & 0.410 & 0.365 & \textbf{0.300} & \textbf{0.388} & \textbf{0.324} & \textbf{0.439} \\ 
  VAR & 0.891 & 0.925 & 1.584 & 0.845 & 0.584 & 1.090 & 0.910 & 1.395 \\ 
  GFT & -- & -- & 0.999 & -- & -- & -- & -- & -- \\ 
  Lu et al. (2019) & -- & -- & -- & -- & -- & -- & -- & -- \\ 
  naive & 0.375 & \textbf{0.305} & \textbf{0.323} & \textbf{0.360} & 0.379 & 0.458 & 0.367 & 0.578 \\ 
   \hline  \multicolumn{1}{l}{Correlation}\\ARGOX & \textbf{0.962} & 0.942 & 0.918 & 0.882 & \textbf{0.955} & \textbf{0.956} & \textbf{0.953} & \textbf{0.976} \\ 
  VAR & 0.699 & 0.443 & 0.222 & 0.674 & 0.855 & 0.785 & 0.767 & 0.822 \\ 
  GFT & -- & -- & 0.774 & -- & -- & -- & -- & -- \\ 
  Lu et al. (2019) & -- & -- & -- & -- & -- & -- & -- & -- \\ 
  naive & 0.955 & \textbf{0.942} & \textbf{0.928} & \textbf{0.887} & 0.925 & 0.938 & 0.934 & 0.960 \\ 
   \hline
\end{tabular}
\caption{Comparison of different methods for state-level \%ILI estimation in Connecticut (CT).  The MSE, MAE, and correlation are reported. The method with the best performance is highlighted in boldface for each metric in each period. \label{tab_state7}} 
\end{table}

\begin{figure}[!h] 
  \centering 
\includegraphics[width=\linewidth, page=7]{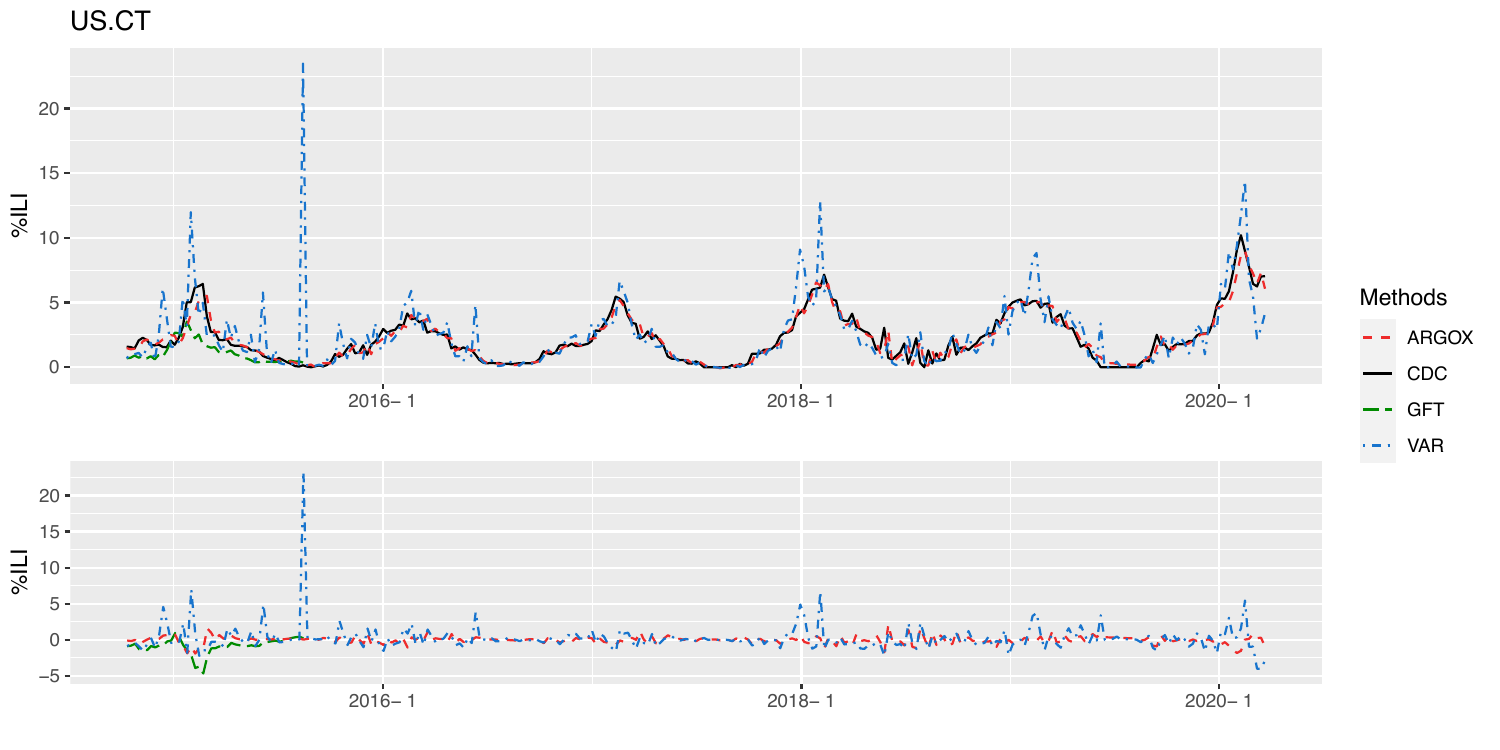} 
\caption{Plots of the \%ILI estimates (top) and the estimation errors (bottom) for Connecticut (CT).}
\end{figure}
\newpage      
\begin{table}[ht]
\centering
\begin{tabular}{crrrrrrrr}
  \hline
  & Whole period '14-'20 & Overall '14-'17 & '14-'15 & '15-'16 & '16-'17 & '17-'18 & '18-'19 & '19-'20 \\ 
  \hline  \multicolumn{1}{l}{MSE}\\ARGOX & \textbf{0.187} & 0.126 & 0.297 & \textbf{0.062} & \textbf{0.041} & \textbf{0.445} & \textbf{0.071} & \textbf{0.725} \\ 
  VAR & 0.644 & 0.313 & 0.495 & 0.116 & 0.486 & 2.306 & 0.239 & 2.213 \\ 
  GFT & -- & -- & 3.978 & -- & -- & -- & -- & -- \\ 
  Lu et al. (2019) & -- & 0.163 & 0.377 & 0.073 & 0.053 & -- & -- & -- \\ 
  naive & 0.211 & \textbf{0.107} & \textbf{0.231} & 0.066 & 0.053 & 0.598 & 0.086 & 0.881 \\ 
   \hline  \multicolumn{1}{l}{MAE}\\ARGOX & \textbf{0.224} & 0.189 & 0.266 & 0.188 & \textbf{0.156} & \textbf{0.340} & 0.209 & \textbf{0.600} \\ 
  VAR & 0.359 & 0.277 & 0.333 & 0.230 & 0.388 & 0.673 & 0.340 & 0.994 \\ 
  GFT & -- & -- & 1.783 & -- & -- & -- & -- & -- \\ 
  Lu et al. (2019) & -- & -- & -- & -- & -- & -- & -- & -- \\ 
  naive & 0.232 & \textbf{0.169} & \textbf{0.217} & \textbf{0.173} & 0.173 & 0.455 & \textbf{0.208} & 0.674 \\ 
   \hline  \multicolumn{1}{l}{Correlation}\\ARGOX & \textbf{0.874} & 0.824 & 0.813 & 0.832 & \textbf{0.769} & \textbf{0.889} & \textbf{0.899} & \textbf{0.781} \\ 
  VAR & 0.748 & 0.718 & 0.760 & 0.659 & 0.730 & 0.848 & 0.779 & 0.457 \\ 
  GFT & -- & -- & 0.747 & -- & -- & -- & -- & -- \\ 
  Lu et al. (2019) & -- & 0.843 & 0.841 & \textbf{0.840} & 0.614 & -- & -- & -- \\ 
  naive & 0.864 & \textbf{0.860} & \textbf{0.864} & 0.832 & 0.739 & 0.850 & 0.882 & 0.748 \\ 
   \hline
\end{tabular}
\caption{Comparison of different methods for state-level \%ILI estimation in Delaware (DE).  The MSE, MAE, and correlation are reported. The method with the best performance is highlighted in boldface for each metric in each period. \label{tab_state8}} 
\end{table}

\begin{figure}[!h] 
  \centering 
\includegraphics[width=\linewidth, page=8]{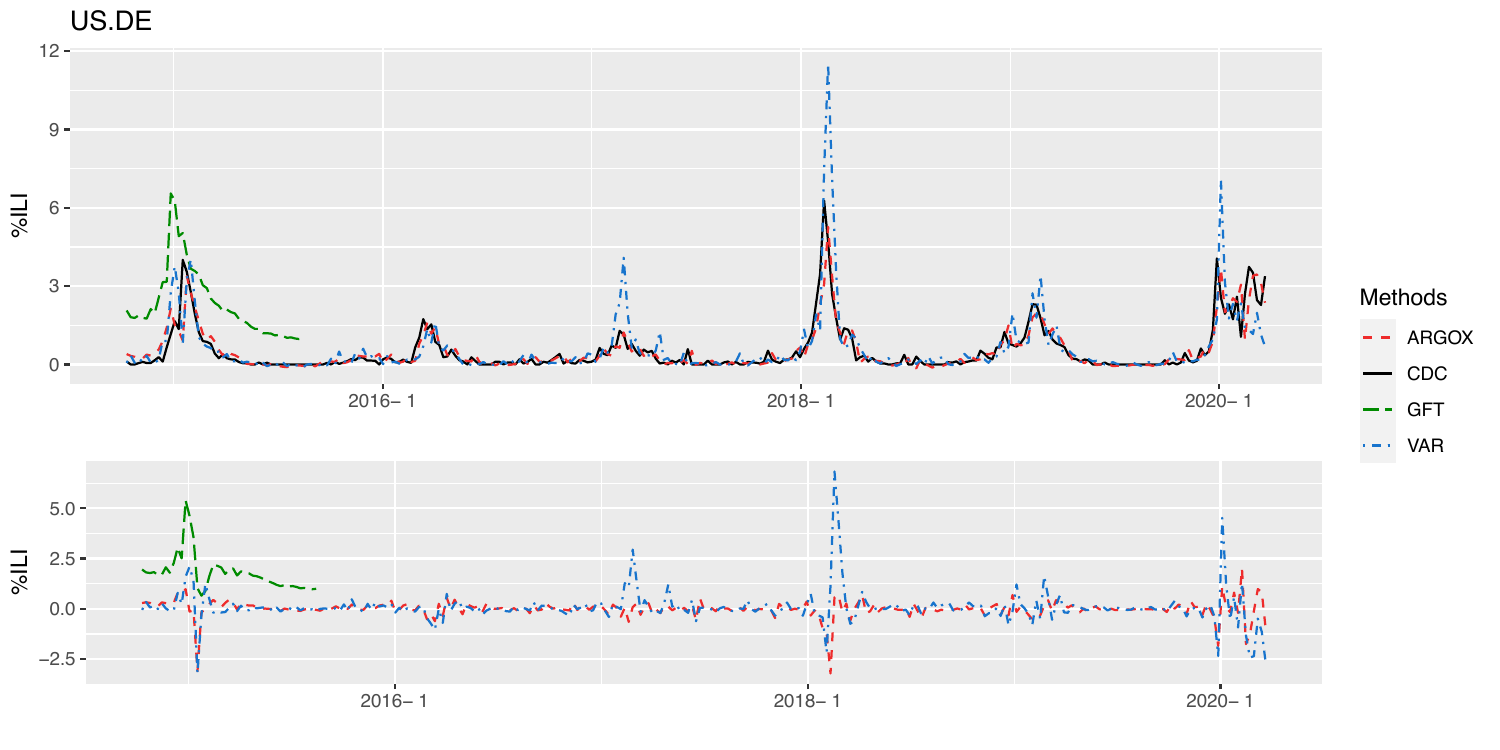} 
\caption{Plots of the \%ILI estimates (top) and the estimation errors (bottom) for Delaware (DE).}
\end{figure}
\newpage      
\begin{table}[ht]
\centering
\begin{tabular}{crrrrrrrr}
  \hline
  & Whole period '14-'20 & Overall '14-'17 & '14-'15 & '15-'16 & '16-'17 & '17-'18 & '18-'19 & '19-'20 \\ 
  \hline  \multicolumn{1}{l}{MSE}\\ARGOX & \textbf{ 2.011} & \textbf{ 2.806} & \textbf{ 2.875} & \textbf{ 2.634} & \textbf{ 4.192} &  1.531 & \textbf{ 0.216} & \textbf{ 0.516} \\ 
  VAR & 15.898 & 10.397 &  5.739 & 15.635 & 17.804 & 14.672 &  7.150 &  1.439 \\ 
  GFT & -- & -- & 25.894 & -- & -- & -- & -- & -- \\ 
  Lu et al. (2019) & -- & -- & -- & -- & -- & -- & -- & -- \\ 
  naive &  2.081 &  2.907 &  3.124 &  2.787 &  4.811 & \textbf{ 0.695} &  0.280 &  0.689 \\ 
   \hline  \multicolumn{1}{l}{MAE}\\ARGOX & 0.968 & 1.227 & \textbf{1.195} & 1.289 & \textbf{1.418} & 1.012 & \textbf{0.320} & \textbf{0.568} \\ 
  VAR & 2.126 & 2.249 & 1.741 & 2.723 & 3.370 & 2.132 & 1.775 & 0.921 \\ 
  GFT & -- & -- & 4.759 & -- & -- & -- & -- & -- \\ 
  Lu et al. (2019) & -- & -- & -- & -- & -- & -- & -- & -- \\ 
  naive & \textbf{0.888} & \textbf{1.194} & 1.297 & \textbf{1.277} & 1.509 & \textbf{0.444} & 0.342 & 0.650 \\ 
   \hline  \multicolumn{1}{l}{Correlation}\\ARGOX & 0.873 & 0.820 & \textbf{0.828} & 0.786 & \textbf{0.609} & 0.666 & \textbf{0.677} & \textbf{0.794} \\ 
  VAR & 0.554 & 0.605 & 0.602 & 0.580 & 0.475 & 0.602 & 0.243 & 0.656 \\ 
  GFT & -- & -- & 0.728 & -- & -- & -- & -- & -- \\ 
  Lu et al. (2019) & -- & -- & -- & -- & -- & -- & -- & -- \\ 
  naive & \textbf{0.873} & \textbf{0.828} & 0.810 & \textbf{0.810} & 0.562 & \textbf{0.728} & 0.605 & 0.741 \\ 
   \hline
\end{tabular}
\caption{Comparison of different methods for state-level \%ILI estimation in District of Columbia (DC).  The MSE, MAE, and correlation are reported. The method with the best performance is highlighted in boldface for each metric in each period. \label{tab_state9}} 
\end{table}

\begin{figure}[!h] 
  \centering 
\includegraphics[width=\linewidth, page=9]{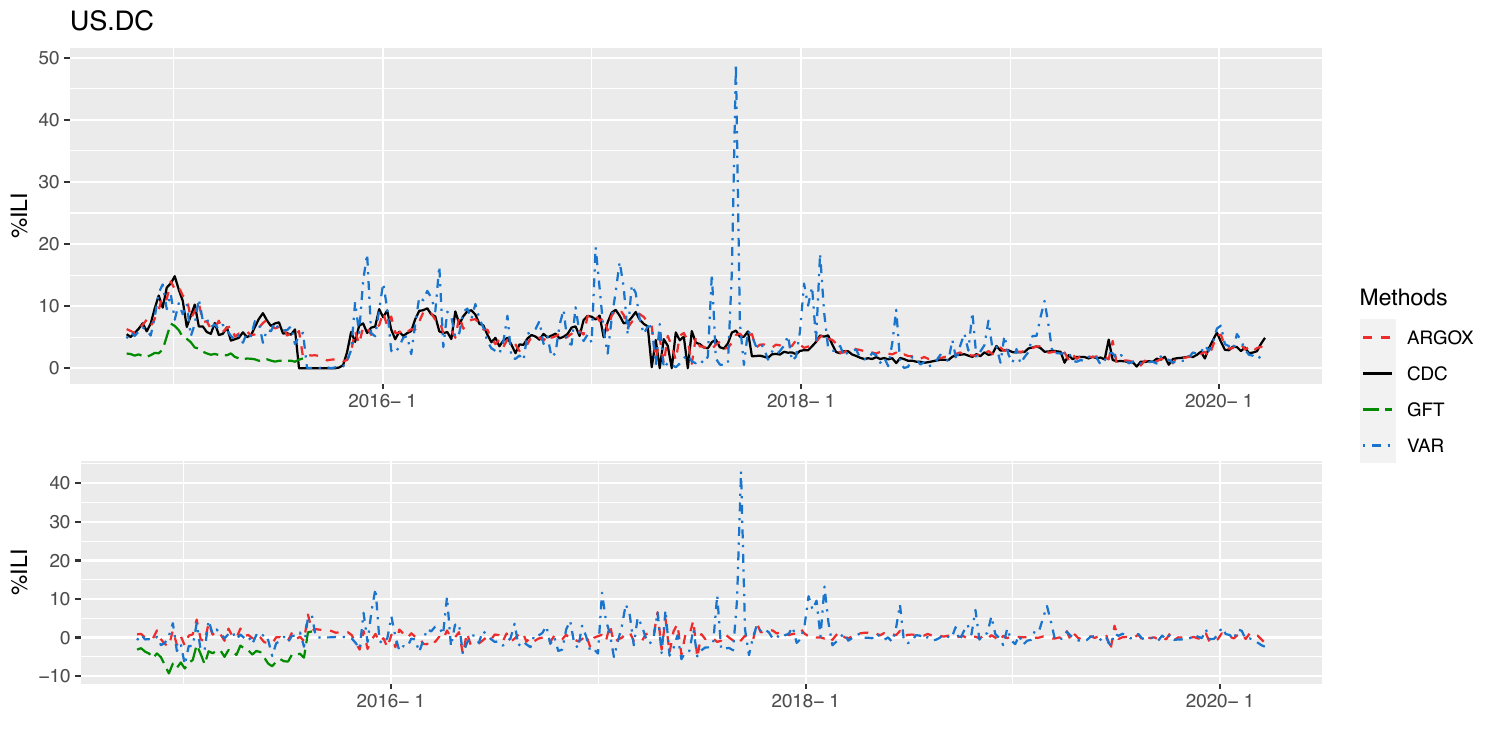} 
\caption{Plots of the \%ILI estimates (top) and the estimation errors (bottom) for District of Columbia (DC).}
\end{figure}
\newpage      
\begin{table}[ht]
\centering
\begin{tabular}{crrrrrrrr}
  \hline
  & Whole period '14-'20 & Overall '14-'17 & '14-'15 & '15-'16 & '16-'17 & '17-'18 & '18-'19 & '19-'20 \\ 
  \hline  \multicolumn{1}{l}{MSE}\\ARGOX & \textbf{0.346} & \textbf{0.282} & 0.606 & \textbf{0.096} & \textbf{0.192} & \textbf{1.057} & \textbf{0.145} & \textbf{0.742} \\ 
  VAR & 0.715 & 0.416 & 0.493 & 0.469 & 0.445 & 1.595 & 0.986 & 2.239 \\ 
  GFT & -- & -- & 0.600 & -- & -- & -- & -- & -- \\ 
  Lu et al. (2019) & -- & 0.283 & \textbf{0.408} & 0.224 & 0.347 & -- & -- & -- \\ 
  naive & 0.688 & 0.410 & 0.888 & 0.129 & 0.318 & 2.188 & 0.452 & 2.045 \\ 
   \hline  \multicolumn{1}{l}{MAE}\\ARGOX & \textbf{0.337} & \textbf{0.313} & \textbf{0.378} & \textbf{0.242} & \textbf{0.373} & \textbf{0.617} & \textbf{0.271} & \textbf{0.658} \\ 
  VAR & 0.527 & 0.454 & 0.451 & 0.484 & 0.518 & 0.834 & 0.666 & 0.954 \\ 
  GFT & -- & -- & 0.470 & -- & -- & -- & -- & -- \\ 
  Lu et al. (2019) & -- & -- & -- & -- & -- & -- & -- & -- \\ 
  naive & 0.442 & 0.347 & 0.442 & 0.275 & 0.418 & 0.837 & 0.488 & 1.085 \\ 
   \hline  \multicolumn{1}{l}{Correlation}\\ARGOX & \textbf{0.970} & 0.928 & 0.882 & \textbf{0.877} & \textbf{0.945} & \textbf{0.968} & \textbf{0.966} & \textbf{0.936} \\ 
  VAR & 0.943 & 0.889 & 0.908 & 0.352 & 0.870 & 0.956 & 0.846 & 0.869 \\ 
  GFT & -- & -- & 0.901 & -- & -- & -- & -- & -- \\ 
  Lu et al. (2019) & -- & \textbf{0.945} & \textbf{0.957} & 0.711 & 0.936 & -- & -- & -- \\ 
  naive & 0.941 & 0.893 & 0.827 & 0.837 & 0.909 & 0.934 & 0.878 & 0.830 \\ 
   \hline
\end{tabular}
\caption{Comparison of different methods for state-level \%ILI estimation in Georgia (GA).  The MSE, MAE, and correlation are reported. The method with the best performance is highlighted in boldface for each metric in each period. \label{tab_state10}} 
\end{table}

\begin{figure}[!h] 
  \centering 
\includegraphics[width=\linewidth, page=10]{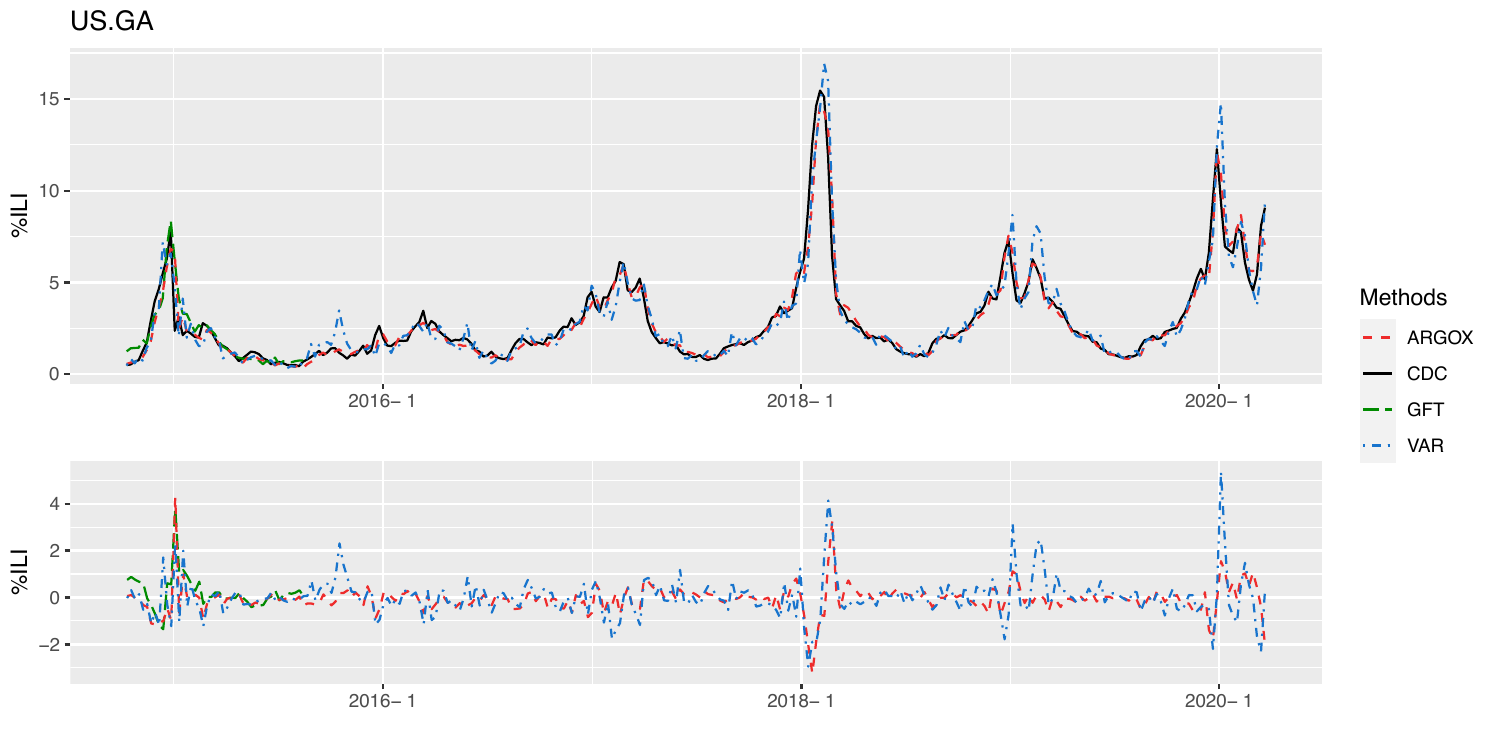} 
\caption{Plots of the \%ILI estimates (top) and the estimation errors (bottom) for Georgia (GA).}
\end{figure}
\newpage      
\begin{table}[ht]
\centering
\begin{tabular}{crrrrrrrr}
  \hline
  & Whole period '14-'20 & Overall '14-'17 & '14-'15 & '15-'16 & '16-'17 & '17-'18 & '18-'19 & '19-'20 \\ 
  \hline  \multicolumn{1}{l}{MSE}\\ARGOX & \textbf{ 0.821} &  1.133 &  2.491 & \textbf{ 0.593} & \textbf{ 0.463} & \textbf{ 1.279} & \textbf{ 0.381} & \textbf{ 0.468} \\ 
  VAR &  2.395 &  2.945 &  7.087 &  1.148 &  0.893 &  4.532 &  0.409 &  3.337 \\ 
  GFT & -- & -- & 15.289 & -- & -- & -- & -- & -- \\ 
  Lu et al. (2019) & -- & -- & -- & -- & -- & -- & -- & -- \\ 
  naive &  0.880 & \textbf{ 1.099} & \textbf{ 2.305} &  0.619 &  0.548 &  1.574 &  0.500 &  0.590 \\ 
   \hline  \multicolumn{1}{l}{MAE}\\ARGOX & \textbf{0.618} & \textbf{0.707} & 1.059 & \textbf{0.619} & \textbf{0.558} & \textbf{0.832} & \textbf{0.469} & \textbf{0.528} \\ 
  VAR & 0.989 & 1.093 & 1.805 & 0.813 & 0.765 & 1.354 & 0.513 & 1.420 \\ 
  GFT & -- & -- & 3.292 & -- & -- & -- & -- & -- \\ 
  Lu et al. (2019) & -- & -- & -- & -- & -- & -- & -- & -- \\ 
  naive & 0.662 & 0.717 & \textbf{1.037} & 0.642 & 0.609 & 0.966 & 0.510 & 0.608 \\ 
   \hline  \multicolumn{1}{l}{Correlation}\\ARGOX & \textbf{0.935} & 0.943 & 0.929 & \textbf{0.906} & \textbf{0.747} & \textbf{0.780} & \textbf{0.868} & \textbf{0.894} \\ 
  VAR & 0.871 & 0.891 & 0.870 & 0.832 & 0.632 & 0.681 & 0.854 & 0.720 \\ 
  GFT & -- & -- & 0.899 & -- & -- & -- & -- & -- \\ 
  Lu et al. (2019) & -- & -- & -- & -- & -- & -- & -- & -- \\ 
  naive & 0.933 & \textbf{0.946} & \textbf{0.936} & 0.905 & 0.716 & 0.730 & 0.833 & 0.874 \\ 
   \hline
\end{tabular}
\caption{Comparison of different methods for state-level \%ILI estimation in Hawaii (HI).  The MSE, MAE, and correlation are reported. The method with the best performance is highlighted in boldface for each metric in each period. \label{tab_state11}} 
\end{table}

\begin{figure}[!h] 
  \centering 
\includegraphics[width=\linewidth, page=11]{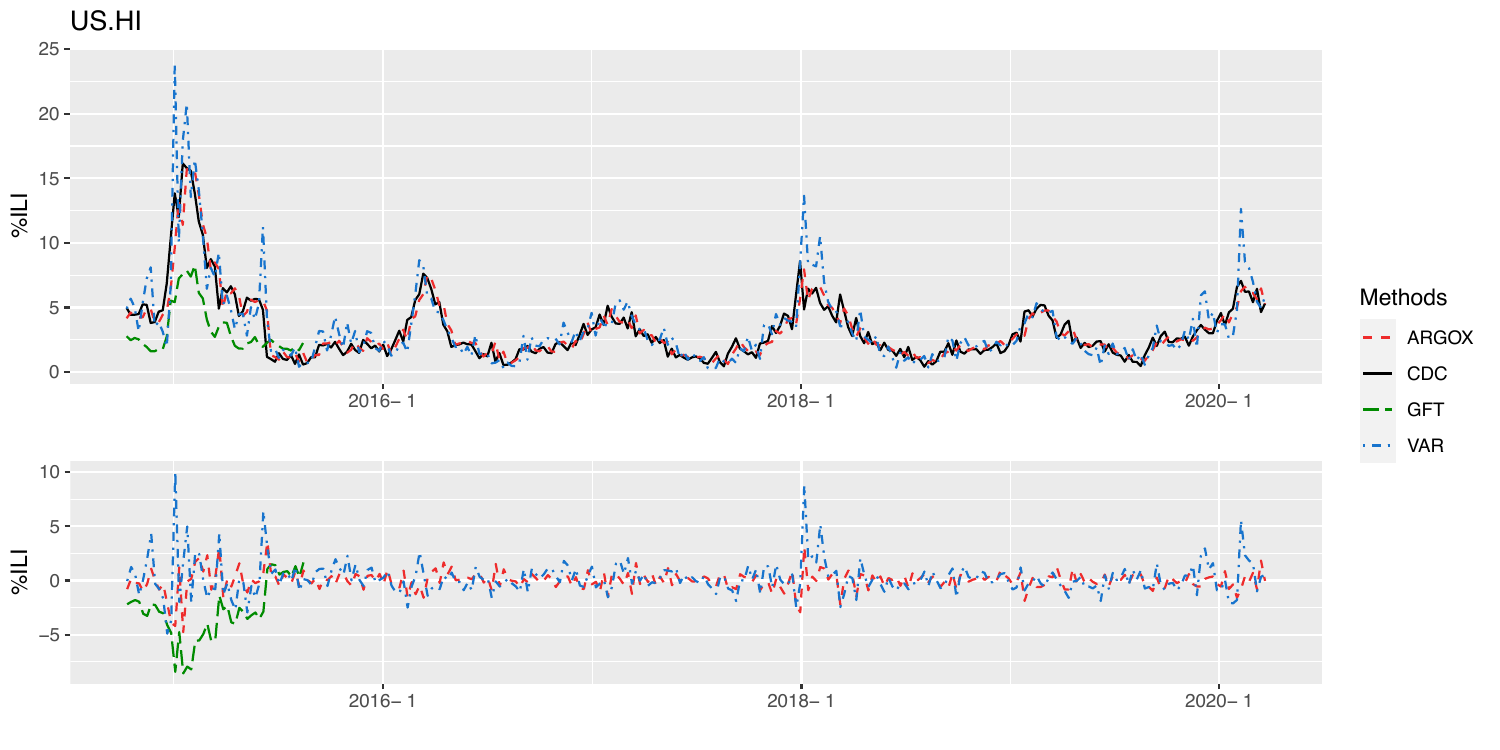} 
\caption{Plots of the \%ILI estimates (top) and the estimation errors (bottom) for Hawaii (HI).}
\end{figure}
\newpage      
\begin{table}[ht]
\centering
\begin{tabular}{crrrrrrrr}
  \hline
  & Whole period '14-'20 & Overall '14-'17 & '14-'15 & '15-'16 & '16-'17 & '17-'18 & '18-'19 & '19-'20 \\ 
  \hline  \multicolumn{1}{l}{MSE}\\ARGOX & \textbf{0.252} & \textbf{0.283} & \textbf{0.433} & 0.250 & \textbf{0.164} & \textbf{0.255} & 0.552 & 0.186 \\ 
  VAR & 1.008 & 1.605 & 2.637 & 0.525 & 0.777 & 1.020 & \textbf{0.539} & 0.541 \\ 
  GFT & -- & -- & 1.395 & -- & -- & -- & -- & -- \\ 
  Lu et al. (2019) & -- & 0.377 & 0.494 & \textbf{0.205} & -- & -- & -- & -- \\ 
  naive & 0.322 & 0.369 & 0.675 & 0.221 & 0.228 & 0.398 & 0.673 & \textbf{0.164} \\ 
   \hline  \multicolumn{1}{l}{MAE}\\ARGOX & \textbf{0.351} & \textbf{0.402} & \textbf{0.473} & 0.397 & \textbf{0.316} & \textbf{0.379} & 0.512 & 0.335 \\ 
  VAR & 0.605 & 0.809 & 1.073 & 0.578 & 0.702 & 0.728 & \textbf{0.501} & 0.557 \\ 
  GFT & -- & -- & 0.965 & -- & -- & -- & -- & -- \\ 
  Lu et al. (2019) & -- & -- & -- & -- & -- & -- & -- & -- \\ 
  naive & 0.383 & 0.452 & 0.629 & \textbf{0.375} & 0.362 & 0.469 & 0.583 & \textbf{0.333} \\ 
   \hline  \multicolumn{1}{l}{Correlation}\\ARGOX & \textbf{0.919} & 0.928 & 0.939 & 0.621 & \textbf{0.803} & \textbf{0.874} & 0.778 & \textbf{0.874} \\ 
  VAR & 0.744 & 0.687 & 0.732 & 0.383 & 0.274 & 0.816 & \textbf{0.818} & 0.494 \\ 
  GFT & -- & -- & \textbf{0.952} & -- & -- & -- & -- & -- \\ 
  Lu et al. (2019) & -- & \textbf{0.929} & 0.931 & \textbf{0.697} & -- & -- & -- & -- \\ 
  naive & 0.900 & 0.909 & 0.904 & 0.692 & 0.750 & 0.818 & 0.754 & 0.863 \\ 
   \hline
\end{tabular}
\caption{Comparison of different methods for state-level \%ILI estimation in Idaho (ID).  The MSE, MAE, and correlation are reported. The method with the best performance is highlighted in boldface for each metric in each period. \label{tab_state12}} 
\end{table}

\begin{figure}[!h] 
  \centering 
\includegraphics[width=\linewidth, page=12]{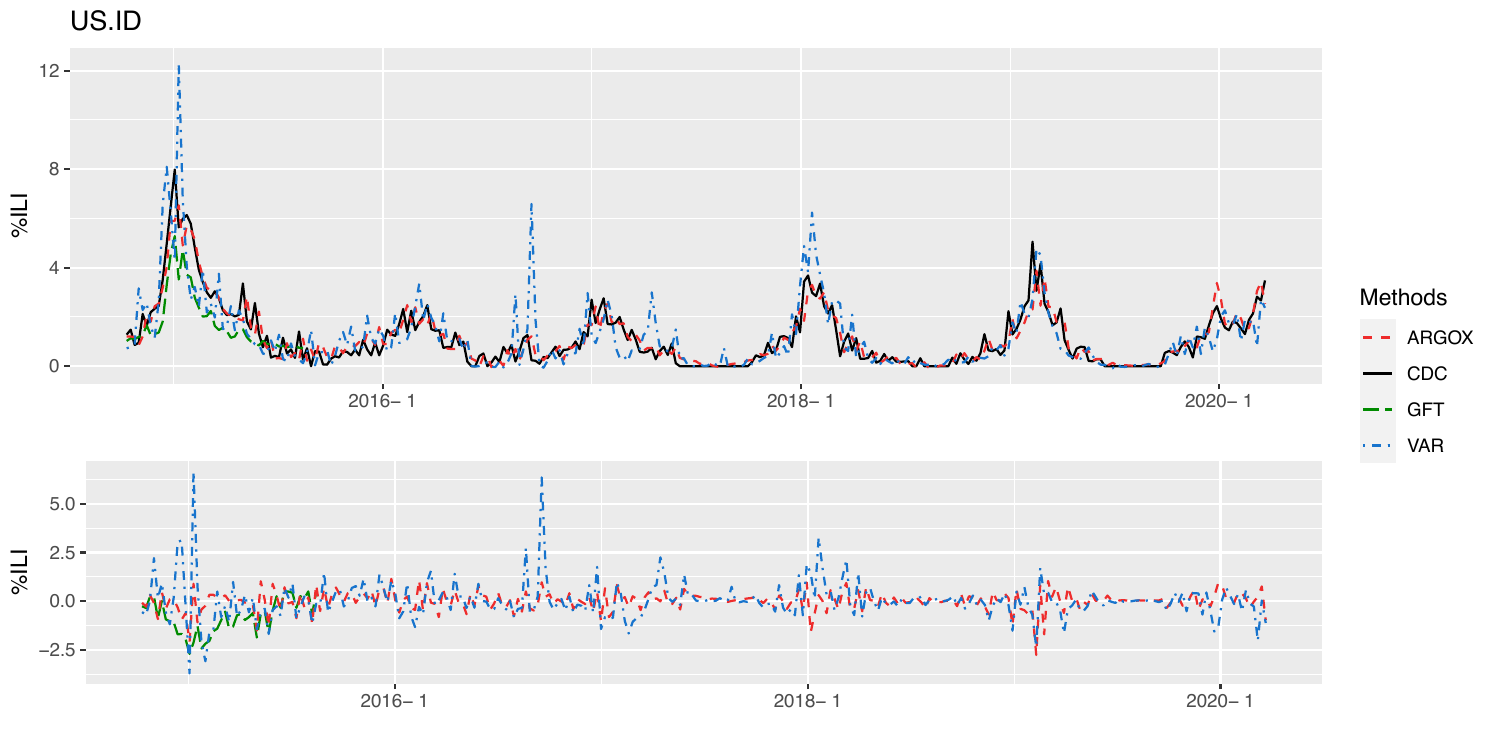} 
\caption{Plots of the \%ILI estimates (top) and the estimation errors (bottom) for Idaho (ID).}
\end{figure}
\newpage      
\begin{table}[ht]
\centering
\begin{tabular}{crrrrrrrr}
  \hline
  & Whole period '14-'20 & Overall '14-'17 & '14-'15 & '15-'16 & '16-'17 & '17-'18 & '18-'19 & '19-'20 \\ 
  \hline  \multicolumn{1}{l}{MSE}\\ARGOX & \textbf{0.085} & \textbf{0.079} & \textbf{0.156} & \textbf{0.050} & \textbf{0.046} & \textbf{0.106} & \textbf{0.077} & \textbf{0.291} \\ 
  VAR & 0.203 & 0.157 & 0.259 & 0.068 & 0.146 & 0.327 & 0.122 & 0.829 \\ 
  GFT & -- & -- & 0.414 & -- & -- & -- & -- & -- \\ 
  Lu et al. (2019) & -- & -- & -- & -- & -- & -- & -- & -- \\ 
  naive & 0.163 & 0.117 & 0.188 & 0.073 & 0.129 & 0.381 & 0.155 & 0.479 \\ 
   \hline  \multicolumn{1}{l}{MAE}\\ARGOX & \textbf{0.185} & \textbf{0.189} & \textbf{0.247} & \textbf{0.179} & \textbf{0.177} & \textbf{0.213} & \textbf{0.194} & \textbf{0.387} \\ 
  VAR & 0.289 & 0.285 & 0.357 & 0.203 & 0.292 & 0.365 & 0.228 & 0.680 \\ 
  GFT & -- & -- & 0.574 & -- & -- & -- & -- & -- \\ 
  Lu et al. (2019) & -- & -- & -- & -- & -- & -- & -- & -- \\ 
  naive & 0.254 & 0.230 & 0.262 & 0.221 & 0.268 & 0.396 & 0.285 & 0.523 \\ 
   \hline  \multicolumn{1}{l}{Correlation}\\ARGOX & \textbf{0.980} & \textbf{0.967} & \textbf{0.951} & \textbf{0.959} & \textbf{0.978} & \textbf{0.984} & \textbf{0.947} & \textbf{0.967} \\ 
  VAR & 0.952 & 0.941 & 0.933 & 0.956 & 0.931 & 0.955 & 0.913 & 0.905 \\ 
  GFT & -- & -- & 0.947 & -- & -- & -- & -- & -- \\ 
  Lu et al. (2019) & -- & -- & -- & -- & -- & -- & -- & -- \\ 
  naive & 0.962 & 0.951 & 0.939 & 0.939 & 0.936 & 0.939 & 0.889 & 0.945 \\ 
   \hline
\end{tabular}
\caption{Comparison of different methods for state-level \%ILI estimation in Illinois (IL).  The MSE, MAE, and correlation are reported. The method with the best performance is highlighted in boldface for each metric in each period. \label{tab_state13}} 
\end{table}

\begin{figure}[!h] 
  \centering 
\includegraphics[width=\linewidth, page=13]{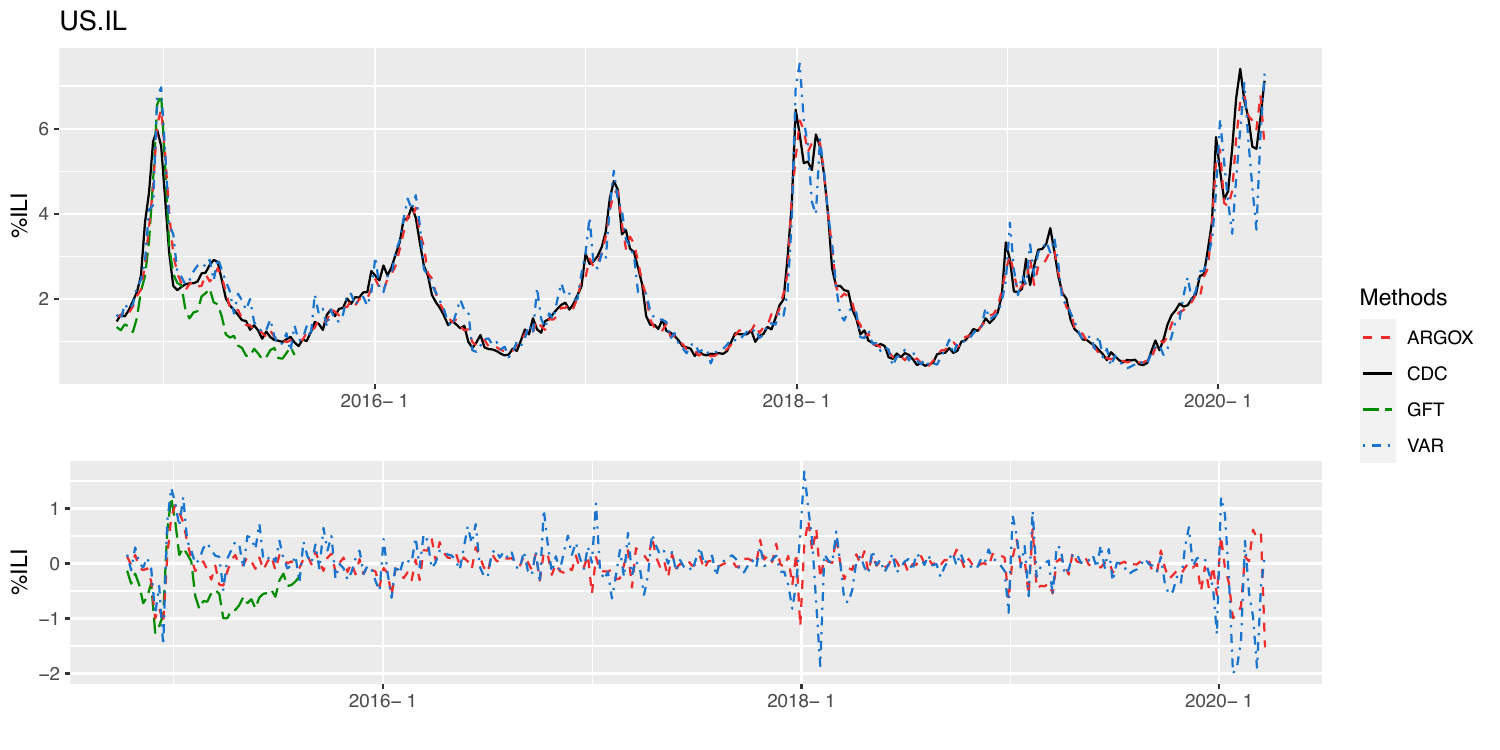} 
\caption{Plots of the \%ILI estimates (top) and the estimation errors (bottom) for Illinois (IL).}
\end{figure}
\newpage      
\begin{table}[ht]
\centering
\begin{tabular}{crrrrrrrr}
  \hline
  & Whole period '14-'20 & Overall '14-'17 & '14-'15 & '15-'16 & '16-'17 & '17-'18 & '18-'19 & '19-'20 \\ 
  \hline  \multicolumn{1}{l}{MSE}\\ARGOX & \textbf{0.326} & \textbf{0.304} & 0.372 & \textbf{0.164} & \textbf{0.493} & \textbf{0.345} & \textbf{0.328} & \textbf{0.998} \\ 
  VAR & 1.266 & 0.761 & 1.410 & 0.463 & 0.682 & 3.456 & 1.959 & 2.641 \\ 
  GFT & -- & -- & \textbf{0.220} & -- & -- & -- & -- & -- \\ 
  Lu et al. (2019) & -- & -- & -- & -- & -- & -- & -- & -- \\ 
  naive & 0.537 & 0.515 & 0.751 & 0.315 & 0.689 & 0.928 & 0.613 & 1.094 \\ 
   \hline  \multicolumn{1}{l}{MAE}\\ARGOX & \textbf{0.372} & \textbf{0.371} & 0.382 & \textbf{0.310} & \textbf{0.485} & \textbf{0.435} & \textbf{0.408} & \textbf{0.576} \\ 
  VAR & 0.636 & 0.531 & 0.638 & 0.499 & 0.640 & 1.056 & 0.940 & 0.992 \\ 
  GFT & -- & -- & \textbf{0.373} & -- & -- & -- & -- & -- \\ 
  Lu et al. (2019) & -- & -- & -- & -- & -- & -- & -- & -- \\ 
  naive & 0.459 & 0.464 & 0.486 & 0.433 & 0.596 & 0.608 & 0.540 & 0.685 \\ 
   \hline  \multicolumn{1}{l}{Correlation}\\ARGOX & \textbf{0.947} & \textbf{0.925} & 0.942 & \textbf{0.878} & \textbf{0.900} & \textbf{0.972} & \textbf{0.922} & \textbf{0.860} \\ 
  VAR & 0.857 & 0.869 & 0.907 & 0.707 & 0.857 & 0.884 & 0.668 & 0.609 \\ 
  GFT & -- & -- & \textbf{0.970} & -- & -- & -- & -- & -- \\ 
  Lu et al. (2019) & -- & -- & -- & -- & -- & -- & -- & -- \\ 
  naive & 0.914 & 0.877 & 0.885 & 0.776 & 0.865 & 0.925 & 0.857 & 0.840 \\ 
   \hline
\end{tabular}
\caption{Comparison of different methods for state-level \%ILI estimation in Indiana (IN).  The MSE, MAE, and correlation are reported. The method with the best performance is highlighted in boldface for each metric in each period. \label{tab_state14}} 
\end{table}

\begin{figure}[!h] 
  \centering 
\includegraphics[width=\linewidth, page=14]{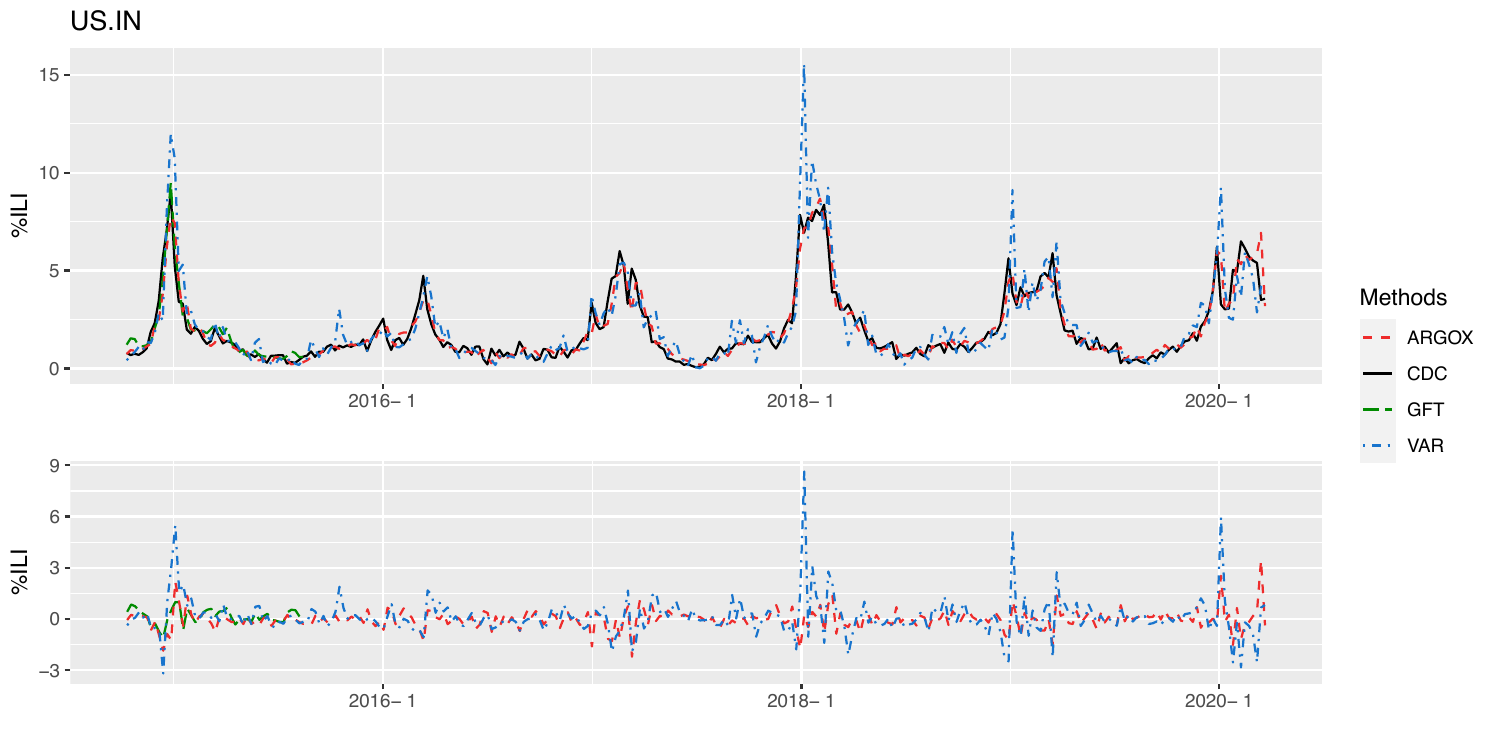} 
\caption{Plots of the \%ILI estimates (top) and the estimation errors (bottom) for Indiana (IN).}
\end{figure}
\newpage      
\begin{table}[ht]
\centering
\begin{tabular}{crrrrrrrr}
  \hline
  & Whole period '14-'20 & Overall '14-'17 & '14-'15 & '15-'16 & '16-'17 & '17-'18 & '18-'19 & '19-'20 \\ 
  \hline  \multicolumn{1}{l}{MSE}\\ARGOX & \textbf{0.166} & \textbf{0.182} & 0.380 & \textbf{0.073} & \textbf{0.122} & \textbf{0.147} & \textbf{0.175} & \textbf{0.444} \\ 
  VAR & 0.446 & 0.201 & \textbf{0.318} & 0.096 & 0.256 & 0.888 & 0.430 & 2.245 \\ 
  GFT & -- & -- & 4.702 & -- & -- & -- & -- & -- \\ 
  Lu et al. (2019) & -- & -- & -- & -- & -- & -- & -- & -- \\ 
  naive & 0.219 & 0.196 & 0.371 & 0.087 & 0.181 & 0.298 & 0.229 & 0.706 \\ 
   \hline  \multicolumn{1}{l}{MAE}\\ARGOX & \textbf{0.241} & \textbf{0.243} & 0.311 & \textbf{0.225} & \textbf{0.250} & \textbf{0.280} & \textbf{0.288} & \textbf{0.442} \\ 
  VAR & 0.350 & 0.279 & 0.301 & 0.249 & 0.381 & 0.516 & 0.450 & 0.933 \\ 
  GFT & -- & -- & 1.237 & -- & -- & -- & -- & -- \\ 
  Lu et al. (2019) & -- & -- & -- & -- & -- & -- & -- & -- \\ 
  naive & 0.265 & 0.250 & \textbf{0.284} & 0.244 & 0.302 & 0.360 & 0.347 & 0.513 \\ 
   \hline  \multicolumn{1}{l}{Correlation}\\ARGOX & \textbf{0.920} & \textbf{0.816} & 0.708 & \textbf{0.720} & \textbf{0.916} & \textbf{0.945} & \textbf{0.920} & \textbf{0.880} \\ 
  VAR & 0.825 & 0.787 & \textbf{0.775} & 0.677 & 0.809 & 0.815 & 0.867 & 0.563 \\ 
  GFT & -- & -- & 0.605 & -- & -- & -- & -- & -- \\ 
  Lu et al. (2019) & -- & -- & -- & -- & -- & -- & -- & -- \\ 
  naive & 0.896 & 0.812 & 0.745 & 0.702 & 0.875 & 0.881 & 0.895 & 0.811 \\ 
   \hline
\end{tabular}
\caption{Comparison of different methods for state-level \%ILI estimation in Iowa (IA).  The MSE, MAE, and correlation are reported. The method with the best performance is highlighted in boldface for each metric in each period. \label{tab_state15}} 
\end{table}

\begin{figure}[!h] 
  \centering 
\includegraphics[width=\linewidth, page=15]{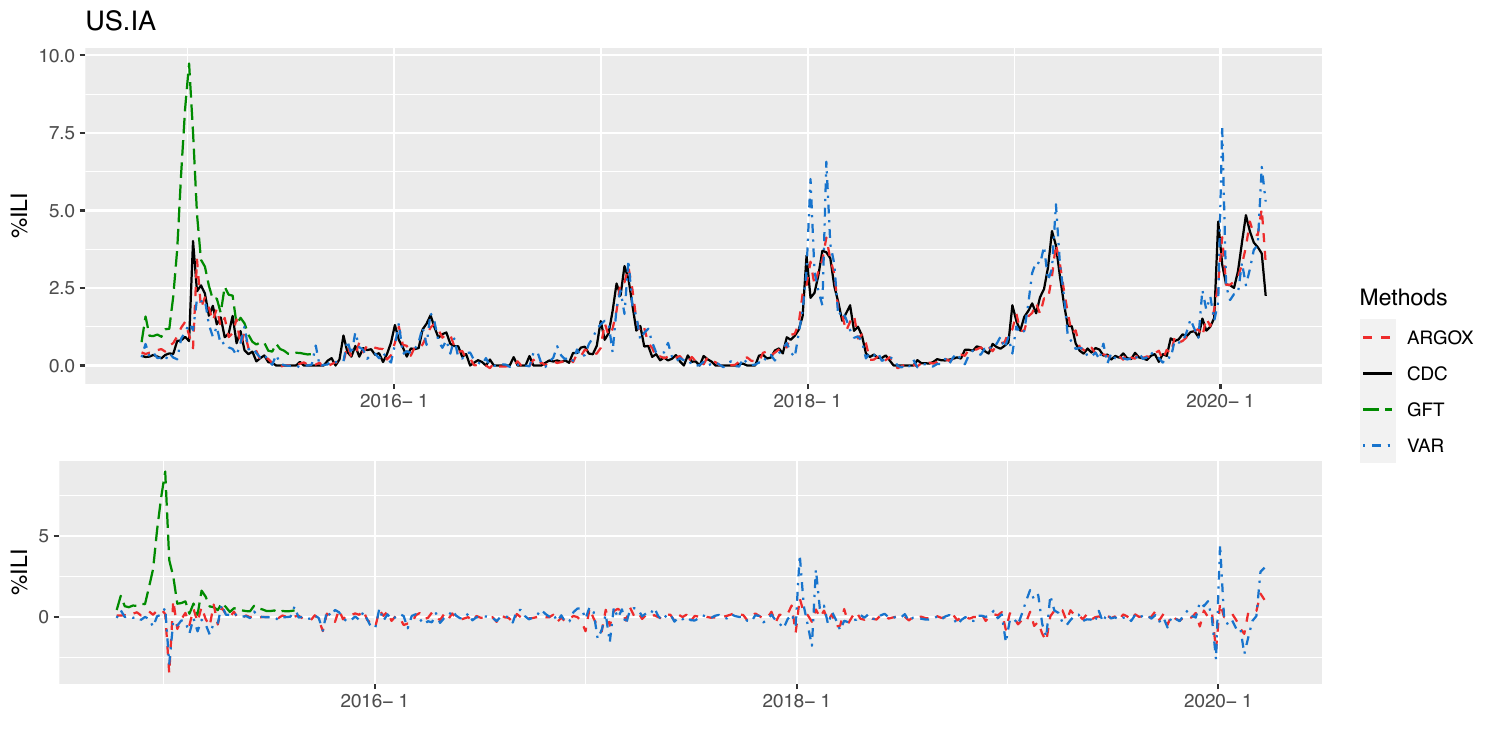} 
\caption{Plots of the \%ILI estimates (top) and the estimation errors (bottom) for Iowa (IA).}
\end{figure}
\newpage      
\begin{table}[ht]
\centering
\begin{tabular}{crrrrrrrr}
  \hline
  & Whole period '14-'20 & Overall '14-'17 & '14-'15 & '15-'16 & '16-'17 & '17-'18 & '18-'19 & '19-'20 \\ 
  \hline  \multicolumn{1}{l}{MSE}\\ARGOX & \textbf{0.269} & \textbf{0.326} & \textbf{0.450} & \textbf{0.069} & 0.640 & \textbf{0.260} & \textbf{0.230} & \textbf{0.463} \\ 
  VAR & 1.489 & 0.909 & 1.504 & 0.616 & 1.045 & 2.389 & 0.866 & 7.667 \\ 
  GFT & -- & -- & 1.929 & -- & -- & -- & -- & -- \\ 
  Lu et al. (2019) & -- & 0.328 & 0.607 & 0.118 & \textbf{0.591} & -- & -- & -- \\ 
  naive & 0.461 & 0.492 & 0.719 & 0.089 & 0.942 & 1.013 & 0.307 & 0.642 \\ 
   \hline  \multicolumn{1}{l}{MAE}\\ARGOX & \textbf{0.328} & \textbf{0.334} & \textbf{0.389} & 0.224 & \textbf{0.527} & \textbf{0.394} & \textbf{0.309} & \textbf{0.496} \\ 
  VAR & 0.664 & 0.562 & 0.733 & 0.530 & 0.675 & 1.038 & 0.692 & 1.673 \\ 
  GFT & -- & -- & 0.983 & -- & -- & -- & -- & -- \\ 
  Lu et al. (2019) & -- & -- & -- & -- & -- & -- & -- & -- \\ 
  naive & 0.405 & 0.397 & 0.481 & \textbf{0.222} & 0.674 & 0.709 & 0.381 & 0.586 \\ 
   \hline  \multicolumn{1}{l}{Correlation}\\ARGOX & \textbf{0.980} & \textbf{0.967} & \textbf{0.969} & \textbf{0.901} & 0.953 & \textbf{0.992} & \textbf{0.955} & \textbf{0.967} \\ 
  VAR & 0.919 & 0.918 & 0.919 & 0.697 & 0.919 & 0.954 & 0.871 & 0.699 \\ 
  GFT & -- & -- & 0.962 & -- & -- & -- & -- & -- \\ 
  Lu et al. (2019) & -- & 0.965 & 0.960 & 0.860 & \textbf{0.961} & -- & -- & -- \\ 
  naive & 0.965 & 0.948 & 0.945 & 0.874 & 0.929 & 0.965 & 0.937 & 0.959 \\ 
   \hline
\end{tabular}
\caption{Comparison of different methods for state-level \%ILI estimation in Kansas (KS).  The MSE, MAE, and correlation are reported. The method with the best performance is highlighted in boldface for each metric in each period. \label{tab_state16}} 
\end{table}

\begin{figure}[!h] 
  \centering 
\includegraphics[width=\linewidth, page=16]{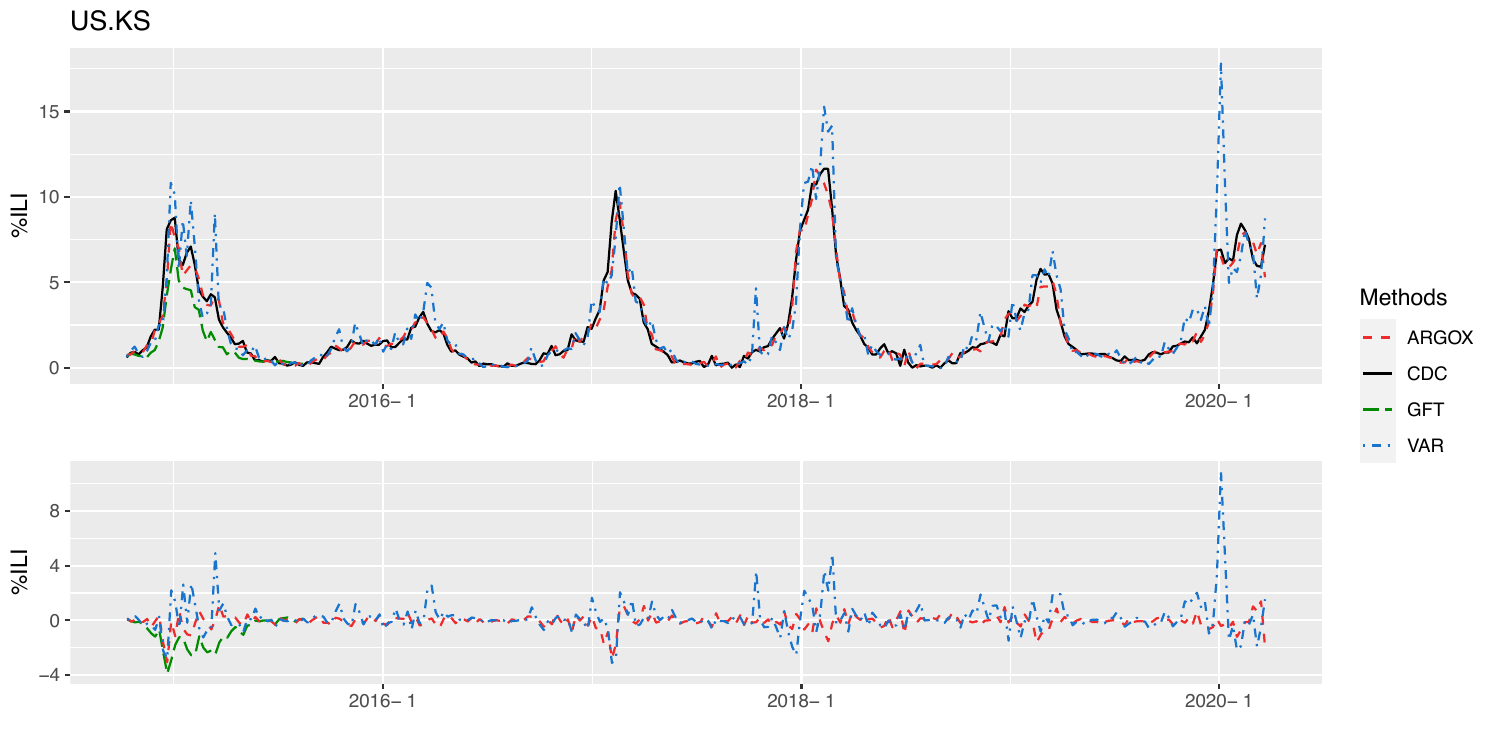} 
\caption{Plots of the \%ILI estimates (top) and the estimation errors (bottom) for Kansas (KS).}
\end{figure}
\newpage      
\begin{table}[ht]
\centering
\begin{tabular}{crrrrrrrr}
  \hline
  & Whole period '14-'20 & Overall '14-'17 & '14-'15 & '15-'16 & '16-'17 & '17-'18 & '18-'19 & '19-'20 \\ 
  \hline  \multicolumn{1}{l}{MSE}\\ARGOX & \textbf{ 0.467} & \textbf{ 0.310} & \textbf{ 0.071} & \textbf{ 0.058} & \textbf{ 1.049} & \textbf{ 0.871} & \textbf{ 0.785} & \textbf{ 1.342} \\ 
  VAR &  3.082 &  0.823 &  0.508 &  0.076 &  2.562 &  9.726 &  3.600 & 13.010 \\ 
  GFT & -- & -- &  7.601 & -- & -- & -- & -- & -- \\ 
  Lu et al. (2019) & -- &  0.415 &  0.106 &  0.076 &  1.433 & -- & -- & -- \\ 
  naive &  0.724 &  0.447 &  0.137 &  0.079 &  1.494 &  1.641 &  1.078 &  2.174 \\ 
   \hline  \multicolumn{1}{l}{MAE}\\ARGOX & \textbf{0.385} & \textbf{0.303} & \textbf{0.158} & \textbf{0.162} & \textbf{0.788} & \textbf{0.612} & \textbf{0.548} & \textbf{0.810} \\ 
  VAR & 0.777 & 0.408 & 0.288 & 0.194 & 1.014 & 1.715 & 1.293 & 2.222 \\ 
  GFT & -- & -- & 1.951 & -- & -- & -- & -- & -- \\ 
  Lu et al. (2019) & -- & -- & -- & -- & -- & -- & -- & -- \\ 
  naive & 0.466 & 0.351 & 0.197 & 0.190 & 0.913 & 0.851 & 0.680 & 1.072 \\ 
   \hline  \multicolumn{1}{l}{Correlation}\\ARGOX & \textbf{0.971} & \textbf{0.957} & 0.954 & \textbf{0.934} & \textbf{0.931} & \textbf{0.971} & \textbf{0.952} & \textbf{0.930} \\ 
  VAR & 0.879 & 0.876 & 0.905 & 0.922 & 0.825 & 0.899 & 0.857 & 0.617 \\ 
  GFT & -- & -- & \textbf{0.964} & -- & -- & -- & -- & -- \\ 
  Lu et al. (2019) & -- & 0.941 & 0.940 & 0.923 & 0.907 & -- & -- & -- \\ 
  naive & 0.955 & 0.936 & 0.910 & 0.908 & 0.898 & 0.943 & 0.934 & 0.891 \\ 
   \hline
\end{tabular}
\caption{Comparison of different methods for state-level \%ILI estimation in Kentucky (KY).  The MSE, MAE, and correlation are reported. The method with the best performance is highlighted in boldface for each metric in each period. \label{tab_state17}} 
\end{table}

\begin{figure}[!h] 
  \centering 
\includegraphics[width=\linewidth, page=17]{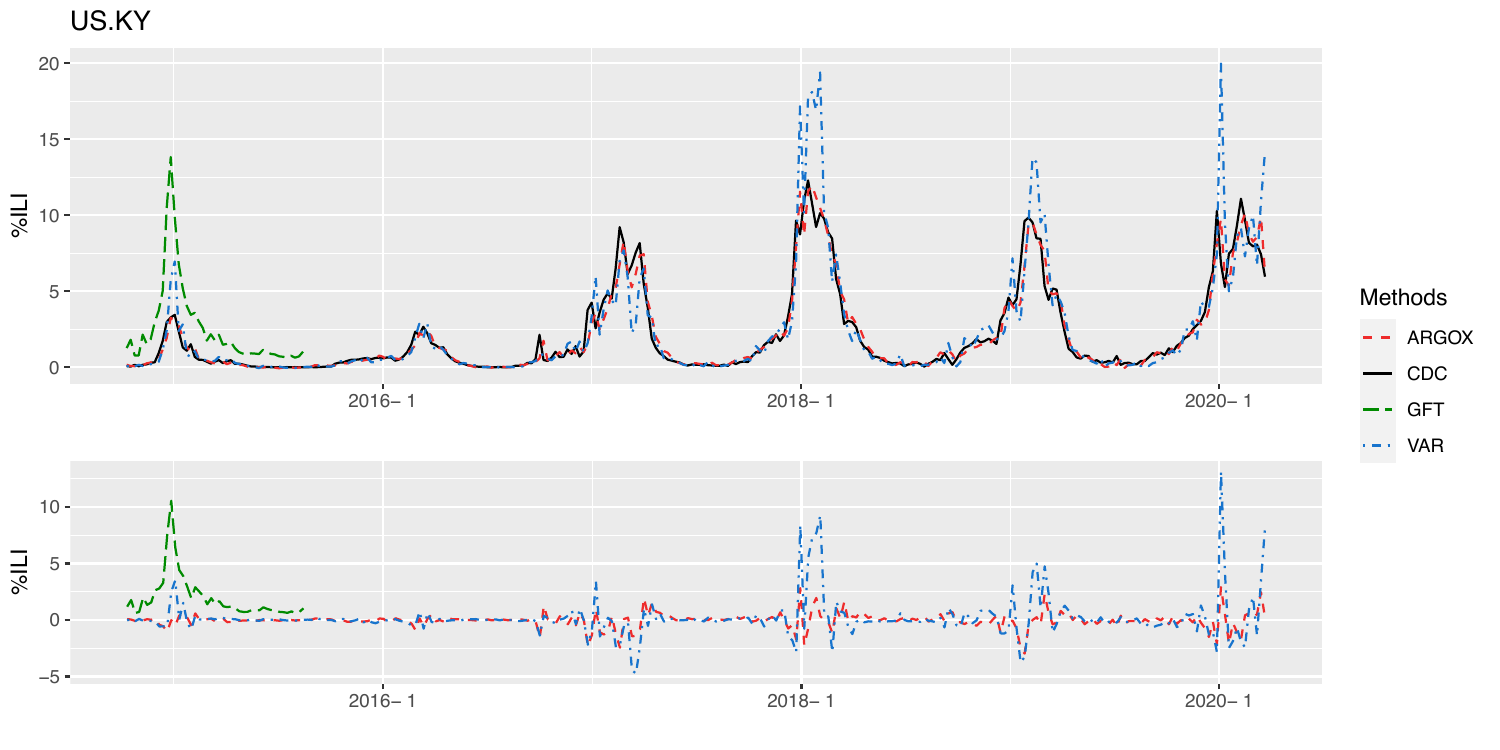} 
\caption{Plots of the \%ILI estimates (top) and the estimation errors (bottom) for Kentucky (KY).}
\end{figure}
\newpage      
\begin{table}[ht]
\centering
\begin{tabular}{crrrrrrrr}
  \hline
  & Whole period '14-'20 & Overall '14-'17 & '14-'15 & '15-'16 & '16-'17 & '17-'18 & '18-'19 & '19-'20 \\ 
  \hline  \multicolumn{1}{l}{MSE}\\ARGOX & \textbf{0.281} & \textbf{0.084} & \textbf{0.162} & \textbf{0.049} & \textbf{0.064} & \textbf{0.403} & 0.770 & \textbf{1.135} \\ 
  VAR & 0.554 & 0.310 & 0.654 & 0.194 & 0.115 & 0.574 & \textbf{0.758} & 2.684 \\ 
  GFT & -- & -- & 0.917 & -- & -- & -- & -- & -- \\ 
  Lu et al. (2019) & -- & 0.149 & 0.281 & 0.132 & 0.154 & -- & -- & -- \\ 
  naive & 0.429 & 0.144 & 0.287 & 0.062 & 0.126 & 0.817 & 0.963 & 1.667 \\ 
   \hline  \multicolumn{1}{l}{MAE}\\ARGOX & \textbf{0.306} & \textbf{0.191} & \textbf{0.258} & \textbf{0.177} & \textbf{0.189} & \textbf{0.444} & \textbf{0.560} & \textbf{0.792} \\ 
  VAR & 0.483 & 0.366 & 0.541 & 0.336 & 0.260 & 0.562 & 0.622 & 1.375 \\ 
  GFT & -- & -- & 0.676 & -- & -- & -- & -- & -- \\ 
  Lu et al. (2019) & -- & -- & -- & -- & -- & -- & -- & -- \\ 
  naive & 0.392 & 0.249 & 0.353 & 0.195 & 0.272 & 0.591 & 0.705 & 1.033 \\ 
   \hline  \multicolumn{1}{l}{Correlation}\\ARGOX & \textbf{0.979} & \textbf{0.977} & 0.975 & \textbf{0.921} & \textbf{0.974} & \textbf{0.979} & 0.948 & \textbf{0.884} \\ 
  VAR & 0.960 & 0.942 & 0.947 & 0.711 & 0.951 & 0.972 & \textbf{0.959} & 0.795 \\ 
  GFT & -- & -- & \textbf{0.984} & -- & -- & -- & -- & -- \\ 
  Lu et al. (2019) & -- & 0.963 & 0.955 & 0.796 & 0.958 & -- & -- & -- \\ 
  naive & 0.968 & 0.960 & 0.954 & 0.876 & 0.948 & 0.954 & 0.927 & 0.833 \\ 
   \hline
\end{tabular}
\caption{Comparison of different methods for state-level \%ILI estimation in Louisiana (LA).  The MSE, MAE, and correlation are reported. The method with the best performance is highlighted in boldface for each metric in each period. \label{tab_state18}} 
\end{table}

\begin{figure}[!h] 
  \centering 
\includegraphics[width=\linewidth, page=18]{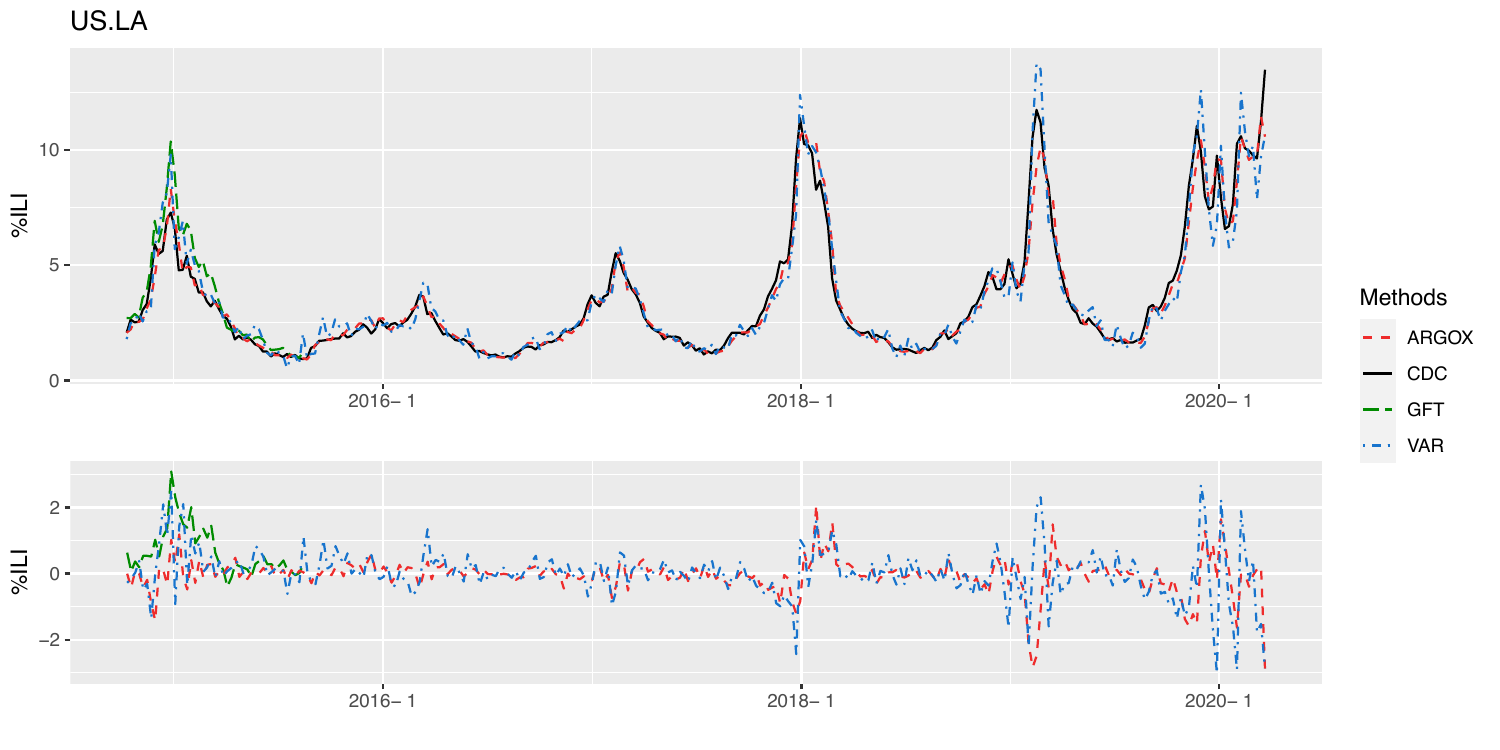} 
\caption{Plots of the \%ILI estimates (top) and the estimation errors (bottom) for Louisiana (LA).}
\end{figure}
\newpage      
\begin{table}[ht]
\centering
\begin{tabular}{crrrrrrrr}
  \hline
  & Whole period '14-'20 & Overall '14-'17 & '14-'15 & '15-'16 & '16-'17 & '17-'18 & '18-'19 & '19-'20 \\ 
  \hline  \multicolumn{1}{l}{MSE}\\ARGOX & \textbf{0.087} & 0.086 & \textbf{0.097} & 0.109 & 0.078 & \textbf{0.062} & 0.090 & \textbf{0.293} \\ 
  VAR & 0.176 & 0.172 & 0.230 & 0.159 & 0.132 & 0.139 & 0.317 & 0.401 \\ 
  GFT & -- & -- & 0.497 & -- & -- & -- & -- & -- \\ 
  Lu et al. (2019) & -- & \textbf{0.073} & 0.102 & \textbf{0.094} & \textbf{0.061} & -- & -- & -- \\ 
  naive & 0.097 & 0.105 & 0.118 & 0.141 & 0.088 & 0.062 & \textbf{0.088} & 0.303 \\ 
   \hline  \multicolumn{1}{l}{MAE}\\ARGOX & \textbf{0.208} & \textbf{0.225} & \textbf{0.237} & \textbf{0.264} & \textbf{0.221} & \textbf{0.192} & 0.247 & \textbf{0.331} \\ 
  VAR & 0.313 & 0.318 & 0.368 & 0.301 & 0.277 & 0.303 & 0.460 & 0.520 \\ 
  GFT & -- & -- & 0.448 & -- & -- & -- & -- & -- \\ 
  Lu et al. (2019) & -- & -- & -- & -- & -- & -- & -- & -- \\ 
  naive & 0.222 & 0.247 & 0.264 & 0.290 & 0.229 & 0.204 & \textbf{0.232} & 0.350 \\ 
   \hline  \multicolumn{1}{l}{Correlation}\\ARGOX & \textbf{0.957} & 0.853 & 0.901 & 0.310 & 0.794 & \textbf{0.806} & \textbf{0.943} & 0.964 \\ 
  VAR & 0.912 & 0.716 & 0.783 & 0.114 & 0.628 & 0.626 & 0.823 & 0.952 \\ 
  GFT & -- & -- & \textbf{0.914} & -- & -- & -- & -- & -- \\ 
  Lu et al. (2019) & -- & \textbf{0.873} & 0.897 & \textbf{0.375} & \textbf{0.854} & -- & -- & -- \\ 
  naive & 0.951 & 0.831 & 0.885 & 0.257 & 0.789 & 0.780 & 0.940 & \textbf{0.964} \\ 
   \hline
\end{tabular}
\caption{Comparison of different methods for state-level \%ILI estimation in Maine (ME).  The MSE, MAE, and correlation are reported. The method with the best performance is highlighted in boldface for each metric in each period. \label{tab_state19}} 
\end{table}

\begin{figure}[!h] 
  \centering 
\includegraphics[width=\linewidth, page=19]{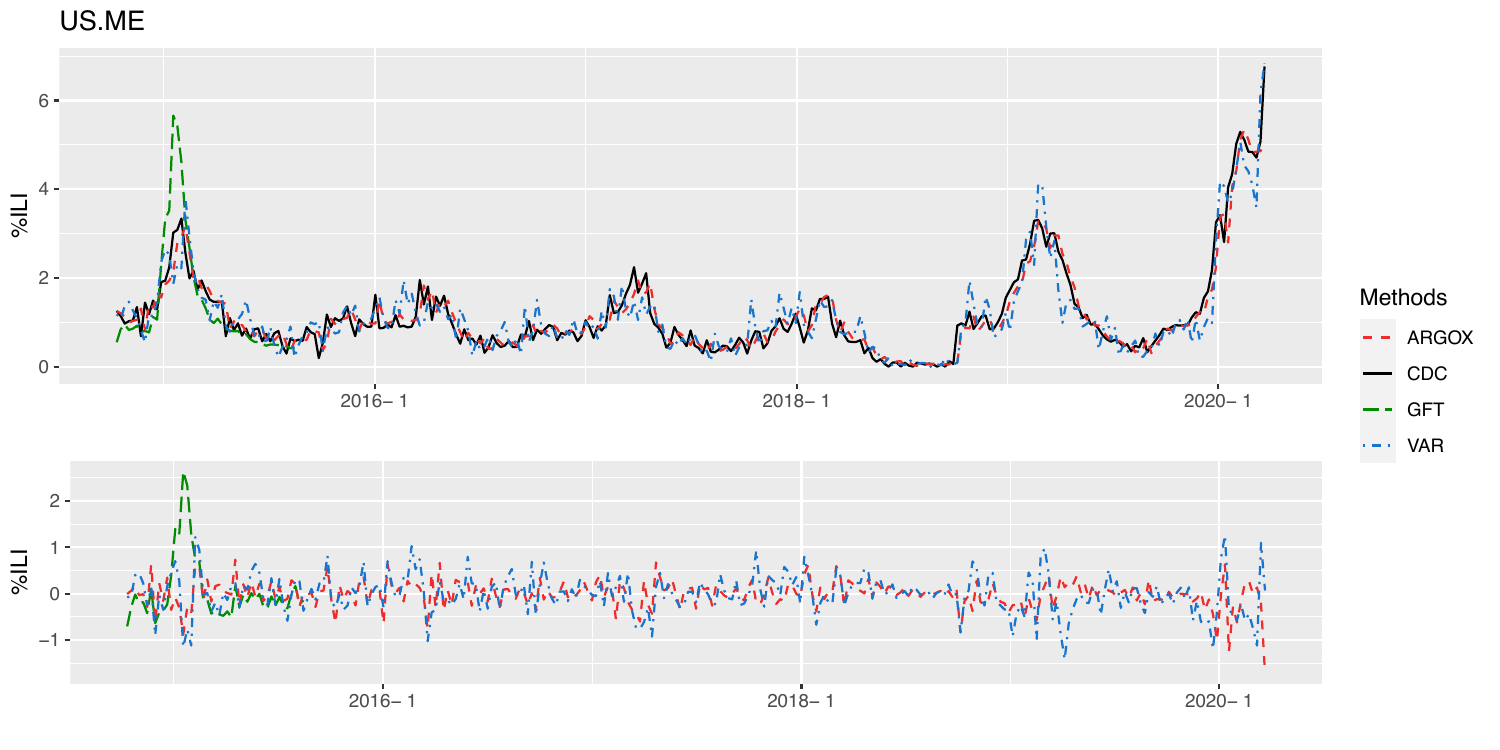} 
\caption{Plots of the \%ILI estimates (top) and the estimation errors (bottom) for Maine (ME).}
\end{figure}
\newpage      
\begin{table}[ht]
\centering
\begin{tabular}{crrrrrrrr}
  \hline
  & Whole period '14-'20 & Overall '14-'17 & '14-'15 & '15-'16 & '16-'17 & '17-'18 & '18-'19 & '19-'20 \\ 
  \hline  \multicolumn{1}{l}{MSE}\\ARGOX & \textbf{0.267} & \textbf{0.309} & \textbf{0.356} & \textbf{0.336} & \textbf{0.361} & \textbf{0.170} & \textbf{0.270} & \textbf{0.574} \\ 
  VAR & 0.831 & 0.846 & 1.053 & 0.645 & 0.846 & 0.463 & 0.620 & 2.937 \\ 
  GFT & -- & -- & 1.504 & -- & -- & -- & -- & -- \\ 
  Lu et al. (2019) & -- & 0.347 & 0.365 & 0.347 & 0.441 & -- & -- & -- \\ 
  naive & 0.376 & 0.433 & 0.507 & 0.432 & 0.520 & 0.376 & 0.389 & 0.678 \\ 
   \hline  \multicolumn{1}{l}{MAE}\\ARGOX & \textbf{0.374} & \textbf{0.413} & \textbf{0.396} & \textbf{0.465} & \textbf{0.487} & \textbf{0.331} & \textbf{0.409} & \textbf{0.517} \\ 
  VAR & 0.595 & 0.639 & 0.640 & 0.597 & 0.661 & 0.503 & 0.567 & 1.131 \\ 
  GFT & -- & -- & 0.968 & -- & -- & -- & -- & -- \\ 
  Lu et al. (2019) & -- & -- & -- & -- & -- & -- & -- & -- \\ 
  naive & 0.423 & 0.452 & 0.456 & 0.494 & 0.510 & 0.441 & 0.496 & 0.594 \\ 
   \hline  \multicolumn{1}{l}{Correlation}\\ARGOX & \textbf{0.940} & \textbf{0.846} & 0.804 & \textbf{0.732} & \textbf{0.918} & \textbf{0.968} & \textbf{0.911} & \textbf{0.947} \\ 
  VAR & 0.837 & 0.653 & 0.741 & 0.585 & 0.788 & 0.905 & 0.788 & 0.819 \\ 
  GFT & -- & -- & 0.816 & -- & -- & -- & -- & -- \\ 
  Lu et al. (2019) & -- & 0.824 & \textbf{0.842} & 0.718 & 0.885 & -- & -- & -- \\ 
  naive & 0.916 & 0.800 & 0.729 & 0.695 & 0.872 & 0.925 & 0.876 & 0.937 \\ 
   \hline
\end{tabular}
\caption{Comparison of different methods for state-level \%ILI estimation in Maryland (MD).  The MSE, MAE, and correlation are reported. The method with the best performance is highlighted in boldface for each metric in each period. \label{tab_state20}} 
\end{table}

\begin{figure}[!h] 
  \centering 
\includegraphics[width=\linewidth, page=20]{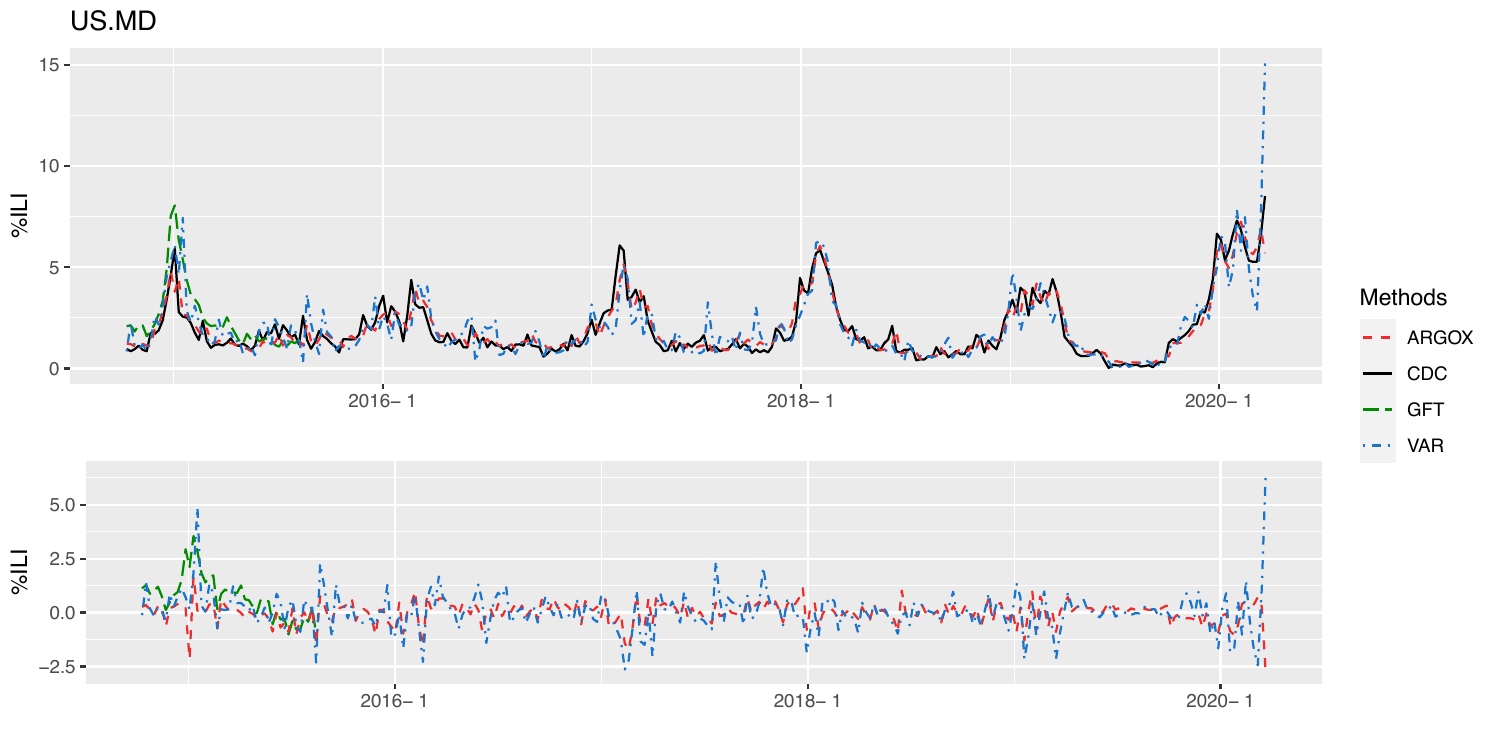} 
\caption{Plots of the \%ILI estimates (top) and the estimation errors (bottom) for Maryland (MD).}
\end{figure}
\newpage      
\begin{table}[ht]
\centering
\begin{tabular}{crrrrrrrr}
  \hline
  & Whole period '14-'20 & Overall '14-'17 & '14-'15 & '15-'16 & '16-'17 & '17-'18 & '18-'19 & '19-'20 \\ 
  \hline  \multicolumn{1}{l}{MSE}\\ARGOX & \textbf{0.080} & 0.040 & 0.059 & 0.042 & 0.032 & \textbf{0.125} & \textbf{0.043} & \textbf{0.469} \\ 
  VAR & 0.155 & 0.101 & 0.133 & 0.076 & 0.137 & 0.315 & 0.160 & 0.510 \\ 
  GFT & -- & -- & 0.112 & -- & -- & -- & -- & -- \\ 
  Lu et al. (2019) & -- & \textbf{0.027} & \textbf{0.038} & \textbf{0.034} & \textbf{0.019} & -- & -- & -- \\ 
  naive & 0.126 & 0.070 & 0.100 & 0.066 & 0.073 & 0.284 & 0.105 & 0.528 \\ 
   \hline  \multicolumn{1}{l}{MAE}\\ARGOX & \textbf{0.161} & \textbf{0.139} & \textbf{0.146} & \textbf{0.160} & \textbf{0.145} & \textbf{0.243} & \textbf{0.152} & \textbf{0.397} \\ 
  VAR & 0.266 & 0.234 & 0.263 & 0.219 & 0.283 & 0.393 & 0.293 & 0.540 \\ 
  GFT & -- & -- & 0.263 & -- & -- & -- & -- & -- \\ 
  Lu et al. (2019) & -- & -- & -- & -- & -- & -- & -- & -- \\ 
  naive & 0.217 & 0.181 & 0.214 & 0.181 & 0.203 & 0.358 & 0.231 & 0.530 \\ 
   \hline  \multicolumn{1}{l}{Correlation}\\ARGOX & \textbf{0.970} & 0.949 & 0.936 & 0.912 & 0.959 & \textbf{0.968} & \textbf{0.966} & \textbf{0.930} \\ 
  VAR & 0.938 & 0.874 & 0.869 & 0.845 & 0.818 & 0.927 & 0.884 & 0.928 \\ 
  GFT & -- & -- & 0.947 & -- & -- & -- & -- & -- \\ 
  Lu et al. (2019) & -- & \textbf{0.967} & \textbf{0.966} & \textbf{0.929} & \textbf{0.976} & -- & -- & -- \\ 
  naive & 0.950 & 0.912 & 0.891 & 0.870 & 0.908 & 0.921 & 0.919 & 0.923 \\ 
   \hline
\end{tabular}
\caption{Comparison of different methods for state-level \%ILI estimation in Massachusetts (MA).  The MSE, MAE, and correlation are reported. The method with the best performance is highlighted in boldface for each metric in each period. \label{tab_state21}} 
\end{table}

\begin{figure}[!h] 
  \centering 
\includegraphics[width=\linewidth, page=21]{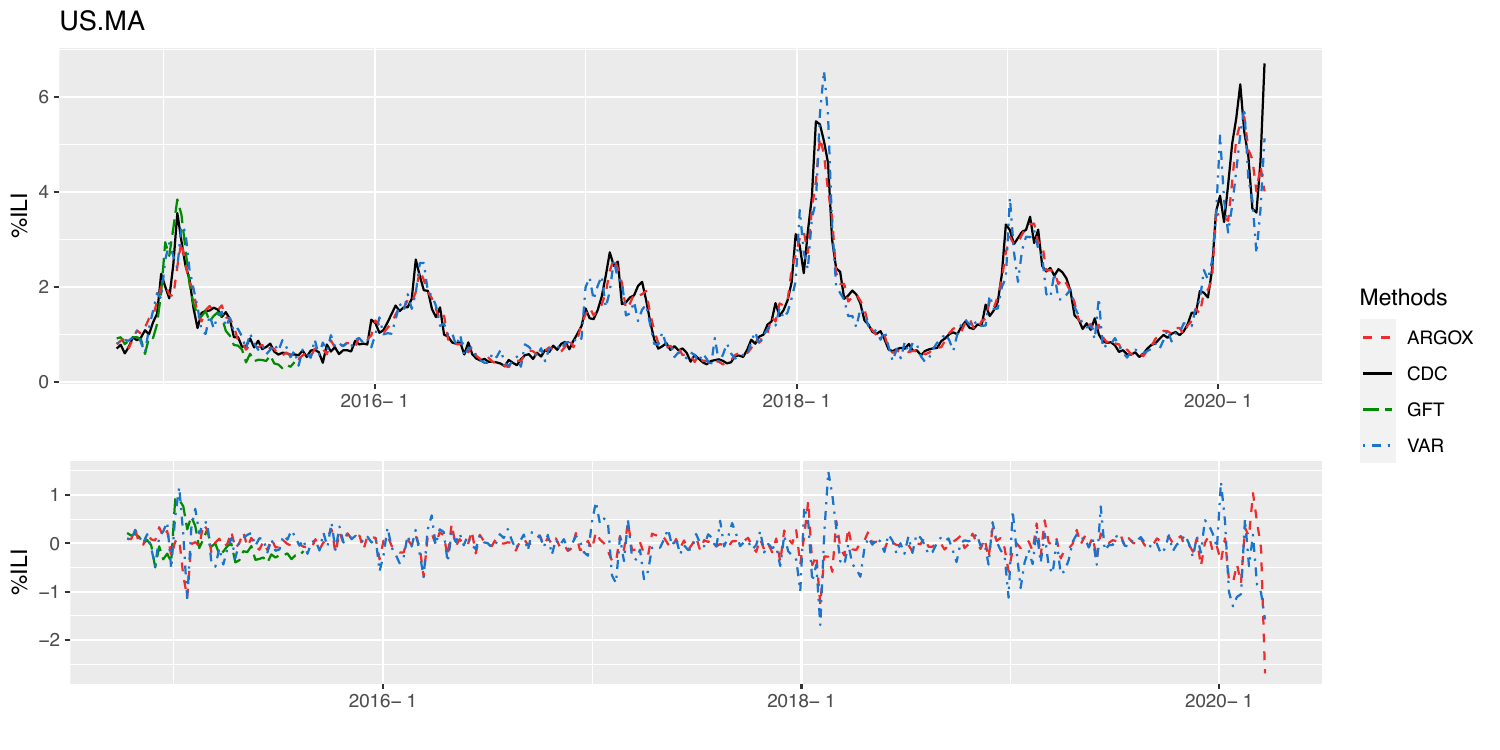} 
\caption{Plots of the \%ILI estimates (top) and the estimation errors (bottom) for Massachusetts (MA).}
\end{figure}
\newpage      
\begin{table}[ht]
\centering
\begin{tabular}{crrrrrrrr}
  \hline
  & Whole period '14-'20 & Overall '14-'17 & '14-'15 & '15-'16 & '16-'17 & '17-'18 & '18-'19 & '19-'20 \\ 
  \hline  \multicolumn{1}{l}{MSE}\\ARGOX & \textbf{0.073} & \textbf{0.077} & \textbf{0.131} & \textbf{0.032} & \textbf{0.093} & \textbf{0.100} & \textbf{0.082} & \textbf{0.119} \\ 
  VAR & 0.177 & 0.168 & 0.174 & 0.174 & 0.249 & 0.318 & 0.181 & 0.317 \\ 
  GFT & -- & -- & 1.313 & -- & -- & -- & -- & -- \\ 
  Lu et al. (2019) & -- & 0.132 & 0.246 & 0.081 & 0.174 & -- & -- & -- \\ 
  naive & 0.130 & 0.146 & 0.260 & 0.066 & 0.169 & 0.232 & 0.108 & 0.177 \\ 
   \hline  \multicolumn{1}{l}{MAE}\\ARGOX & \textbf{0.187} & \textbf{0.183} & \textbf{0.237} & \textbf{0.134} & \textbf{0.221} & \textbf{0.256} & \textbf{0.222} & \textbf{0.253} \\ 
  VAR & 0.292 & 0.293 & 0.284 & 0.309 & 0.403 & 0.413 & 0.314 & 0.380 \\ 
  GFT & -- & -- & 0.957 & -- & -- & -- & -- & -- \\ 
  Lu et al. (2019) & -- & -- & -- & -- & -- & -- & -- & -- \\ 
  naive & 0.236 & 0.227 & 0.284 & 0.184 & 0.299 & 0.392 & 0.255 & 0.308 \\ 
   \hline  \multicolumn{1}{l}{Correlation}\\ARGOX & \textbf{0.969} & \textbf{0.962} & 0.935 & \textbf{0.963} & \textbf{0.955} & \textbf{0.977} & \textbf{0.873} & \textbf{0.949} \\ 
  VAR & 0.934 & 0.928 & 0.938 & 0.905 & 0.874 & 0.940 & 0.795 & 0.871 \\ 
  GFT & -- & -- & \textbf{0.979} & -- & -- & -- & -- & -- \\ 
  Lu et al. (2019) & -- & 0.936 & 0.919 & 0.908 & 0.928 & -- & -- & -- \\ 
  naive & 0.945 & 0.928 & 0.868 & 0.917 & 0.917 & 0.939 & 0.837 & 0.930 \\ 
   \hline
\end{tabular}
\caption{Comparison of different methods for state-level \%ILI estimation in Michigan (MI).  The MSE, MAE, and correlation are reported. The method with the best performance is highlighted in boldface for each metric in each period. \label{tab_state22}} 
\end{table}

\begin{figure}[!h] 
  \centering 
\includegraphics[width=\linewidth, page=22]{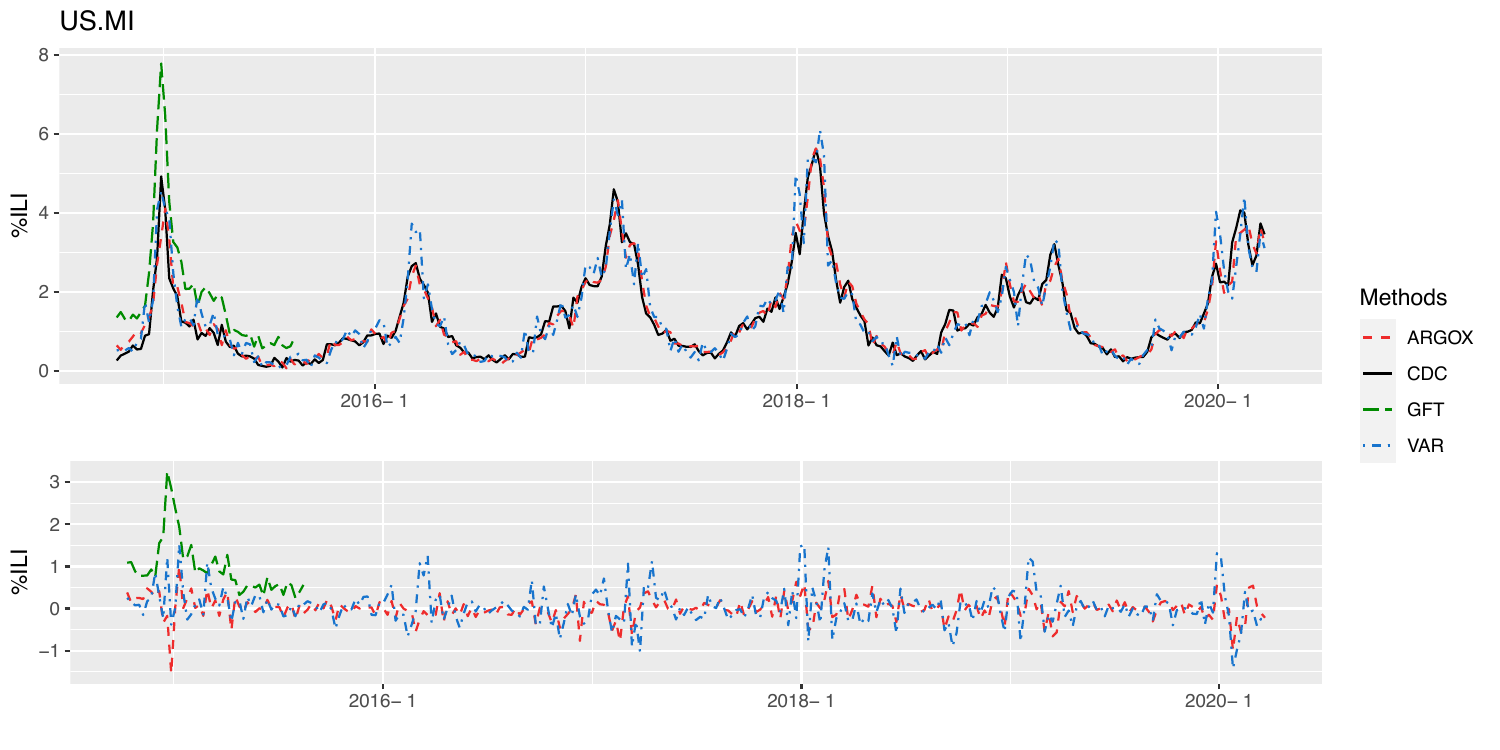} 
\caption{Plots of the \%ILI estimates (top) and the estimation errors (bottom) for Michigan (MI).}
\end{figure}
\newpage      
\begin{table}[ht]
\centering
\begin{tabular}{crrrrrrrr}
  \hline
  & Whole period '14-'20 & Overall '14-'17 & '14-'15 & '15-'16 & '16-'17 & '17-'18 & '18-'19 & '19-'20 \\ 
  \hline  \multicolumn{1}{l}{MSE}\\ARGOX & \textbf{0.309} & \textbf{0.251} & \textbf{0.388} & \textbf{0.283} & \textbf{0.174} & \textbf{0.148} & \textbf{0.266} & \textbf{1.447} \\ 
  VAR & 0.953 & 1.049 & 1.552 & 0.634 & 1.414 & 0.387 & 1.047 & 2.487 \\ 
  GFT & -- & -- & 0.544 & -- & -- & -- & -- & -- \\ 
  Lu et al. (2019) & -- & 0.287 & 0.472 & 0.370 & 0.215 & -- & -- & -- \\ 
  naive & 0.499 & 0.443 & 0.722 & 0.510 & 0.277 & 0.247 & 0.289 & 2.395 \\ 
   \hline  \multicolumn{1}{l}{MAE}\\ARGOX & \textbf{0.371} & \textbf{0.374} & \textbf{0.436} & \textbf{0.460} & \textbf{0.333} & \textbf{0.320} & \textbf{0.364} & \textbf{0.773} \\ 
  VAR & 0.644 & 0.640 & 0.713 & 0.608 & 0.779 & 0.475 & 0.779 & 1.143 \\ 
  GFT & -- & -- & 0.527 & -- & -- & -- & -- & -- \\ 
  Lu et al. (2019) & -- & -- & -- & -- & -- & -- & -- & -- \\ 
  naive & 0.441 & 0.440 & 0.535 & 0.544 & 0.378 & 0.388 & 0.402 & 0.965 \\ 
   \hline  \multicolumn{1}{l}{Correlation}\\ARGOX & \textbf{0.929} & \textbf{0.922} & 0.923 & \textbf{0.770} & \textbf{0.935} & \textbf{0.978} & \textbf{0.825} & \textbf{0.726} \\ 
  VAR & 0.815 & 0.779 & 0.839 & 0.576 & 0.536 & 0.948 & 0.662 & 0.602 \\ 
  GFT & -- & -- & \textbf{0.931} & -- & -- & -- & -- & -- \\ 
  Lu et al. (2019) & -- & 0.907 & 0.910 & 0.703 & 0.915 & -- & -- & -- \\ 
  naive & 0.888 & 0.865 & 0.855 & 0.629 & 0.888 & 0.960 & 0.807 & 0.575 \\ 
   \hline
\end{tabular}
\caption{Comparison of different methods for state-level \%ILI estimation in Minnesota (MN).  The MSE, MAE, and correlation are reported. The method with the best performance is highlighted in boldface for each metric in each period. \label{tab_state23}} 
\end{table}

\begin{figure}[!h] 
  \centering 
\includegraphics[width=\linewidth, page=23]{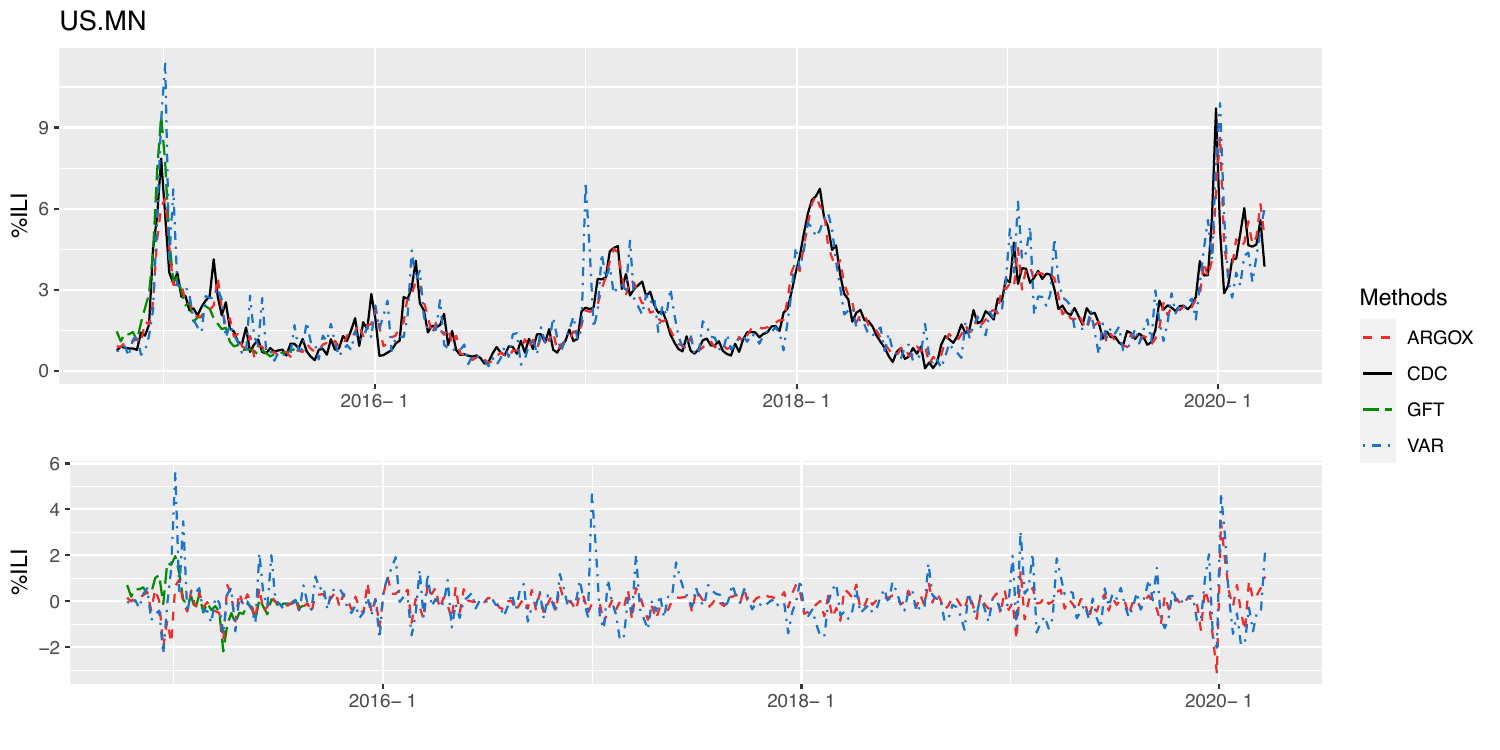} 
\caption{Plots of the \%ILI estimates (top) and the estimation errors (bottom) for Minnesota (MN).}
\end{figure}
\newpage      
\begin{table}[ht]
\centering
\begin{tabular}{crrrrrrrr}
  \hline
  & Whole period '14-'20 & Overall '14-'17 & '14-'15 & '15-'16 & '16-'17 & '17-'18 & '18-'19 & '19-'20 \\ 
  \hline  \multicolumn{1}{l}{MSE}\\ARGOX & \textbf{0.334} & \textbf{0.224} & \textbf{0.381} & \textbf{0.092} & \textbf{0.241} & \textbf{0.961} & \textbf{0.358} & \textbf{0.640} \\ 
  VAR & 0.821 & 0.462 & 0.663 & 0.439 & 0.428 & 2.412 & 0.989 & 1.619 \\ 
  GFT & -- & -- & 1.936 & -- & -- & -- & -- & -- \\ 
  Lu et al. (2019) & -- & -- & -- & -- & -- & -- & -- & -- \\ 
  naive & 0.622 & 0.476 & 0.931 & 0.200 & 0.422 & 1.690 & 0.706 & 1.043 \\ 
   \hline  \multicolumn{1}{l}{MAE}\\ARGOX & \textbf{0.384} & \textbf{0.335} & \textbf{0.413} & \textbf{0.261} & \textbf{0.392} & \textbf{0.677} & \textbf{0.415} & \textbf{0.665} \\ 
  VAR & 0.614 & 0.510 & 0.571 & 0.533 & 0.531 & 1.032 & 0.724 & 0.939 \\ 
  GFT & -- & -- & 1.220 & -- & -- & -- & -- & -- \\ 
  Lu et al. (2019) & -- & -- & -- & -- & -- & -- & -- & -- \\ 
  naive & 0.508 & 0.450 & 0.606 & 0.344 & 0.519 & 0.882 & 0.601 & 0.851 \\ 
   \hline  \multicolumn{1}{l}{Correlation}\\ARGOX & \textbf{0.967} & \textbf{0.965} & \textbf{0.969} & \textbf{0.921} & \textbf{0.932} & \textbf{0.963} & \textbf{0.962} & \textbf{0.894} \\ 
  VAR & 0.932 & 0.932 & 0.947 & 0.731 & 0.879 & 0.931 & 0.925 & 0.778 \\ 
  GFT & -- & -- & 0.960 & -- & -- & -- & -- & -- \\ 
  Lu et al. (2019) & -- & -- & -- & -- & -- & -- & -- & -- \\ 
  naive & 0.940 & 0.926 & 0.923 & 0.835 & 0.883 & 0.930 & 0.912 & 0.822 \\ 
   \hline
\end{tabular}
\caption{Comparison of different methods for state-level \%ILI estimation in Mississippi (MS).  The MSE, MAE, and correlation are reported. The method with the best performance is highlighted in boldface for each metric in each period. \label{tab_state24}} 
\end{table}

\begin{figure}[!h] 
  \centering 
\includegraphics[width=\linewidth, page=24]{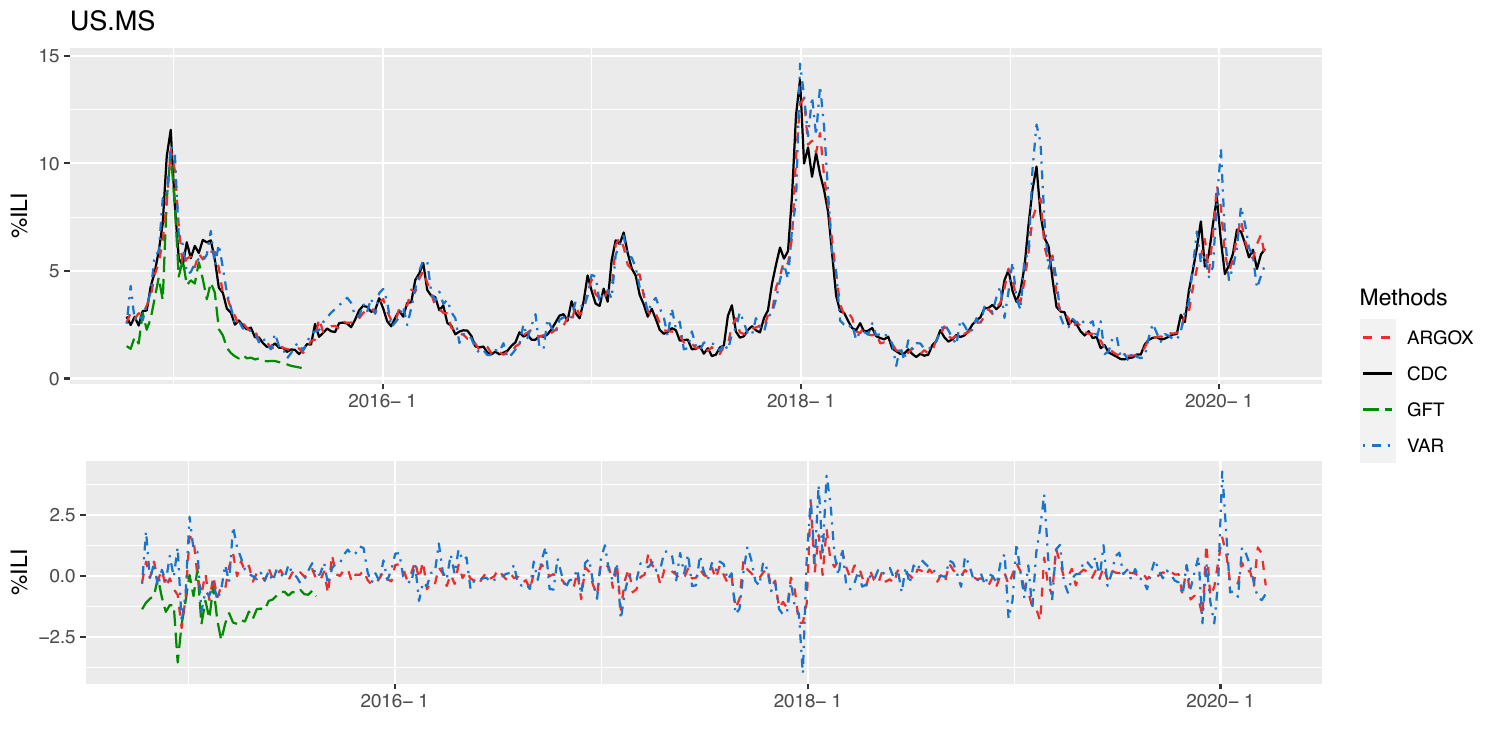} 
\caption{Plots of the \%ILI estimates (top) and the estimation errors (bottom) for Mississippi (MS).}
\end{figure}
\newpage      
\begin{table}[ht]
\centering
\begin{tabular}{crrrrrrrr}
  \hline
  & Whole period '14-'20 & Overall '14-'17 & '14-'15 & '15-'16 & '16-'17 & '17-'18 & '18-'19 & '19-'20 \\ 
  \hline  \multicolumn{1}{l}{MSE}\\ARGOX & \textbf{0.442} & \textbf{0.241} & \textbf{0.336} & \textbf{0.139} & \textbf{0.353} & \textbf{0.817} & \textbf{0.654} & \textbf{1.709} \\ 
  VAR & 2.090 & 0.934 & 0.843 & 0.969 & 1.633 & 6.039 & 1.068 & 9.347 \\ 
  GFT & -- & -- & 0.482 & -- & -- & -- & -- & -- \\ 
  Lu et al. (2019) & -- & -- & -- & -- & -- & -- & -- & -- \\ 
  naive & 0.767 & 0.412 & 0.741 & 0.155 & 0.478 & 1.818 & 1.017 & 2.692 \\ 
   \hline  \multicolumn{1}{l}{MAE}\\ARGOX & \textbf{0.392} & \textbf{0.319} & \textbf{0.358} & 0.314 & \textbf{0.379} & \textbf{0.571} & \textbf{0.559} & \textbf{0.918} \\ 
  VAR & 0.712 & 0.611 & 0.637 & 0.639 & 0.795 & 1.068 & 0.691 & 2.100 \\ 
  GFT & -- & -- & 0.570 & -- & -- & -- & -- & -- \\ 
  Lu et al. (2019) & -- & -- & -- & -- & -- & -- & -- & -- \\ 
  naive & 0.499 & 0.392 & 0.486 & \textbf{0.311} & 0.466 & 0.883 & 0.718 & 1.114 \\ 
   \hline  \multicolumn{1}{l}{Correlation}\\ARGOX & \textbf{0.963} & \textbf{0.951} & 0.956 & \textbf{0.775} & \textbf{0.939} & \textbf{0.972} & \textbf{0.936} & \textbf{0.929} \\ 
  VAR & 0.866 & 0.852 & 0.883 & 0.626 & 0.844 & 0.897 & 0.901 & 0.658 \\ 
  GFT & -- & -- & \textbf{0.959} & -- & -- & -- & -- & -- \\ 
  Lu et al. (2019) & -- & -- & -- & -- & -- & -- & -- & -- \\ 
  naive & 0.935 & 0.916 & 0.898 & 0.761 & 0.919 & 0.938 & 0.890 & 0.895 \\ 
   \hline
\end{tabular}
\caption{Comparison of different methods for state-level \%ILI estimation in Missouri (MO).  The MSE, MAE, and correlation are reported. The method with the best performance is highlighted in boldface for each metric in each period. \label{tab_state25}} 
\end{table}

\begin{figure}[!h] 
  \centering 
\includegraphics[width=\linewidth, page=25]{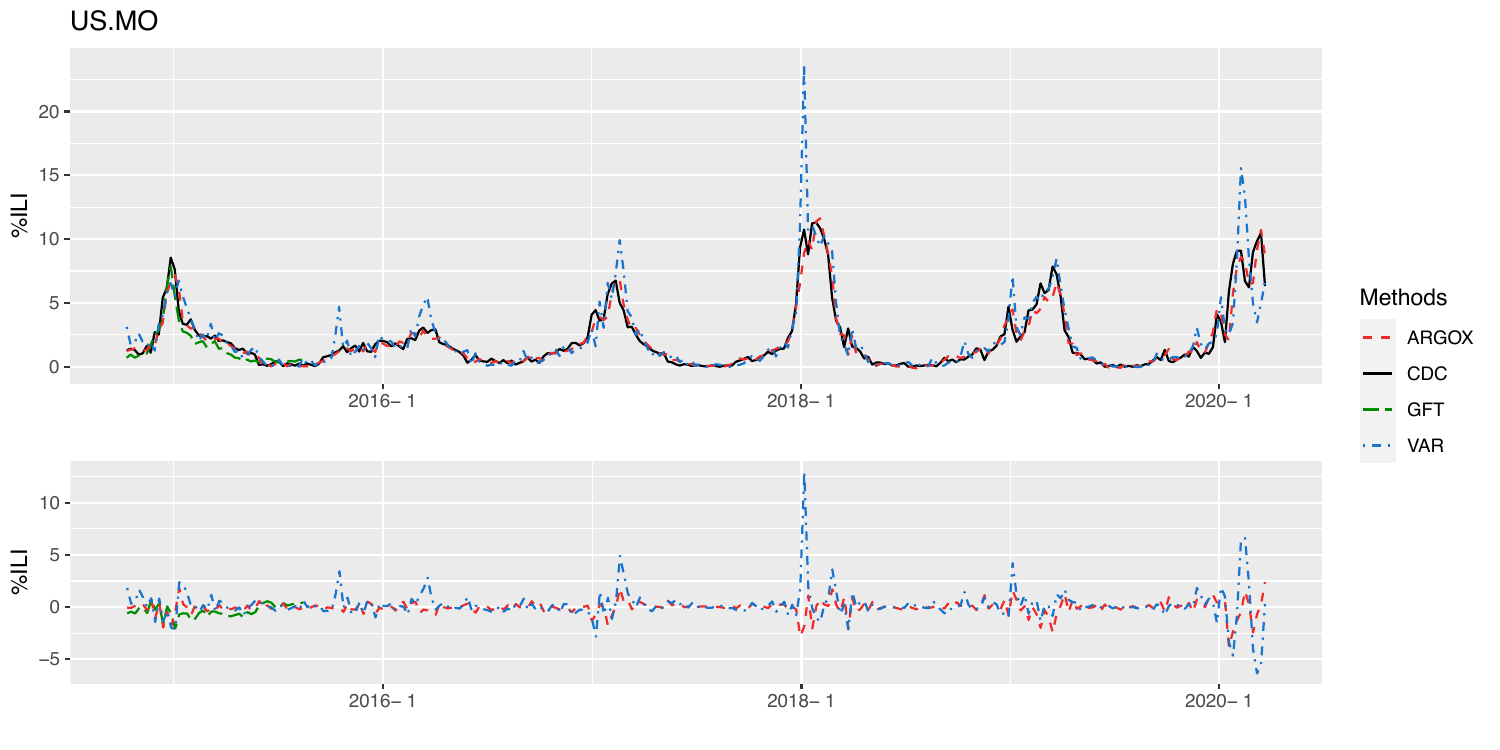} 
\caption{Plots of the \%ILI estimates (top) and the estimation errors (bottom) for Missouri (MO).}
\end{figure}
\newpage      
\begin{table}[ht]
\centering
\begin{tabular}{crrrrrrrr}
  \hline
  & Whole period '14-'20 & Overall '14-'17 & '14-'15 & '15-'16 & '16-'17 & '17-'18 & '18-'19 & '19-'20 \\ 
  \hline  \multicolumn{1}{l}{MSE}\\ARGOX & \textbf{0.055} & \textbf{0.039} & \textbf{0.061} & \textbf{0.055} & \textbf{0.020} & \textbf{0.022} & 0.194 & \textbf{0.121} \\ 
  VAR & 0.198 & 0.157 & 0.275 & 0.080 & 0.136 & 0.037 & 0.351 & 0.885 \\ 
  GFT & -- & -- & 1.095 & -- & -- & -- & -- & -- \\ 
  Lu et al. (2019) & -- & -- & -- & -- & -- & -- & -- & -- \\ 
  naive & 0.061 & 0.049 & 0.079 & 0.067 & 0.024 & 0.029 & \textbf{0.184} & 0.142 \\ 
   \hline  \multicolumn{1}{l}{MAE}\\ARGOX & \textbf{0.143} & \textbf{0.131} & 0.181 & \textbf{0.164} & \textbf{0.100} & \textbf{0.110} & 0.322 & \textbf{0.254} \\ 
  VAR & 0.248 & 0.232 & 0.330 & 0.216 & 0.212 & 0.140 & 0.449 & 0.598 \\ 
  GFT & -- & -- & 0.870 & -- & -- & -- & -- & -- \\ 
  Lu et al. (2019) & -- & -- & -- & -- & -- & -- & -- & -- \\ 
  naive & 0.144 & 0.132 & \textbf{0.179} & 0.169 & 0.112 & 0.120 & \textbf{0.320} & 0.271 \\ 
   \hline  \multicolumn{1}{l}{Correlation}\\ARGOX & \textbf{0.967} & \textbf{0.854} & 0.893 & \textbf{0.419} & \textbf{0.580} & \textbf{0.779} & \textbf{0.919} & \textbf{0.972} \\ 
  VAR & 0.877 & 0.552 & 0.585 & 0.367 & 0.398 & 0.668 & 0.897 & 0.802 \\ 
  GFT & -- & -- & \textbf{0.939} & -- & -- & -- & -- & -- \\ 
  Lu et al. (2019) & -- & -- & -- & -- & -- & -- & -- & -- \\ 
  naive & 0.963 & 0.821 & 0.865 & 0.341 & 0.499 & 0.736 & 0.916 & 0.969 \\ 
   \hline
\end{tabular}
\caption{Comparison of different methods for state-level \%ILI estimation in Montana (MT).  The MSE, MAE, and correlation are reported. The method with the best performance is highlighted in boldface for each metric in each period. \label{tab_state26}} 
\end{table}

\begin{figure}[!h] 
  \centering 
\includegraphics[width=\linewidth, page=26]{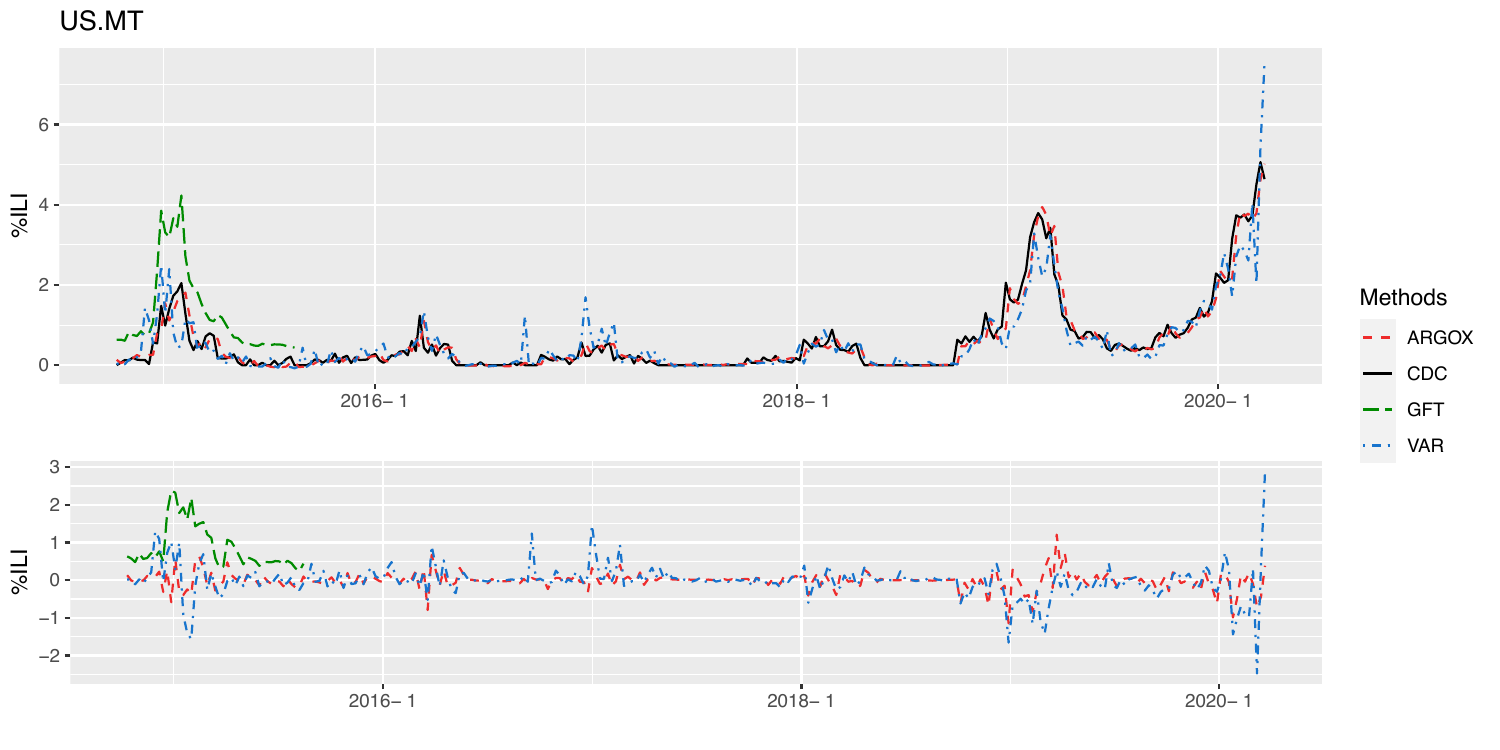} 
\caption{Plots of the \%ILI estimates (top) and the estimation errors (bottom) for Montana (MT).}
\end{figure}
\newpage      
\begin{table}[ht]
\centering
\begin{tabular}{crrrrrrrr}
  \hline
  & Whole period '14-'20 & Overall '14-'17 & '14-'15 & '15-'16 & '16-'17 & '17-'18 & '18-'19 & '19-'20 \\ 
  \hline  \multicolumn{1}{l}{MSE}\\ARGOX & \textbf{ 0.380} & \textbf{ 0.231} & \textbf{ 0.470} &  0.086 & \textbf{ 0.195} & \textbf{ 0.684} & \textbf{ 0.755} & \textbf{ 0.840} \\ 
  VAR &  2.520 &  1.172 &  3.148 &  0.093 &  0.353 &  2.474 &  4.081 & 13.168 \\ 
  GFT & -- & -- &  0.678 & -- & -- & -- & -- & -- \\ 
  Lu et al. (2019) & -- &  0.265 &  0.605 &  0.141 &  0.251 & -- & -- & -- \\ 
  naive &  0.497 &  0.303 &  0.636 & \textbf{ 0.082} &  0.272 &  0.822 &  0.944 &  1.341 \\ 
   \hline  \multicolumn{1}{l}{MAE}\\ARGOX & \textbf{0.392} & \textbf{0.287} & \textbf{0.389} & 0.234 & \textbf{0.303} & \textbf{0.617} & \textbf{0.574} & \textbf{0.704} \\ 
  VAR & 0.779 & 0.490 & 0.881 & 0.253 & 0.373 & 1.175 & 1.295 & 1.846 \\ 
  GFT & -- & -- & 0.501 & -- & -- & -- & -- & -- \\ 
  Lu et al. (2019) & -- & -- & -- & -- & -- & -- & -- & -- \\ 
  naive & 0.434 & 0.318 & 0.460 & \textbf{0.211} & 0.353 & 0.692 & 0.684 & 0.736 \\ 
   \hline  \multicolumn{1}{l}{Correlation}\\ARGOX & \textbf{0.950} & 0.866 & 0.835 & 0.631 & \textbf{0.897} & \textbf{0.925} & \textbf{0.918} & \textbf{0.900} \\ 
  VAR & 0.769 & 0.495 & 0.255 & 0.498 & 0.818 & 0.723 & 0.628 & 0.735 \\ 
  GFT & -- & -- & 0.846 & -- & -- & -- & -- & -- \\ 
  Lu et al. (2019) & -- & \textbf{0.878} & \textbf{0.861} & \textbf{0.680} & 0.863 & -- & -- & -- \\ 
  naive & 0.936 & 0.836 & 0.788 & 0.608 & 0.859 & 0.902 & 0.901 & 0.842 \\ 
   \hline
\end{tabular}
\caption{Comparison of different methods for state-level \%ILI estimation in Nebraska (NE).  The MSE, MAE, and correlation are reported. The method with the best performance is highlighted in boldface for each metric in each period. \label{tab_state27}} 
\end{table}

\begin{figure}[!h] 
  \centering 
\includegraphics[width=\linewidth, page=27]{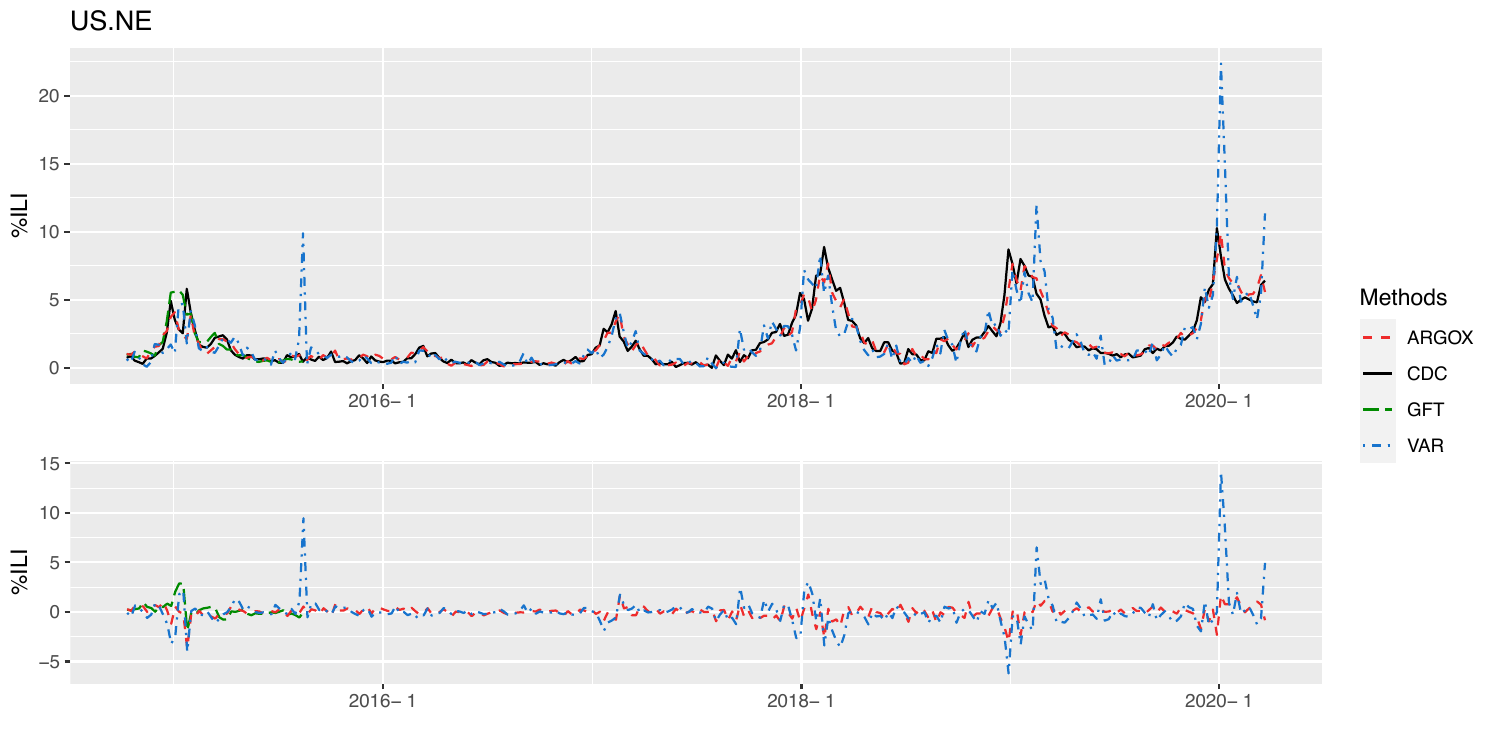} 
\caption{Plots of the \%ILI estimates (top) and the estimation errors (bottom) for Nebraska (NE).}
\end{figure}
\newpage      
\begin{table}[ht]
\centering
\begin{tabular}{crrrrrrrr}
  \hline
  & Whole period '14-'20 & Overall '14-'17 & '14-'15 & '15-'16 & '16-'17 & '17-'18 & '18-'19 & '19-'20 \\ 
  \hline  \multicolumn{1}{l}{MSE}\\ARGOX & \textbf{0.096} & \textbf{0.089} & \textbf{0.132} & \textbf{0.096} & 0.066 & \textbf{0.046} & \textbf{0.142} & 0.322 \\ 
  VAR & 0.232 & 0.163 & 0.185 & 0.199 & 0.187 & 0.395 & 0.342 & 0.687 \\ 
  GFT & -- & -- & 2.565 & -- & -- & -- & -- & -- \\ 
  Lu et al. (2019) & -- & 0.112 & 0.261 & 0.112 & \textbf{0.053} & -- & -- & -- \\ 
  naive & 0.115 & 0.109 & 0.179 & 0.131 & 0.060 & 0.122 & 0.184 & \textbf{0.256} \\ 
   \hline  \multicolumn{1}{l}{MAE}\\ARGOX & \textbf{0.208} & \textbf{0.210} & \textbf{0.257} & \textbf{0.224} & 0.197 & \textbf{0.170} & \textbf{0.231} & 0.477 \\ 
  VAR & 0.318 & 0.295 & 0.328 & 0.330 & 0.335 & 0.374 & 0.423 & 0.607 \\ 
  GFT & -- & -- & 1.509 & -- & -- & -- & -- & -- \\ 
  Lu et al. (2019) & -- & -- & -- & -- & -- & -- & -- & -- \\ 
  naive & 0.224 & 0.218 & 0.288 & 0.251 & \textbf{0.185} & 0.262 & 0.283 & \textbf{0.378} \\ 
   \hline  \multicolumn{1}{l}{Correlation}\\ARGOX & \textbf{0.940} & \textbf{0.932} & \textbf{0.948} & \textbf{0.901} & 0.876 & \textbf{0.977} & \textbf{0.920} & 0.699 \\ 
  VAR & 0.874 & 0.882 & 0.927 & 0.829 & 0.769 & 0.944 & 0.863 & 0.705 \\ 
  GFT & -- & -- & 0.936 & -- & -- & -- & -- & -- \\ 
  Lu et al. (2019) & -- & 0.915 & 0.896 & 0.881 & \textbf{0.884} & -- & -- & -- \\ 
  naive & 0.929 & 0.919 & 0.929 & 0.859 & 0.865 & 0.933 & 0.898 & \textbf{0.766} \\ 
   \hline
\end{tabular}
\caption{Comparison of different methods for state-level \%ILI estimation in Nevada (NV).  The MSE, MAE, and correlation are reported. The method with the best performance is highlighted in boldface for each metric in each period. \label{tab_state28}} 
\end{table}

\begin{figure}[!h] 
  \centering 
\includegraphics[width=\linewidth, page=28]{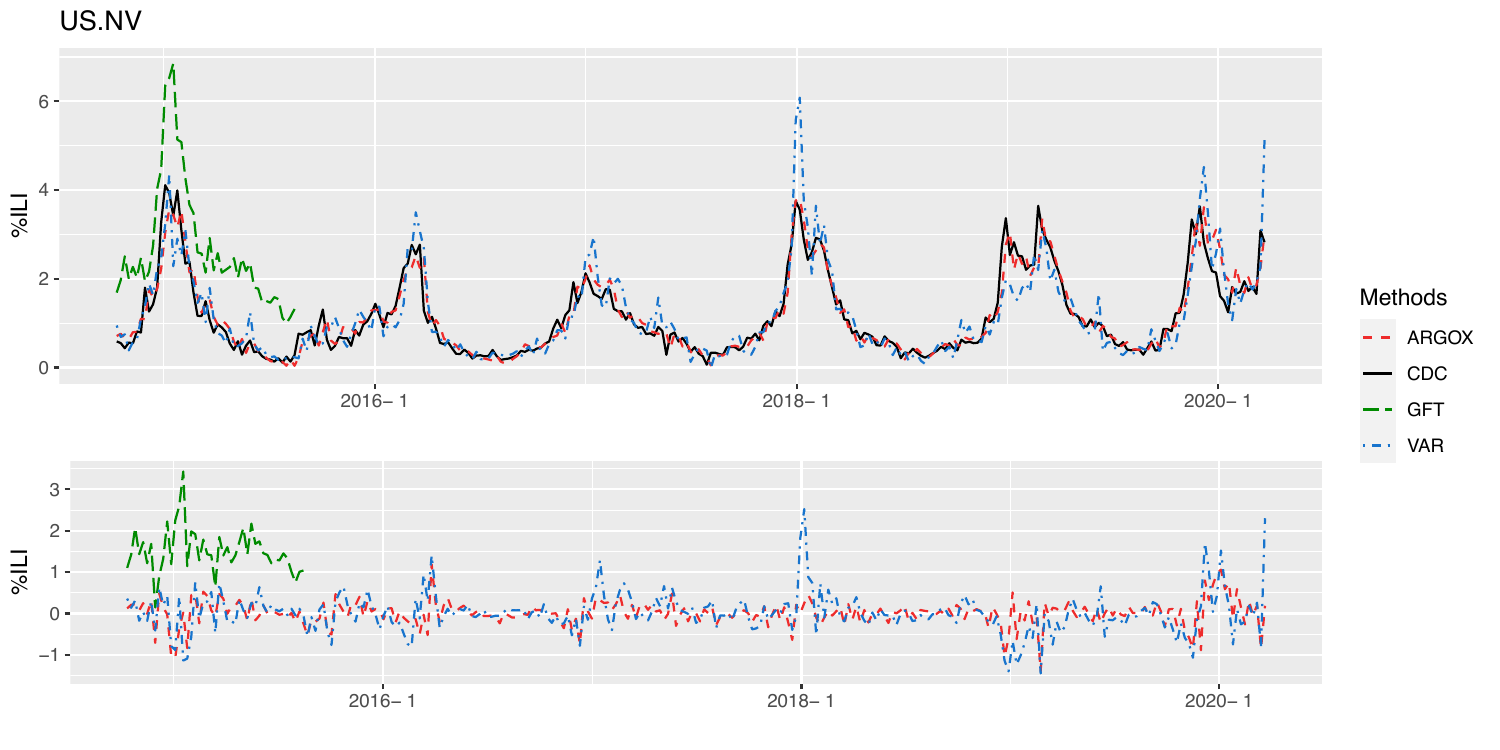} 
\caption{Plots of the \%ILI estimates (top) and the estimation errors (bottom) for Nevada (NV).}
\end{figure}
\newpage      
\begin{table}[ht]
\centering
\begin{tabular}{crrrrrrrr}
  \hline
  & Whole period '14-'20 & Overall '14-'17 & '14-'15 & '15-'16 & '16-'17 & '17-'18 & '18-'19 & '19-'20 \\ 
  \hline  \multicolumn{1}{l}{MSE}\\ARGOX & \textbf{0.175} & 0.054 & 0.058 & \textbf{0.021} & 0.111 & 0.470 & \textbf{0.334} & \textbf{0.476} \\ 
  VAR & 0.363 & 0.081 & 0.113 & 0.048 & 0.086 & \textbf{0.447} & 0.617 & 1.991 \\ 
  GFT & -- & -- & 1.113 & -- & -- & -- & -- & -- \\ 
  Lu et al. (2019) & -- & \textbf{0.045} & 0.061 & 0.030 & \textbf{0.071} & -- & -- & -- \\ 
  naive & 0.201 & 0.056 & \textbf{0.052} & 0.027 & 0.120 & 0.616 & 0.374 & 0.477 \\ 
   \hline  \multicolumn{1}{l}{MAE}\\ARGOX & \textbf{0.247} & \textbf{0.152} & 0.150 & \textbf{0.110} & 0.232 & \textbf{0.452} & \textbf{0.337} & \textbf{0.462} \\ 
  VAR & 0.357 & 0.198 & 0.226 & 0.168 & \textbf{0.222} & 0.466 & 0.539 & 0.989 \\ 
  GFT & -- & -- & 0.844 & -- & -- & -- & -- & -- \\ 
  Lu et al. (2019) & -- & -- & -- & -- & -- & -- & -- & -- \\ 
  naive & 0.266 & 0.153 & \textbf{0.142} & 0.116 & 0.242 & 0.464 & 0.384 & 0.529 \\ 
   \hline  \multicolumn{1}{l}{Correlation}\\ARGOX & \textbf{0.913} & 0.846 & 0.887 & \textbf{0.912} & 0.672 & 0.916 & \textbf{0.750} & 0.909 \\ 
  VAR & 0.817 & 0.776 & 0.766 & 0.778 & 0.783 & \textbf{0.930} & 0.554 & 0.572 \\ 
  GFT & -- & -- & \textbf{0.915} & -- & -- & -- & -- & -- \\ 
  Lu et al. (2019) & -- & \textbf{0.877} & 0.887 & 0.894 & \textbf{0.813} & -- & -- & -- \\ 
  naive & 0.904 & 0.854 & 0.902 & 0.883 & 0.692 & 0.887 & 0.743 & \textbf{0.914} \\ 
   \hline
\end{tabular}
\caption{Comparison of different methods for state-level \%ILI estimation in New Hampshire (NH).  The MSE, MAE, and correlation are reported. The method with the best performance is highlighted in boldface for each metric in each period. \label{tab_state29}} 
\end{table}

\begin{figure}[!h] 
  \centering 
\includegraphics[width=\linewidth, page=29]{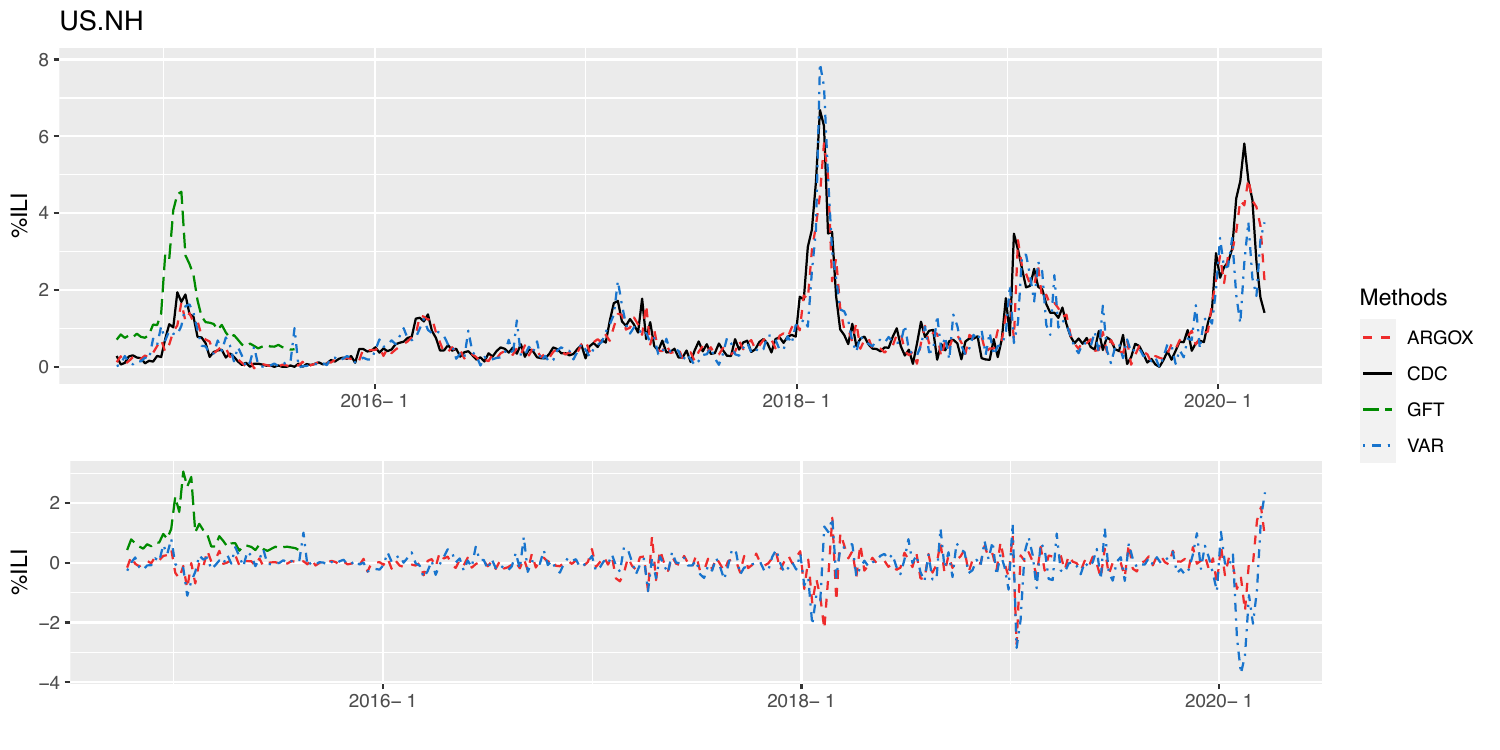} 
\caption{Plots of the \%ILI estimates (top) and the estimation errors (bottom) for New Hampshire (NH).}
\end{figure}
\newpage      
\begin{table}[ht]
\centering
\begin{tabular}{crrrrrrrr}
  \hline
  & Whole period '14-'20 & Overall '14-'17 & '14-'15 & '15-'16 & '16-'17 & '17-'18 & '18-'19 & '19-'20 \\ 
  \hline  \multicolumn{1}{l}{MSE}\\ARGOX & \textbf{0.301} & 0.244 & \textbf{0.218} & 0.228 & 0.408 & \textbf{0.427} & \textbf{0.041} & \textbf{1.491} \\ 
  VAR & 0.757 & 0.797 & 1.205 & 0.440 & 1.098 & 1.405 & 0.257 & 1.542 \\ 
  GFT & -- & -- & 0.644 & -- & -- & -- & -- & -- \\ 
  Lu et al. (2019) & -- & \textbf{0.243} & 0.250 & \textbf{0.228} & 0.466 & -- & -- & -- \\ 
  naive & 0.419 & 0.275 & 0.318 & 0.253 & \textbf{0.367} & 0.979 & 0.186 & 1.689 \\ 
   \hline  \multicolumn{1}{l}{MAE}\\ARGOX & \textbf{0.325} & \textbf{0.381} & \textbf{0.354} & \textbf{0.373} & 0.519 & \textbf{0.452} & \textbf{0.160} & \textbf{0.584} \\ 
  VAR & 0.562 & 0.621 & 0.775 & 0.462 & 0.836 & 0.727 & 0.350 & 0.811 \\ 
  GFT & -- & -- & 0.672 & -- & -- & -- & -- & -- \\ 
  Lu et al. (2019) & -- & -- & -- & -- & -- & -- & -- & -- \\ 
  naive & 0.387 & 0.384 & 0.412 & 0.376 & \textbf{0.456} & 0.594 & 0.303 & 0.797 \\ 
   \hline  \multicolumn{1}{l}{Correlation}\\ARGOX & \textbf{0.964} & 0.944 & \textbf{0.896} & \textbf{0.924} & \textbf{0.926} & \textbf{0.976} & \textbf{0.987} & \textbf{0.903} \\ 
  VAR & 0.907 & 0.826 & 0.638 & 0.835 & 0.766 & 0.911 & 0.926 & 0.888 \\ 
  GFT & -- & -- & 0.896 & -- & -- & -- & -- & -- \\ 
  Lu et al. (2019) & -- & \textbf{0.944} & 0.887 & 0.922 & 0.922 & -- & -- & -- \\ 
  naive & 0.948 & 0.938 & 0.851 & 0.909 & 0.917 & 0.930 & 0.940 & 0.894 \\ 
   \hline
\end{tabular}
\caption{Comparison of different methods for state-level \%ILI estimation in New Jersey (NJ).  The MSE, MAE, and correlation are reported. The method with the best performance is highlighted in boldface for each metric in each period. \label{tab_state30}} 
\end{table}

\begin{figure}[!h] 
  \centering 
\includegraphics[width=\linewidth, page=30]{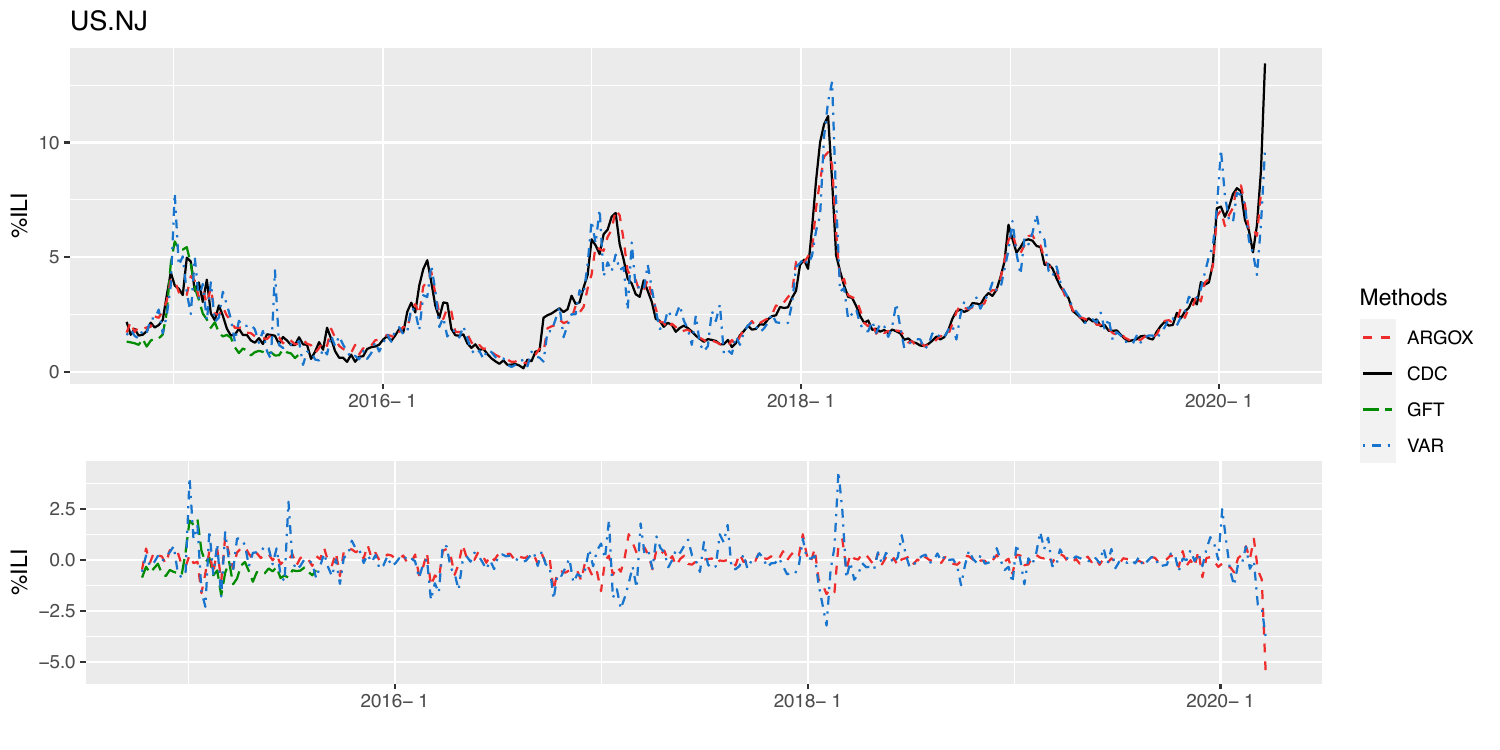} 
\caption{Plots of the \%ILI estimates (top) and the estimation errors (bottom) for New Jersey (NJ).}
\end{figure}
\newpage      
\begin{table}[ht]
\centering
\begin{tabular}{crrrrrrrr}
  \hline
  & Whole period '14-'20 & Overall '14-'17 & '14-'15 & '15-'16 & '16-'17 & '17-'18 & '18-'19 & '19-'20 \\ 
  \hline  \multicolumn{1}{l}{MSE}\\ARGOX & \textbf{0.263} & \textbf{0.146} & \textbf{0.193} & \textbf{0.180} & 0.100 & \textbf{0.487} & \textbf{0.315} & \textbf{1.031} \\ 
  VAR & 0.598 & 0.477 & 0.700 & 0.401 & 0.519 & 0.751 & 0.825 & 1.958 \\ 
  GFT & -- & -- & 1.944 & -- & -- & -- & -- & -- \\ 
  Lu et al. (2019) & -- & 0.164 & 0.293 & 0.213 & \textbf{0.093} & -- & -- & -- \\ 
  naive & 0.417 & 0.247 & 0.399 & 0.234 & 0.167 & 0.807 & 0.560 & 1.499 \\ 
   \hline  \multicolumn{1}{l}{MAE}\\ARGOX & \textbf{0.325} & \textbf{0.274} & \textbf{0.299} & \textbf{0.321} & \textbf{0.241} & \textbf{0.401} & \textbf{0.358} & \textbf{0.785} \\ 
  VAR & 0.478 & 0.454 & 0.475 & 0.487 & 0.516 & 0.538 & 0.611 & 0.928 \\ 
  GFT & -- & -- & 1.222 & -- & -- & -- & -- & -- \\ 
  Lu et al. (2019) & -- & -- & -- & -- & -- & -- & -- & -- \\ 
  naive & 0.389 & 0.329 & 0.368 & 0.368 & 0.310 & 0.580 & 0.501 & 0.806 \\ 
   \hline  \multicolumn{1}{l}{Correlation}\\ARGOX & \textbf{0.966} & \textbf{0.948} & 0.945 & \textbf{0.935} & 0.962 & \textbf{0.957} & \textbf{0.959} & \textbf{0.946} \\ 
  VAR & 0.926 & 0.871 & 0.895 & 0.843 & 0.845 & 0.942 & 0.888 & 0.903 \\ 
  GFT & -- & -- & \textbf{0.951} & -- & -- & -- & -- & -- \\ 
  Lu et al. (2019) & -- & 0.943 & 0.923 & 0.927 & \textbf{0.966} & -- & -- & -- \\ 
  naive & 0.946 & 0.912 & 0.881 & 0.916 & 0.935 & 0.926 & 0.924 & 0.925 \\ 
   \hline
\end{tabular}
\caption{Comparison of different methods for state-level \%ILI estimation in New Mexico (NM).  The MSE, MAE, and correlation are reported. The method with the best performance is highlighted in boldface for each metric in each period. \label{tab_state31}} 
\end{table}

\begin{figure}[!h] 
  \centering 
\includegraphics[width=\linewidth, page=31]{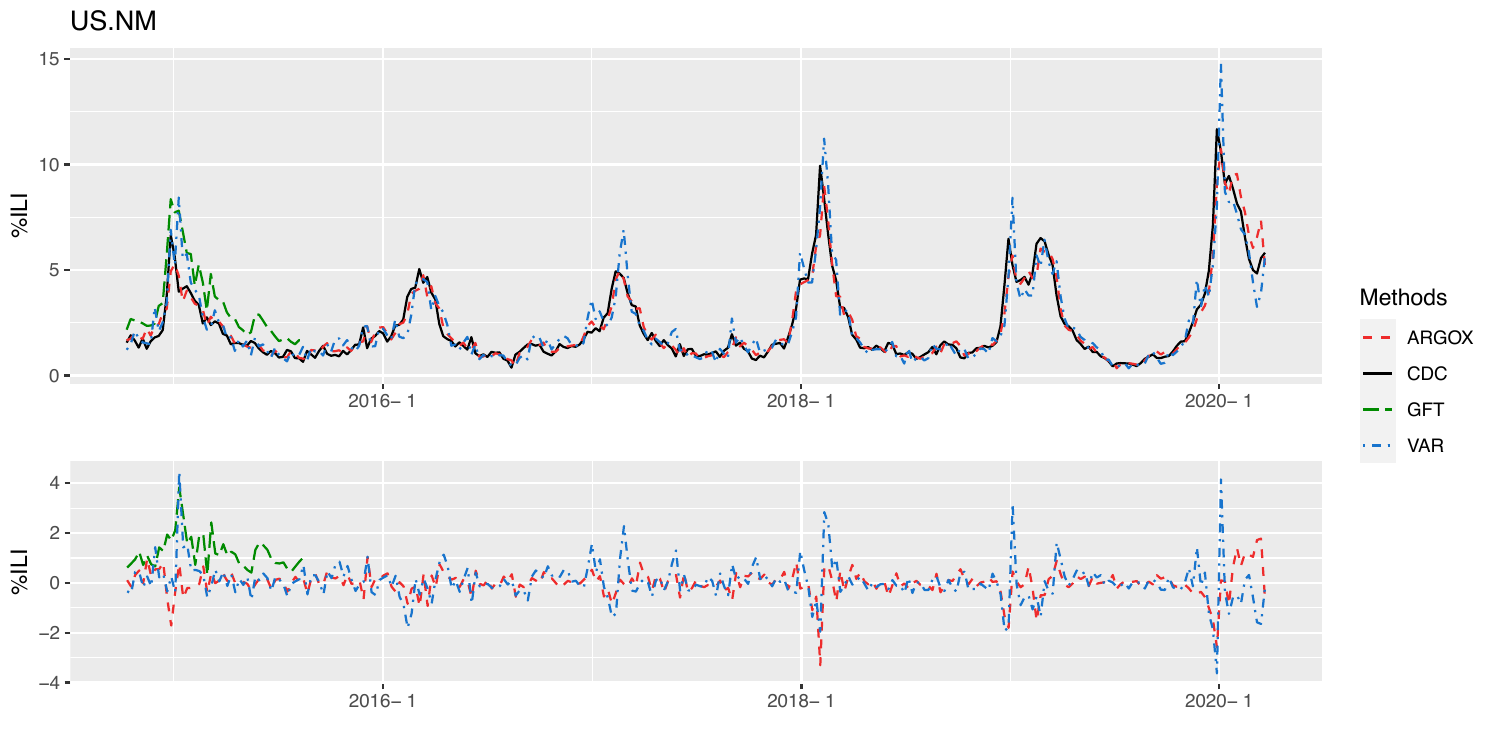} 
\caption{Plots of the \%ILI estimates (top) and the estimation errors (bottom) for New Mexico (NM).}
\end{figure}
\newpage      
\begin{table}[ht]
\centering
\begin{tabular}{crrrrrrrr}
  \hline
  & Whole period '14-'20 & Overall '14-'17 & '14-'15 & '15-'16 & '16-'17 & '17-'18 & '18-'19 & '19-'20 \\ 
  \hline  \multicolumn{1}{l}{MSE}\\ARGOX & \textbf{0.363} & \textbf{0.469} & 0.425 & \textbf{0.254} & \textbf{1.028} & \textbf{0.532} & \textbf{0.080} & \textbf{0.668} \\ 
  VAR & 1.863 & 2.812 & 0.792 & 0.522 & 9.831 & 1.794 & 0.431 & 2.434 \\ 
  GFT & -- & -- & 0.686 & -- & -- & -- & -- & -- \\ 
  Lu et al. (2019) & -- & 1.228 & \textbf{0.349} & 0.461 & 3.932 & -- & -- & -- \\ 
  naive & 0.501 & 0.577 & 0.631 & 0.336 & 1.124 & 0.956 & 0.143 & 1.026 \\ 
   \hline  \multicolumn{1}{l}{MAE}\\ARGOX & \textbf{0.395} & \textbf{0.482} & \textbf{0.480} & \textbf{0.393} & 0.755 & \textbf{0.524} & \textbf{0.212} & \textbf{0.556} \\ 
  VAR & 0.690 & 0.755 & 0.594 & 0.544 & 1.581 & 0.952 & 0.486 & 1.003 \\ 
  GFT & -- & -- & 0.630 & -- & -- & -- & -- & -- \\ 
  Lu et al. (2019) & -- & -- & -- & -- & -- & -- & -- & -- \\ 
  naive & 0.426 & 0.489 & 0.514 & 0.432 & \textbf{0.732} & 0.648 & 0.291 & 0.735 \\ 
   \hline  \multicolumn{1}{l}{Correlation}\\ARGOX & \textbf{0.959} & \textbf{0.939} & 0.925 & \textbf{0.837} & \textbf{0.925} & \textbf{0.970} & \textbf{0.965} & \textbf{0.939} \\ 
  VAR & 0.849 & 0.769 & 0.859 & 0.756 & 0.676 & 0.921 & 0.817 & 0.856 \\ 
  GFT & -- & -- & 0.903 & -- & -- & -- & -- & -- \\ 
  Lu et al. (2019) & -- & 0.856 & \textbf{0.948} & 0.751 & 0.782 & -- & -- & -- \\ 
  naive & 0.943 & 0.926 & 0.890 & 0.794 & 0.915 & 0.943 & 0.940 & 0.904 \\ 
   \hline
\end{tabular}
\caption{Comparison of different methods for state-level \%ILI estimation in New York (NY).  The MSE, MAE, and correlation are reported. The method with the best performance is highlighted in boldface for each metric in each period. \label{tab_state32}} 
\end{table}

\begin{figure}[!h] 
  \centering 
\includegraphics[width=\linewidth, page=32]{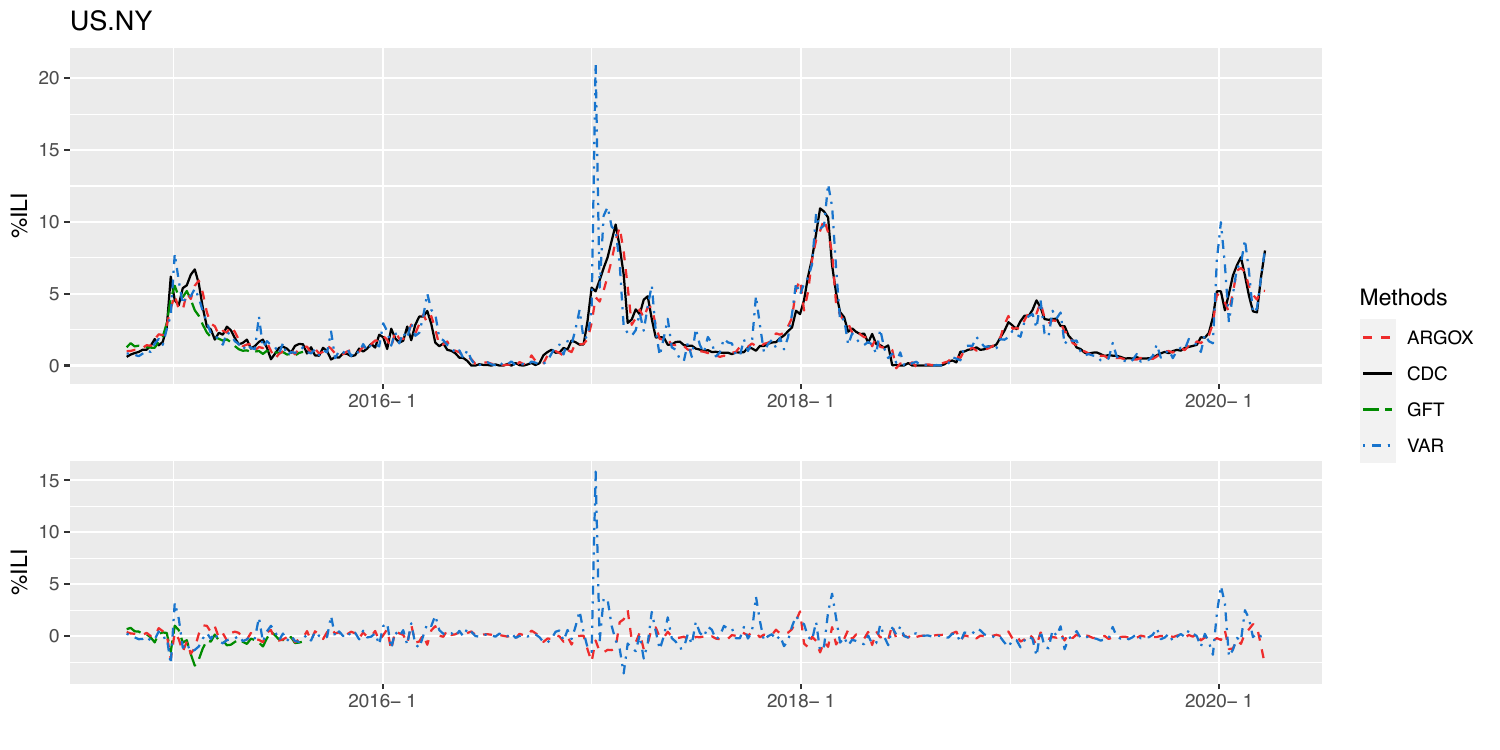} 
\caption{Plots of the \%ILI estimates (top) and the estimation errors (bottom) for New York (NY).}
\end{figure}
\newpage      
\begin{table}[ht]
\centering
\begin{tabular}{crrrrrrrr}
  \hline
  & Whole period '14-'20 & Overall '14-'17 & '14-'15 & '15-'16 & '16-'17 & '17-'18 & '18-'19 & '19-'20 \\ 
  \hline  \multicolumn{1}{l}{MSE}\\ARGOX & \textbf{0.246} & 0.298 & 0.446 & \textbf{0.129} & \textbf{0.449} & \textbf{0.566} & \textbf{0.129} & \textbf{0.134} \\ 
  VAR & 1.006 & 1.072 & 2.404 & 0.297 & 0.740 & 2.260 & 1.597 & 0.279 \\ 
  GFT & -- & -- & 0.374 & -- & -- & -- & -- & -- \\ 
  Lu et al. (2019) & -- & \textbf{0.293} & \textbf{0.309} & 0.297 & 0.554 & -- & -- & -- \\ 
  naive & 0.477 & 0.576 & 0.935 & 0.208 & 0.841 & 1.102 & 0.293 & 0.333 \\ 
   \hline  \multicolumn{1}{l}{MAE}\\ARGOX & \textbf{0.303} & \textbf{0.345} & \textbf{0.395} & \textbf{0.279} & \textbf{0.453} & \textbf{0.480} & \textbf{0.236} & \textbf{0.247} \\ 
  VAR & 0.547 & 0.607 & 0.887 & 0.456 & 0.625 & 0.927 & 0.617 & 0.422 \\ 
  GFT & -- & -- & 0.420 & -- & -- & -- & -- & -- \\ 
  Lu et al. (2019) & -- & -- & -- & -- & -- & -- & -- & -- \\ 
  naive & 0.408 & 0.446 & 0.474 & 0.348 & 0.680 & 0.707 & 0.374 & 0.431 \\ 
   \hline  \multicolumn{1}{l}{Correlation}\\ARGOX & \textbf{0.962} & \textbf{0.943} & 0.938 & \textbf{0.925} & \textbf{0.937} & \textbf{0.956} & \textbf{0.977} & \textbf{0.974} \\ 
  VAR & 0.874 & 0.769 & 0.569 & 0.831 & 0.890 & 0.954 & 0.911 & 0.957 \\ 
  GFT & -- & -- & \textbf{0.975} & -- & -- & -- & -- & -- \\ 
  Lu et al. (2019) & -- & 0.942 & 0.958 & 0.890 & 0.919 & -- & -- & -- \\ 
  naive & 0.925 & 0.887 & 0.860 & 0.881 & 0.881 & 0.910 & 0.940 & 0.941 \\ 
   \hline
\end{tabular}
\caption{Comparison of different methods for state-level \%ILI estimation in North Carolina (NC).  The MSE, MAE, and correlation are reported. The method with the best performance is highlighted in boldface for each metric in each period. \label{tab_state33}} 
\end{table}

\begin{figure}[!h] 
  \centering 
\includegraphics[width=\linewidth, page=33]{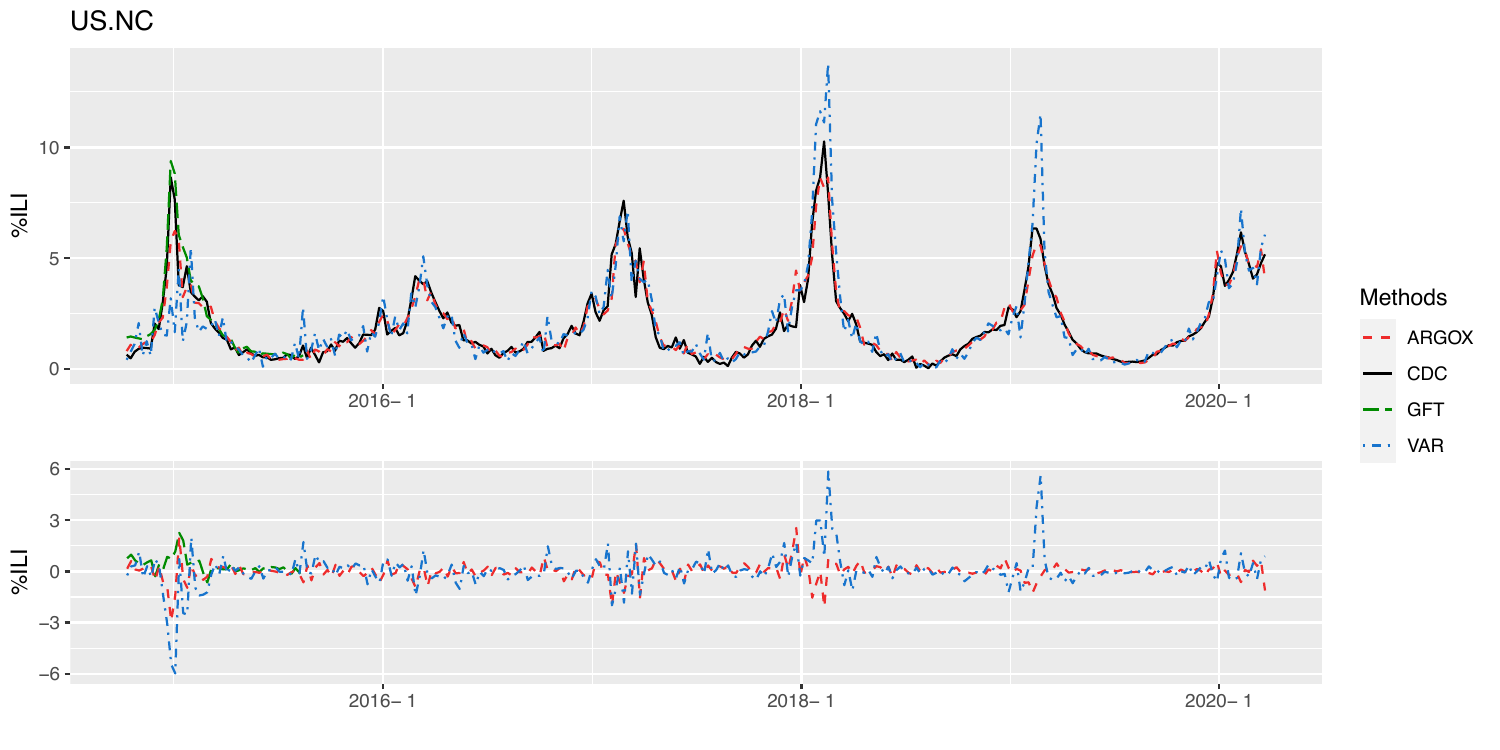} 
\caption{Plots of the \%ILI estimates (top) and the estimation errors (bottom) for North Carolina (NC).}
\end{figure}
\newpage      
\begin{table}[ht]
\centering
\begin{tabular}{crrrrrrrr}
  \hline
  & Whole period '14-'20 & Overall '14-'17 & '14-'15 & '15-'16 & '16-'17 & '17-'18 & '18-'19 & '19-'20 \\ 
  \hline  \multicolumn{1}{l}{MSE}\\ARGOX & \textbf{0.634} & \textbf{0.578} & \textbf{0.717} & \textbf{0.185} & 1.197 & \textbf{0.604} & \textbf{0.440} & \textbf{2.297} \\ 
  VAR & 3.213 & 1.877 & 2.947 & 0.544 & 3.125 & 5.011 & 1.006 & 8.291 \\ 
  GFT & -- & -- & 0.938 & -- & -- & -- & -- & -- \\ 
  Lu et al. (2019) & -- & 0.806 & 1.496 & 0.230 & \textbf{0.927} & -- & -- & -- \\ 
  naive & 0.816 & 0.776 & 0.819 & 0.245 & 1.802 & 0.992 & 0.618 & 2.408 \\ 
   \hline  \multicolumn{1}{l}{MAE}\\ARGOX & \textbf{0.495} & \textbf{0.438} & \textbf{0.454} & \textbf{0.351} & \textbf{0.761} & \textbf{0.609} & \textbf{0.491} & \textbf{1.121} \\ 
  VAR & 0.874 & 0.724 & 0.923 & 0.521 & 1.080 & 1.165 & 0.796 & 1.894 \\ 
  GFT & -- & -- & 0.669 & -- & -- & -- & -- & -- \\ 
  Lu et al. (2019) & -- & -- & -- & -- & -- & -- & -- & -- \\ 
  naive & 0.567 & 0.499 & 0.495 & 0.405 & 0.899 & 0.735 & 0.607 & 1.210 \\ 
   \hline  \multicolumn{1}{l}{Correlation}\\ARGOX & \textbf{0.876} & \textbf{0.839} & 0.879 & \textbf{0.735} & 0.673 & \textbf{0.846} & \textbf{0.867} & \textbf{0.778} \\ 
  VAR & 0.671 & 0.799 & \textbf{0.917} & 0.545 & 0.479 & 0.612 & 0.696 & 0.516 \\ 
  GFT & -- & -- & 0.847 & -- & -- & -- & -- & -- \\ 
  Lu et al. (2019) & -- & 0.809 & 0.808 & 0.658 & \textbf{0.761} & -- & -- & -- \\ 
  naive & 0.846 & 0.800 & 0.869 & 0.687 & 0.575 & 0.740 & 0.808 & 0.769 \\ 
   \hline
\end{tabular}
\caption{Comparison of different methods for state-level \%ILI estimation in North Dakota (ND).  The MSE, MAE, and correlation are reported. The method with the best performance is highlighted in boldface for each metric in each period. \label{tab_state34}} 
\end{table}

\begin{figure}[!h] 
  \centering 
\includegraphics[width=\linewidth, page=34]{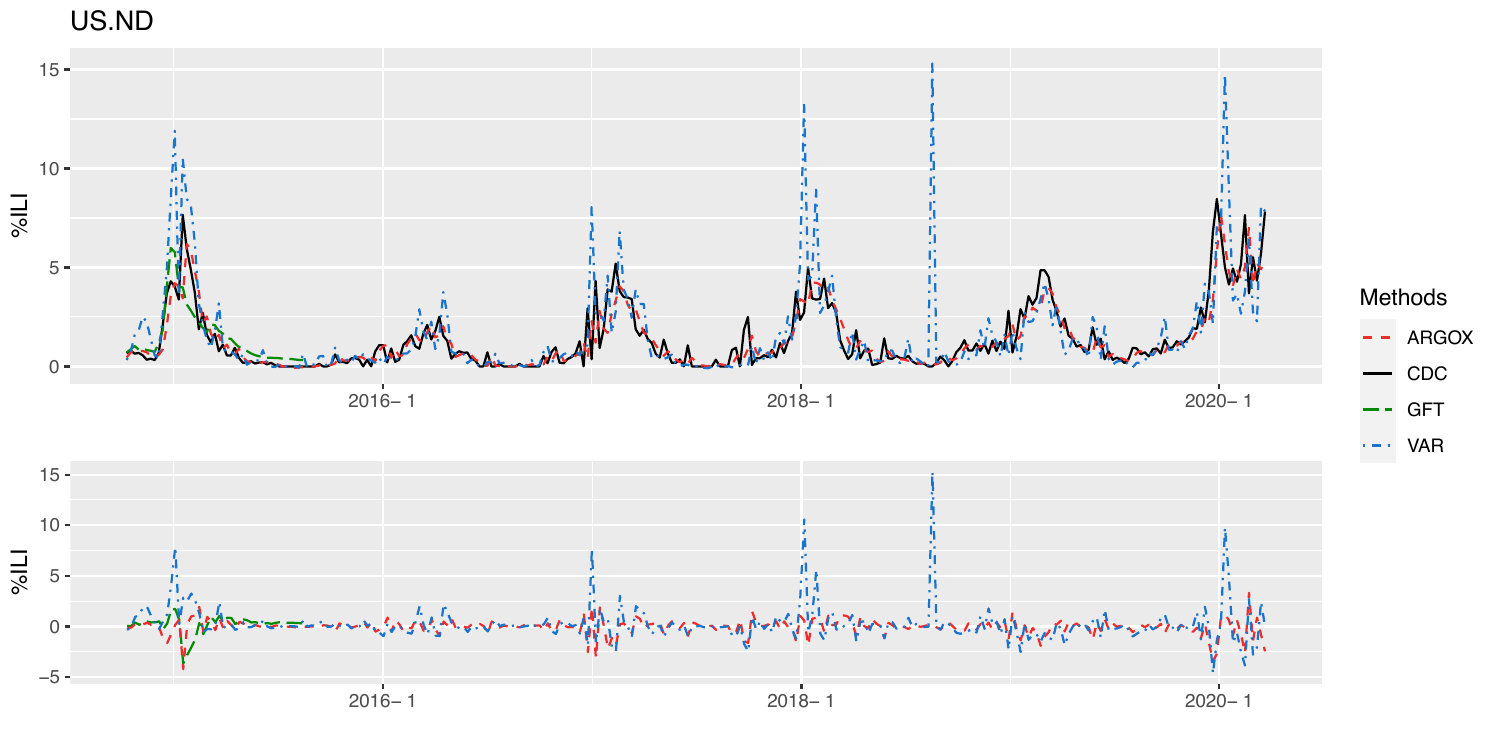} 
\caption{Plots of the \%ILI estimates (top) and the estimation errors (bottom) for North Dakota (ND).}
\end{figure}
\newpage      
\begin{table}[ht]
\centering
\begin{tabular}{crrrrrrrr}
  \hline
  & Whole period '14-'20 & Overall '14-'17 & '14-'15 & '15-'16 & '16-'17 & '17-'18 & '18-'19 & '19-'20 \\ 
  \hline  \multicolumn{1}{l}{MSE}\\ARGOX & \textbf{0.108} & 0.121 & 0.291 & \textbf{0.032} & \textbf{0.057} & \textbf{0.127} & \textbf{0.108} & \textbf{0.232} \\ 
  VAR & 0.248 & 0.312 & 0.784 & 0.063 & 0.119 & 0.232 & 0.110 & 0.596 \\ 
  GFT & -- & -- & 0.819 & -- & -- & -- & -- & -- \\ 
  Lu et al. (2019) & -- & \textbf{0.094} & \textbf{0.236} & 0.042 & 0.095 & -- & -- & -- \\ 
  naive & 0.181 & 0.224 & 0.547 & 0.065 & 0.104 & 0.270 & 0.144 & 0.269 \\ 
   \hline  \multicolumn{1}{l}{MAE}\\ARGOX & \textbf{0.193} & \textbf{0.179} & \textbf{0.248} & \textbf{0.136} & \textbf{0.177} & \textbf{0.253} & 0.240 & \textbf{0.342} \\ 
  VAR & 0.280 & 0.280 & 0.436 & 0.203 & 0.237 & 0.344 & \textbf{0.233} & 0.554 \\ 
  GFT & -- & -- & 0.719 & -- & -- & -- & -- & -- \\ 
  Lu et al. (2019) & -- & -- & -- & -- & -- & -- & -- & -- \\ 
  naive & 0.234 & 0.230 & 0.355 & 0.192 & 0.225 & 0.334 & 0.268 & 0.383 \\ 
   \hline  \multicolumn{1}{l}{Correlation}\\ARGOX & \textbf{0.943} & 0.928 & 0.911 & \textbf{0.915} & \textbf{0.958} & \textbf{0.959} & 0.890 & \textbf{0.922} \\ 
  VAR & 0.885 & 0.846 & 0.835 & 0.829 & 0.919 & 0.957 & \textbf{0.905} & 0.784 \\ 
  GFT & -- & -- & \textbf{0.971} & -- & -- & -- & -- & -- \\ 
  Lu et al. (2019) & -- & \textbf{0.943} & 0.941 & 0.897 & 0.932 & -- & -- & -- \\ 
  naive & 0.906 & 0.869 & 0.833 & 0.835 & 0.922 & 0.912 & 0.860 & 0.915 \\ 
   \hline
\end{tabular}
\caption{Comparison of different methods for state-level \%ILI estimation in Ohio (OH).  The MSE, MAE, and correlation are reported. The method with the best performance is highlighted in boldface for each metric in each period. \label{tab_state35}} 
\end{table}

\begin{figure}[!h] 
  \centering 
\includegraphics[width=\linewidth, page=35]{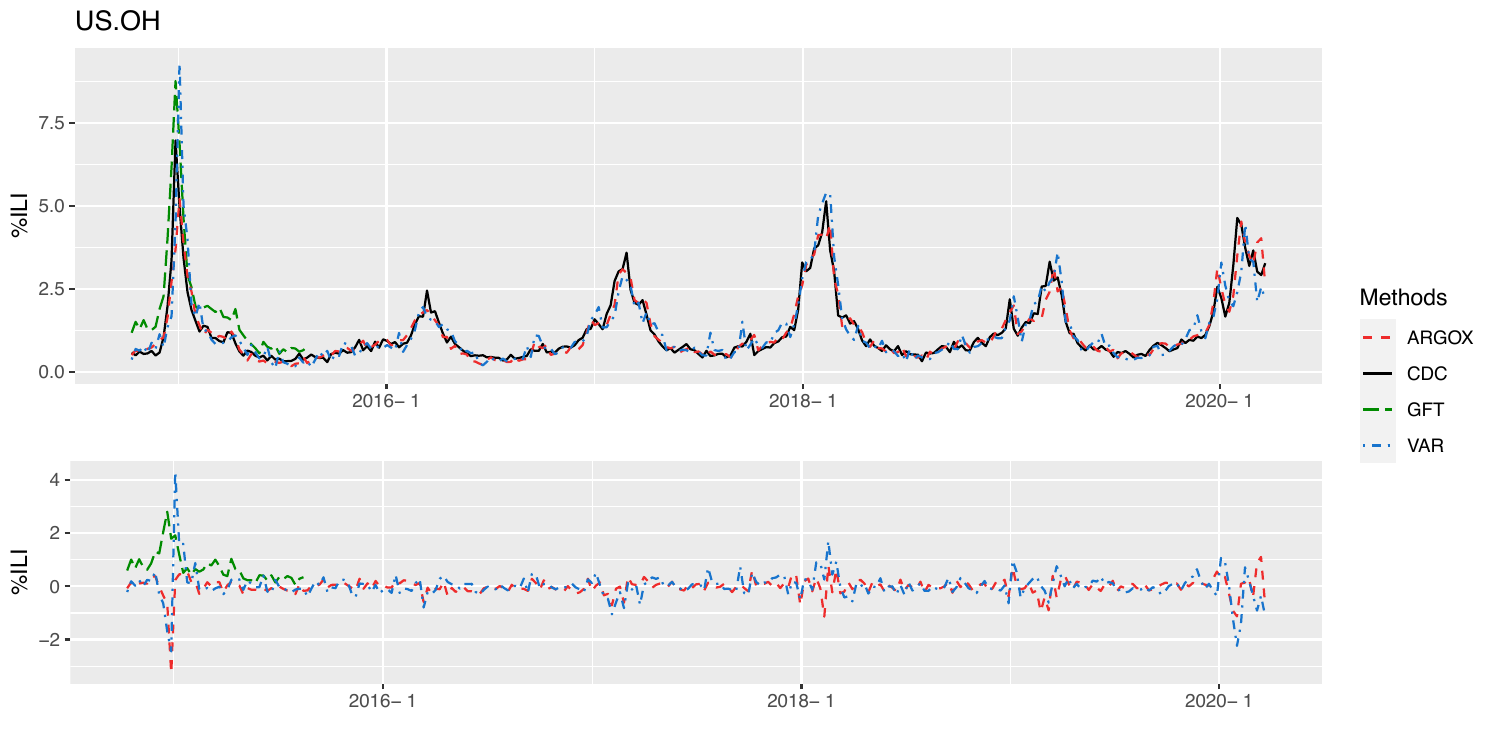} 
\caption{Plots of the \%ILI estimates (top) and the estimation errors (bottom) for Ohio (OH).}
\end{figure}
\newpage      
\begin{table}[ht]
\centering
\begin{tabular}{crrrrrrrr}
  \hline
  & Whole period '14-'20 & Overall '14-'17 & '14-'15 & '15-'16 & '16-'17 & '17-'18 & '18-'19 & '19-'20 \\ 
  \hline  \multicolumn{1}{l}{MSE}\\ARGOX & \textbf{ 0.859} &  1.360 &  2.003 & \textbf{ 0.726} & \textbf{ 2.002} & \textbf{ 0.554} & \textbf{ 0.539} & \textbf{ 0.701} \\ 
  VAR &  8.691 &  4.993 &  3.314 &  2.556 & 13.327 &  5.961 &  9.482 & 52.931 \\ 
  GFT & -- & -- &  3.906 & -- & -- & -- & -- & -- \\ 
  Lu et al. (2019) & -- & -- & -- & -- & -- & -- & -- & -- \\ 
  naive &  0.965 & \textbf{ 1.291} & \textbf{ 1.692} &  0.786 &  2.124 &  1.000 &  0.988 &  1.063 \\ 
   \hline  \multicolumn{1}{l}{MAE}\\ARGOX & \textbf{0.586} & 0.788 & 0.924 & \textbf{0.695} & \textbf{1.052} & \textbf{0.516} & \textbf{0.501} & \textbf{0.573} \\ 
  VAR & 1.415 & 1.282 & 1.186 & 1.253 & 2.139 & 1.349 & 1.956 & 4.176 \\ 
  GFT & -- & -- & 1.390 & -- & -- & -- & -- & -- \\ 
  Lu et al. (2019) & -- & -- & -- & -- & -- & -- & -- & -- \\ 
  naive & 0.646 & \textbf{0.775} & \textbf{0.889} & 0.716 & 1.091 & 0.745 & 0.706 & 0.742 \\ 
   \hline  \multicolumn{1}{l}{Correlation}\\ARGOX & \textbf{0.956} & 0.933 & 0.917 & 0.532 & \textbf{0.935} & \textbf{0.980} & \textbf{0.971} & \textbf{0.959} \\ 
  VAR & 0.808 & 0.803 & 0.858 & 0.493 & 0.761 & 0.914 & 0.890 & 0.610 \\ 
  GFT & -- & -- & 0.856 & -- & -- & -- & -- & -- \\ 
  Lu et al. (2019) & -- & -- & -- & -- & -- & -- & -- & -- \\ 
  naive & 0.951 & \textbf{0.938} & \textbf{0.930} & \textbf{0.535} & 0.931 & 0.943 & 0.940 & 0.943 \\ 
   \hline
\end{tabular}
\caption{Comparison of different methods for state-level \%ILI estimation in Oklahoma (OK).  The MSE, MAE, and correlation are reported. The method with the best performance is highlighted in boldface for each metric in each period. \label{tab_state36}} 
\end{table}

\begin{figure}[!h] 
  \centering 
\includegraphics[width=\linewidth, page=36]{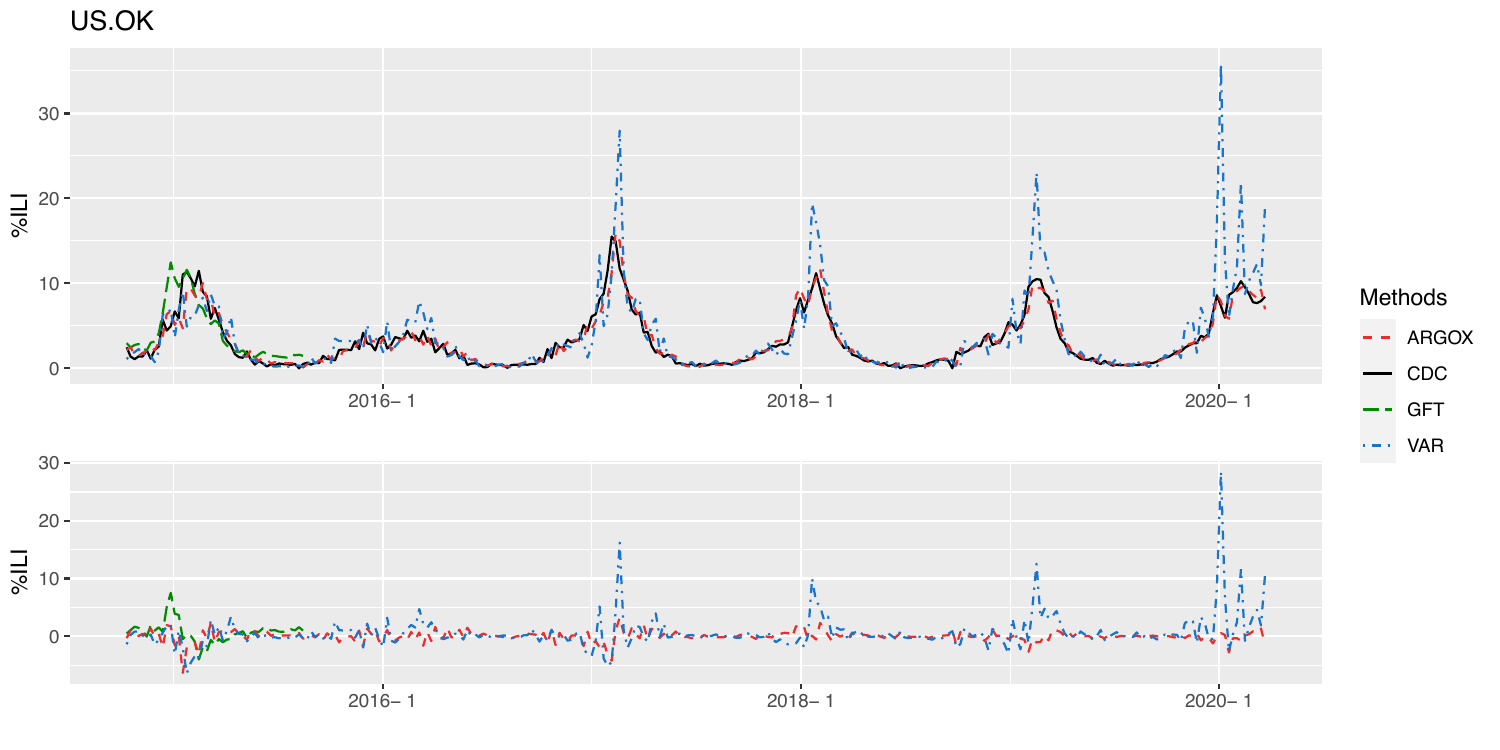} 
\caption{Plots of the \%ILI estimates (top) and the estimation errors (bottom) for Oklahoma (OK).}
\end{figure}
\newpage      
\begin{table}[ht]
\centering
\begin{tabular}{crrrrrrrr}
  \hline
  & Whole period '14-'20 & Overall '14-'17 & '14-'15 & '15-'16 & '16-'17 & '17-'18 & '18-'19 & '19-'20 \\ 
  \hline  \multicolumn{1}{l}{MSE}\\ARGOX & \textbf{0.187} & \textbf{0.286} & 0.123 & 0.358 & \textbf{0.624} & \textbf{0.062} & \textbf{0.157} & \textbf{0.240} \\ 
  VAR & 1.188 & 0.616 & 0.281 & 0.559 & 1.463 & 6.474 & 0.349 & 0.717 \\ 
  GFT & -- & -- & 0.372 & -- & -- & -- & -- & -- \\ 
  Lu et al. (2019) & -- & 0.408 & 0.253 & \textbf{0.350} & 0.869 & -- & -- & -- \\ 
  naive & 0.260 & 0.352 & \textbf{0.094} & 0.400 & 0.894 & 0.190 & 0.203 & 0.470 \\ 
   \hline  \multicolumn{1}{l}{MAE}\\ARGOX & \textbf{0.264} & \textbf{0.327} & 0.254 & 0.435 & \textbf{0.457} & \textbf{0.187} & \textbf{0.273} & \textbf{0.364} \\ 
  VAR & 0.522 & 0.480 & 0.356 & 0.598 & 0.716 & 1.156 & 0.436 & 0.636 \\ 
  GFT & -- & -- & 0.516 & -- & -- & -- & -- & -- \\ 
  Lu et al. (2019) & -- & -- & -- & -- & -- & -- & -- & -- \\ 
  naive & 0.300 & 0.341 & \textbf{0.237} & \textbf{0.428} & 0.560 & 0.310 & 0.275 & 0.534 \\ 
   \hline  \multicolumn{1}{l}{Correlation}\\ARGOX & \textbf{0.945} & \textbf{0.856} & 0.803 & 0.715 & \textbf{0.796} & \textbf{0.988} & \textbf{0.941} & \textbf{0.958} \\ 
  VAR & 0.794 & 0.739 & 0.745 & 0.591 & 0.702 & 0.825 & 0.878 & 0.872 \\ 
  GFT & -- & -- & 0.739 & -- & -- & -- & -- & -- \\ 
  Lu et al. (2019) & -- & 0.802 & 0.779 & \textbf{0.750} & 0.745 & -- & -- & -- \\ 
  naive & 0.924 & 0.832 & \textbf{0.829} & 0.713 & 0.728 & 0.952 & 0.919 & 0.920 \\ 
   \hline
\end{tabular}
\caption{Comparison of different methods for state-level \%ILI estimation in Oregon (OR).  The MSE, MAE, and correlation are reported. The method with the best performance is highlighted in boldface for each metric in each period. \label{tab_state37}} 
\end{table}

\begin{figure}[!h] 
  \centering 
\includegraphics[width=\linewidth, page=37]{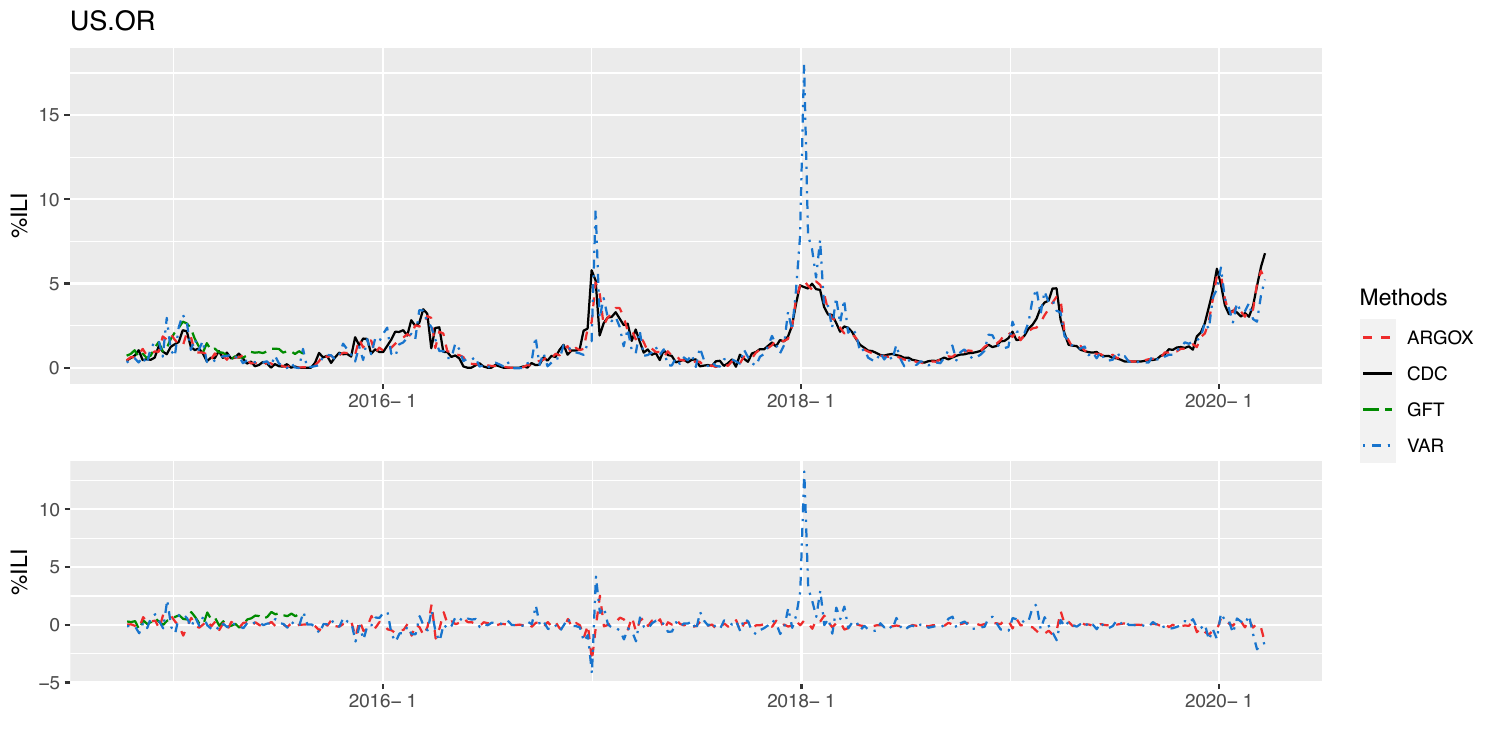} 
\caption{Plots of the \%ILI estimates (top) and the estimation errors (bottom) for Oregon (OR).}
\end{figure}
\newpage      
\begin{table}[ht]
\centering
\begin{tabular}{crrrrrrrr}
  \hline
  & Whole period '14-'20 & Overall '14-'17 & '14-'15 & '15-'16 & '16-'17 & '17-'18 & '18-'19 & '19-'20 \\ 
  \hline  \multicolumn{1}{l}{MSE}\\ARGOX & \textbf{0.124} & 0.157 & 0.236 & \textbf{0.092} & \textbf{0.202} & \textbf{0.165} & \textbf{0.035} & \textbf{0.274} \\ 
  VAR & 0.311 & 0.511 & 1.013 & 0.224 & 0.446 & 0.167 & 0.106 & 0.342 \\ 
  GFT & -- & -- & 0.329 & -- & -- & -- & -- & -- \\ 
  Lu et al. (2019) & -- & \textbf{0.138} & \textbf{0.214} & 0.110 & 0.218 & -- & -- & -- \\ 
  naive & 0.227 & 0.307 & 0.523 & 0.206 & 0.318 & 0.319 & 0.082 & 0.357 \\ 
   \hline  \multicolumn{1}{l}{MAE}\\ARGOX & \textbf{0.215} & \textbf{0.266} & \textbf{0.314} & \textbf{0.224} & \textbf{0.325} & \textbf{0.231} & \textbf{0.149} & \textbf{0.305} \\ 
  VAR & 0.322 & 0.420 & 0.561 & 0.324 & 0.473 & 0.302 & 0.230 & 0.421 \\ 
  GFT & -- & -- & 0.473 & -- & -- & -- & -- & -- \\ 
  Lu et al. (2019) & -- & -- & -- & -- & -- & -- & -- & -- \\ 
  naive & 0.295 & 0.361 & 0.467 & 0.336 & 0.396 & 0.358 & 0.212 & 0.440 \\ 
   \hline  \multicolumn{1}{l}{Correlation}\\ARGOX & \textbf{0.967} & 0.947 & 0.954 & 0.899 & 0.934 & \textbf{0.978} & \textbf{0.977} & \textbf{0.950} \\ 
  VAR & 0.925 & 0.871 & 0.866 & 0.810 & 0.866 & 0.973 & 0.934 & 0.935 \\ 
  GFT & -- & -- & 0.950 & -- & -- & -- & -- & -- \\ 
  Lu et al. (2019) & -- & \textbf{0.954} & \textbf{0.963} & \textbf{0.905} & \textbf{0.935} & -- & -- & -- \\ 
  naive & 0.938 & 0.898 & 0.889 & 0.793 & 0.900 & 0.945 & 0.943 & 0.938 \\ 
   \hline
\end{tabular}
\caption{Comparison of different methods for state-level \%ILI estimation in Pennsylvania (PA).  The MSE, MAE, and correlation are reported. The method with the best performance is highlighted in boldface for each metric in each period. \label{tab_state38}} 
\end{table}

\begin{figure}[!h] 
  \centering 
\includegraphics[width=\linewidth, page=38]{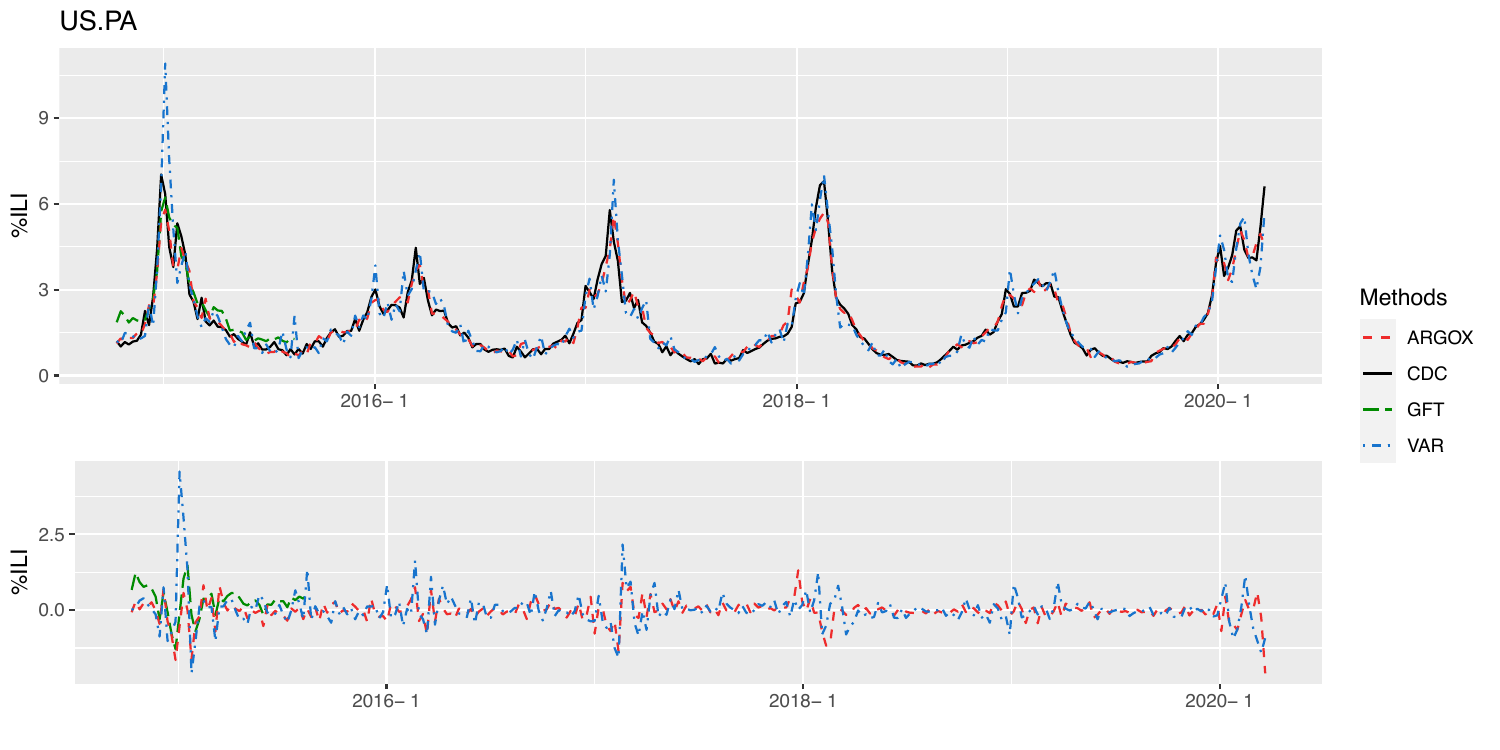} 
\caption{Plots of the \%ILI estimates (top) and the estimation errors (bottom) for Pennsylvania (PA).}
\end{figure}
\newpage      
\begin{table}[ht]
\centering
\begin{tabular}{crrrrrrrr}
  \hline
  & Whole period '14-'20 & Overall '14-'17 & '14-'15 & '15-'16 & '16-'17 & '17-'18 & '18-'19 & '19-'20 \\ 
  \hline  \multicolumn{1}{l}{MSE}\\ARGOX & \textbf{0.235} & 0.144 & 0.194 & 0.056 & 0.212 & \textbf{0.390} & \textbf{0.283} & \textbf{0.998} \\ 
  VAR & 0.858 & 0.272 & 0.349 & 0.129 & 0.423 & 2.802 & 2.542 & 1.130 \\ 
  GFT & -- & -- & 0.417 & -- & -- & -- & -- & -- \\ 
  Lu et al. (2019) & -- & \textbf{0.067} & \textbf{0.117} & \textbf{0.025} & \textbf{0.018} & -- & -- & -- \\ 
  naive & 0.307 & 0.166 & 0.201 & 0.057 & 0.294 & 0.661 & 0.428 & 1.150 \\ 
   \hline  \multicolumn{1}{l}{MAE}\\ARGOX & \textbf{0.276} & \textbf{0.223} & \textbf{0.243} & 0.178 & \textbf{0.314} & \textbf{0.438} & \textbf{0.368} & \textbf{0.682} \\ 
  VAR & 0.433 & 0.313 & 0.313 & 0.248 & 0.494 & 0.818 & 0.894 & 0.737 \\ 
  GFT & -- & -- & 0.565 & -- & -- & -- & -- & -- \\ 
  Lu et al. (2019) & -- & -- & -- & -- & -- & -- & -- & -- \\ 
  naive & 0.314 & 0.232 & 0.257 & \textbf{0.153} & 0.365 & 0.564 & 0.490 & 0.750 \\ 
   \hline  \multicolumn{1}{l}{Correlation}\\ARGOX & \textbf{0.961} & 0.932 & 0.923 & 0.905 & \textbf{0.940} & \textbf{0.954} & \textbf{0.959} & 0.942 \\ 
  VAR & 0.901 & 0.882 & 0.862 & 0.858 & 0.880 & 0.873 & 0.752 & \textbf{0.952} \\ 
  GFT & -- & -- & 0.944 & -- & -- & -- & -- & -- \\ 
  Lu et al. (2019) & -- & \textbf{0.949} & \textbf{0.956} & \textbf{0.961} & 0.047 & -- & -- & -- \\ 
  naive & 0.949 & 0.919 & 0.916 & 0.907 & 0.904 & 0.918 & 0.936 & 0.936 \\ 
   \hline
\end{tabular}
\caption{Comparison of different methods for state-level \%ILI estimation in Rhode Island (RI).  The MSE, MAE, and correlation are reported. The method with the best performance is highlighted in boldface for each metric in each period. \label{tab_state39}} 
\end{table}

\begin{figure}[!h] 
  \centering 
\includegraphics[width=\linewidth, page=39]{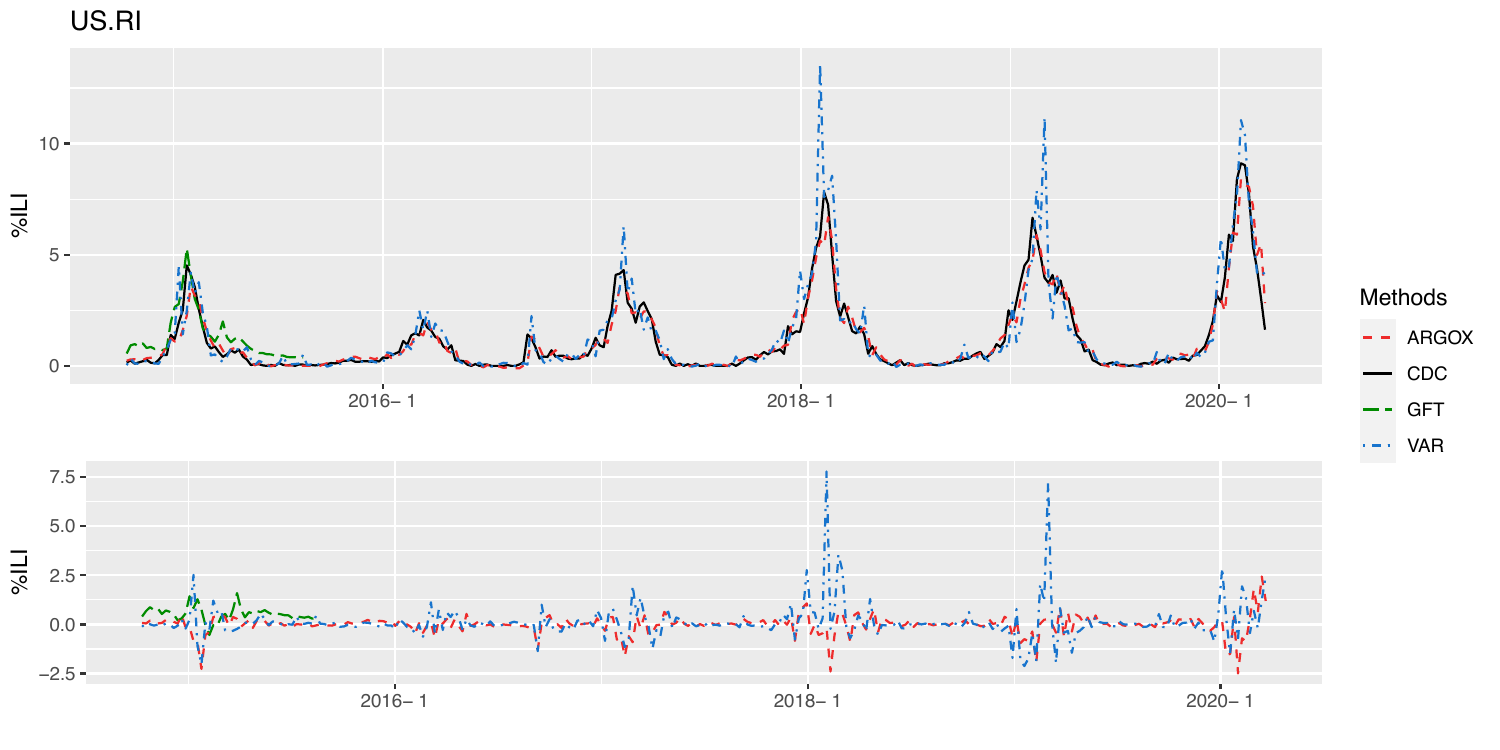} 
\caption{Plots of the \%ILI estimates (top) and the estimation errors (bottom) for Rhode Island (RI).}
\end{figure}
\newpage      
\begin{table}[ht]
\centering
\begin{tabular}{crrrrrrrr}
  \hline
  & Whole period '14-'20 & Overall '14-'17 & '14-'15 & '15-'16 & '16-'17 & '17-'18 & '18-'19 & '19-'20 \\ 
  \hline  \multicolumn{1}{l}{MSE}\\ARGOX & \textbf{0.351} & \textbf{0.336} & 0.317 & \textbf{0.097} & \textbf{0.864} & \textbf{0.472} & \textbf{0.427} & \textbf{0.835} \\ 
  VAR & 1.932 & 0.912 & 0.963 & 0.342 & 2.115 & 8.423 & 1.260 & 3.850 \\ 
  GFT & -- & -- & 2.269 & -- & -- & -- & -- & -- \\ 
  Lu et al. (2019) & -- & 0.383 & \textbf{0.130} & 0.452 & 0.955 & -- & -- & -- \\ 
  naive & 0.767 & 0.465 & 0.565 & 0.141 & 1.013 & 1.790 & 0.871 & 2.547 \\ 
   \hline  \multicolumn{1}{l}{MAE}\\ARGOX & \textbf{0.367} & \textbf{0.348} & \textbf{0.285} & \textbf{0.234} & \textbf{0.759} & \textbf{0.455} & \textbf{0.432} & \textbf{0.726} \\ 
  VAR & 0.687 & 0.531 & 0.436 & 0.459 & 1.085 & 1.772 & 0.741 & 1.029 \\ 
  GFT & -- & -- & 1.235 & -- & -- & -- & -- & -- \\ 
  Lu et al. (2019) & -- & -- & -- & -- & -- & -- & -- & -- \\ 
  naive & 0.516 & 0.403 & 0.387 & 0.292 & 0.795 & 0.880 & 0.683 & 1.172 \\ 
   \hline  \multicolumn{1}{l}{Correlation}\\ARGOX & \textbf{0.979} & \textbf{0.953} & 0.925 & \textbf{0.873} & 0.909 & \textbf{0.985} & \textbf{0.963} & \textbf{0.956} \\ 
  VAR & 0.906 & 0.882 & 0.816 & 0.574 & 0.811 & 0.754 & 0.922 & 0.918 \\ 
  GFT & -- & -- & 0.978 & -- & -- & -- & -- & -- \\ 
  Lu et al. (2019) & -- & 0.951 & \textbf{0.983} & 0.639 & \textbf{0.918} & -- & -- & -- \\ 
  naive & 0.954 & 0.934 & 0.860 & 0.823 & 0.888 & 0.941 & 0.915 & 0.874 \\ 
   \hline
\end{tabular}
\caption{Comparison of different methods for state-level \%ILI estimation in South Carolina (SC).  The MSE, MAE, and correlation are reported. The method with the best performance is highlighted in boldface for each metric in each period. \label{tab_state40}} 
\end{table}

\begin{figure}[!h] 
  \centering 
\includegraphics[width=\linewidth, page=40]{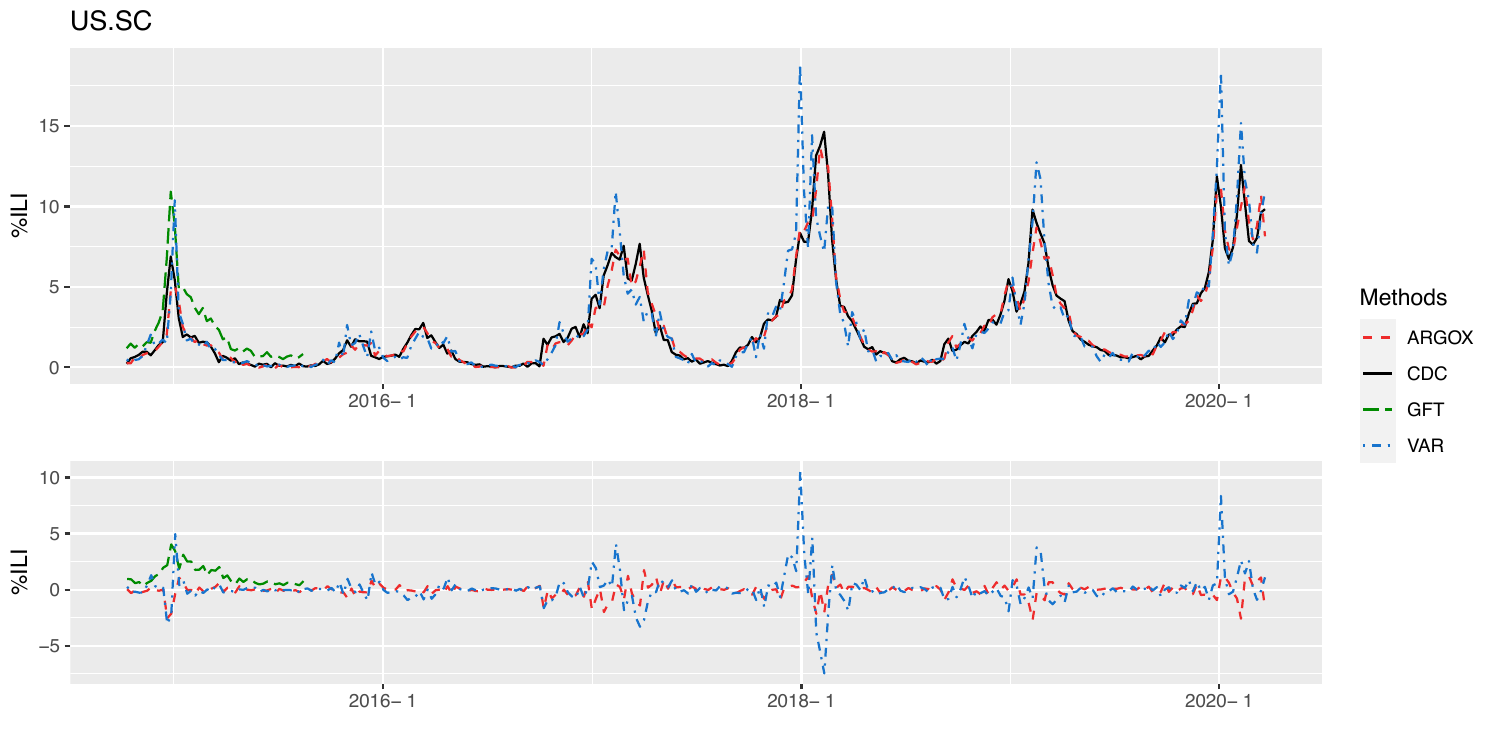} 
\caption{Plots of the \%ILI estimates (top) and the estimation errors (bottom) for South Carolina (SC).}
\end{figure}
\newpage      
\begin{table}[ht]
\centering
\begin{tabular}{crrrrrrrr}
  \hline
  & Whole period '14-'20 & Overall '14-'17 & '14-'15 & '15-'16 & '16-'17 & '17-'18 & '18-'19 & '19-'20 \\ 
  \hline  \multicolumn{1}{l}{MSE}\\ARGOX & \textbf{0.104} & 0.099 & \textbf{0.087} & \textbf{0.087} & 0.166 & \textbf{0.224} & \textbf{0.126} & \textbf{0.081} \\ 
  VAR & 0.331 & 0.372 & 0.363 & 0.371 & 0.594 & 0.250 & 0.228 & 0.992 \\ 
  GFT & -- & -- & 0.953 & -- & -- & -- & -- & -- \\ 
  Lu et al. (2019) & -- & \textbf{0.082} & 0.093 & 0.109 & \textbf{0.102} & -- & -- & -- \\ 
  naive & 0.124 & 0.117 & 0.117 & 0.095 & 0.182 & 0.285 & 0.141 & 0.095 \\ 
   \hline  \multicolumn{1}{l}{MAE}\\ARGOX & \textbf{0.231} & \textbf{0.226} & \textbf{0.215} & \textbf{0.215} & \textbf{0.289} & \textbf{0.347} & \textbf{0.280} & \textbf{0.218} \\ 
  VAR & 0.373 & 0.397 & 0.405 & 0.421 & 0.497 & 0.395 & 0.369 & 0.629 \\ 
  GFT & -- & -- & 0.476 & -- & -- & -- & -- & -- \\ 
  Lu et al. (2019) & -- & -- & -- & -- & -- & -- & -- & -- \\ 
  naive & 0.256 & 0.252 & 0.239 & 0.244 & 0.318 & 0.416 & 0.284 & 0.220 \\ 
   \hline  \multicolumn{1}{l}{Correlation}\\ARGOX & \textbf{0.952} & 0.929 & \textbf{0.932} & \textbf{0.849} & 0.929 & \textbf{0.930} & \textbf{0.884} & \textbf{0.973} \\ 
  VAR & 0.880 & 0.828 & 0.828 & 0.674 & 0.832 & 0.922 & 0.725 & 0.843 \\ 
  GFT & -- & -- & 0.906 & -- & -- & -- & -- & -- \\ 
  Lu et al. (2019) & -- & \textbf{0.941} & 0.926 & 0.805 & \textbf{0.956} & -- & -- & -- \\ 
  naive & 0.943 & 0.918 & 0.907 & 0.840 & 0.924 & 0.908 & 0.844 & 0.970 \\ 
   \hline
\end{tabular}
\caption{Comparison of different methods for state-level \%ILI estimation in South Dakota (SD).  The MSE, MAE, and correlation are reported. The method with the best performance is highlighted in boldface for each metric in each period. \label{tab_state41}} 
\end{table}

\begin{figure}[!h] 
  \centering 
\includegraphics[width=\linewidth, page=41]{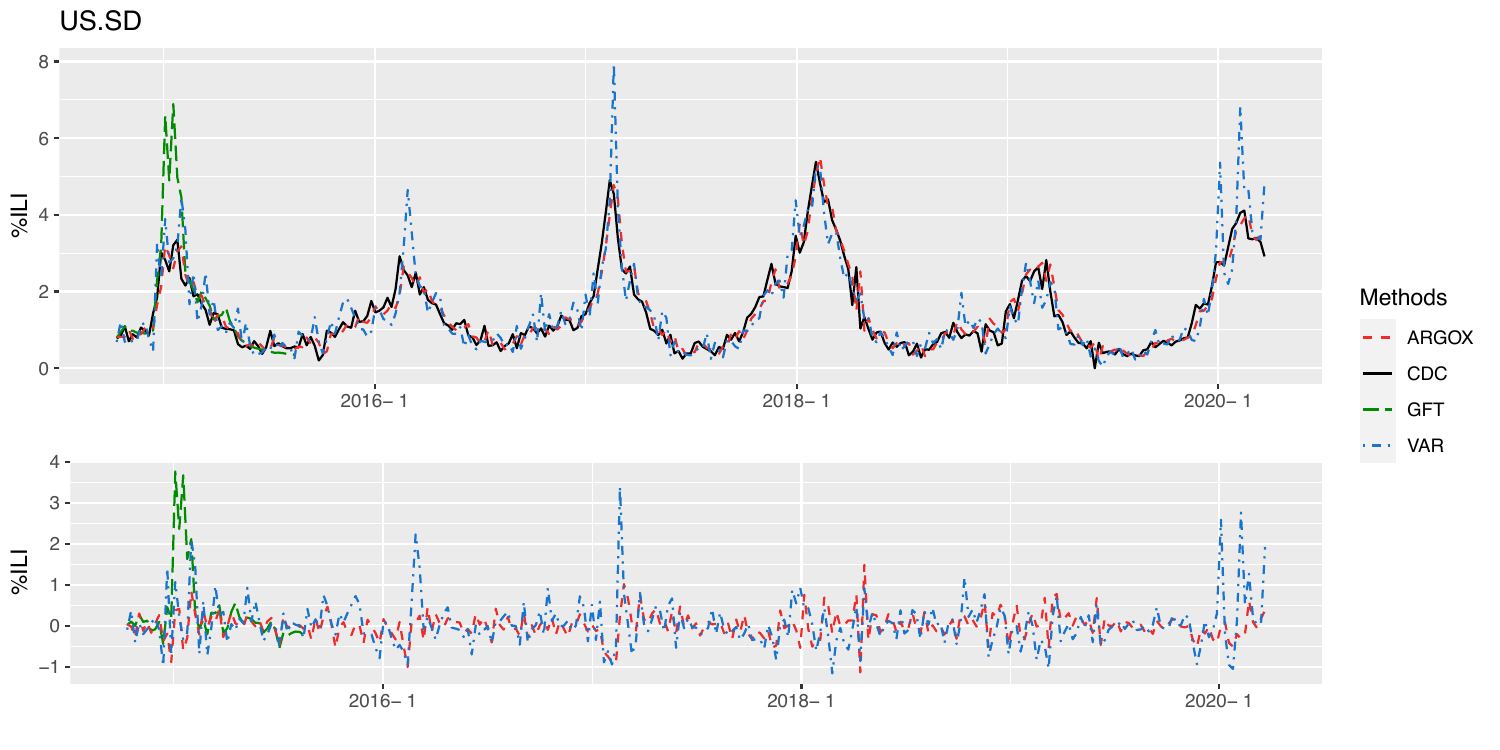} 
\caption{Plots of the \%ILI estimates (top) and the estimation errors (bottom) for South Dakota (SD).}
\end{figure}
\newpage      
\begin{table}[ht]
\centering
\begin{tabular}{crrrrrrrr}
  \hline
  & Whole period '14-'20 & Overall '14-'17 & '14-'15 & '15-'16 & '16-'17 & '17-'18 & '18-'19 & '19-'20 \\ 
  \hline  \multicolumn{1}{l}{MSE}\\ARGOX & \textbf{0.522} & \textbf{0.369} & \textbf{0.321} & \textbf{0.204} & 0.740 & \textbf{1.020} & \textbf{0.239} & \textbf{2.122} \\ 
  VAR & 1.752 & 1.695 & 2.354 & 0.767 & 2.342 & 3.022 & 1.295 & 4.688 \\ 
  GFT & -- & -- & 1.407 & -- & -- & -- & -- & -- \\ 
  Lu et al. (2019) & -- & 0.391 & 0.440 & 0.282 & \textbf{0.717} & -- & -- & -- \\ 
  naive & 0.748 & 0.568 & 0.705 & 0.257 & 0.983 & 1.427 & 0.500 & 2.811 \\ 
   \hline  \multicolumn{1}{l}{MAE}\\ARGOX & \textbf{0.471} & \textbf{0.435} & \textbf{0.377} & 0.378 & \textbf{0.644} & \textbf{0.613} & \textbf{0.379} & \textbf{1.138} \\ 
  VAR & 0.726 & 0.776 & 0.729 & 0.633 & 1.079 & 0.787 & 0.713 & 1.469 \\ 
  GFT & -- & -- & 0.947 & -- & -- & -- & -- & -- \\ 
  Lu et al. (2019) & -- & -- & -- & -- & -- & -- & -- & -- \\ 
  naive & 0.509 & 0.481 & 0.480 & \textbf{0.358} & 0.711 & 0.711 & 0.522 & 1.142 \\ 
   \hline  \multicolumn{1}{l}{Correlation}\\ARGOX & \textbf{0.950} & \textbf{0.936} & 0.956 & \textbf{0.867} & 0.864 & \textbf{0.916} & \textbf{0.941} & \textbf{0.908} \\ 
  VAR & 0.853 & 0.797 & 0.867 & 0.684 & 0.671 & 0.813 & 0.824 & 0.803 \\ 
  GFT & -- & -- & \textbf{0.968} & -- & -- & -- & -- & -- \\ 
  Lu et al. (2019) & -- & 0.936 & 0.953 & 0.841 & \textbf{0.890} & -- & -- & -- \\ 
  naive & 0.929 & 0.900 & 0.898 & 0.836 & 0.822 & 0.869 & 0.869 & 0.881 \\ 
   \hline
\end{tabular}
\caption{Comparison of different methods for state-level \%ILI estimation in Tennessee (TN).  The MSE, MAE, and correlation are reported. The method with the best performance is highlighted in boldface for each metric in each period. \label{tab_state42}} 
\end{table}

\begin{figure}[!h] 
  \centering 
\includegraphics[width=\linewidth, page=42]{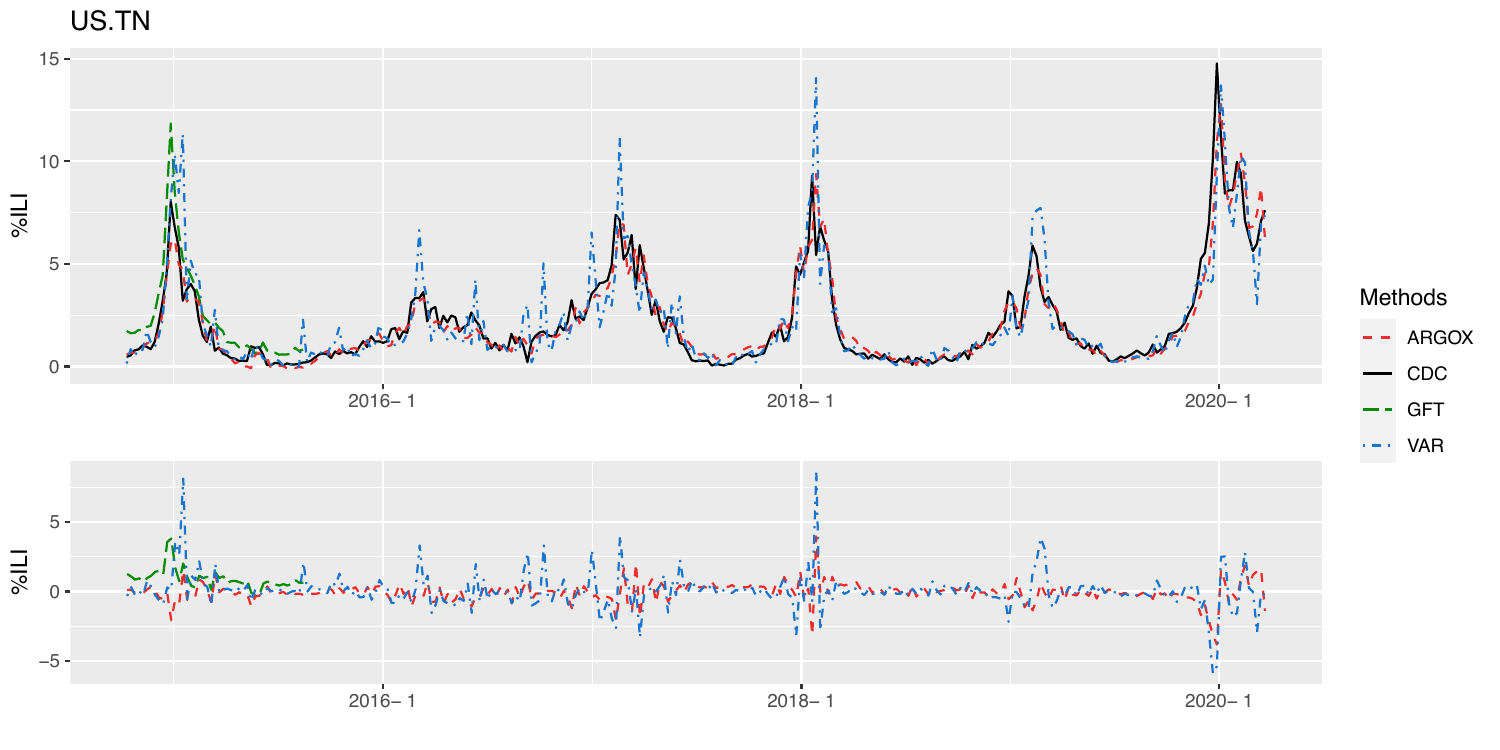} 
\caption{Plots of the \%ILI estimates (top) and the estimation errors (bottom) for Tennessee (TN).}
\end{figure}
\newpage      
\begin{table}[ht]
\centering
\begin{tabular}{crrrrrrrr}
  \hline
  & Whole period '14-'20 & Overall '14-'17 & '14-'15 & '15-'16 & '16-'17 & '17-'18 & '18-'19 & '19-'20 \\ 
  \hline  \multicolumn{1}{l}{MSE}\\ARGOX & \textbf{0.755} & 0.831 & 1.824 & 0.347 & \textbf{0.518} & \textbf{1.066} & \textbf{0.364} & \textbf{2.081} \\ 
  VAR & 1.783 & 1.736 & 3.191 & 1.162 & 1.448 & 3.269 & 2.243 & 3.249 \\ 
  GFT & -- & -- & \textbf{1.403} & -- & -- & -- & -- & -- \\ 
  Lu et al. (2019) & -- & \textbf{0.783} & 2.074 & 0.432 & 0.562 & -- & -- & -- \\ 
  naive & 0.997 & 0.958 & 2.121 & \textbf{0.333} & 0.633 & 1.898 & 0.676 & 2.585 \\ 
   \hline  \multicolumn{1}{l}{MAE}\\ARGOX & \textbf{0.510} & \textbf{0.537} & \textbf{0.759} & 0.466 & \textbf{0.540} & \textbf{0.635} & \textbf{0.419} & \textbf{1.054} \\ 
  VAR & 0.738 & 0.757 & 0.961 & 0.709 & 0.829 & 0.959 & 0.858 & 1.152 \\ 
  GFT & -- & -- & 0.772 & -- & -- & -- & -- & -- \\ 
  Lu et al. (2019) & -- & -- & -- & -- & -- & -- & -- & -- \\ 
  naive & 0.598 & 0.563 & 0.801 & \textbf{0.454} & 0.577 & 0.936 & 0.634 & 1.177 \\ 
   \hline  \multicolumn{1}{l}{Correlation}\\ARGOX & \textbf{0.959} & 0.914 & 0.882 & 0.812 & \textbf{0.959} & \textbf{0.972} & \textbf{0.976} & \textbf{0.908} \\ 
  VAR & 0.920 & 0.839 & 0.827 & 0.641 & 0.866 & 0.937 & 0.931 & 0.876 \\ 
  GFT & -- & -- & \textbf{0.936} & -- & -- & -- & -- & -- \\ 
  Lu et al. (2019) & -- & \textbf{0.919} & 0.870 & 0.791 & 0.954 & -- & -- & -- \\ 
  naive & 0.947 & 0.905 & 0.869 & \textbf{0.816} & 0.945 & 0.951 & 0.946 & 0.887 \\ 
   \hline
\end{tabular}
\caption{Comparison of different methods for state-level \%ILI estimation in Texas (TX).  The MSE, MAE, and correlation are reported. The method with the best performance is highlighted in boldface for each metric in each period. \label{tab_state43}} 
\end{table}

\begin{figure}[!h] 
  \centering 
\includegraphics[width=\linewidth, page=43]{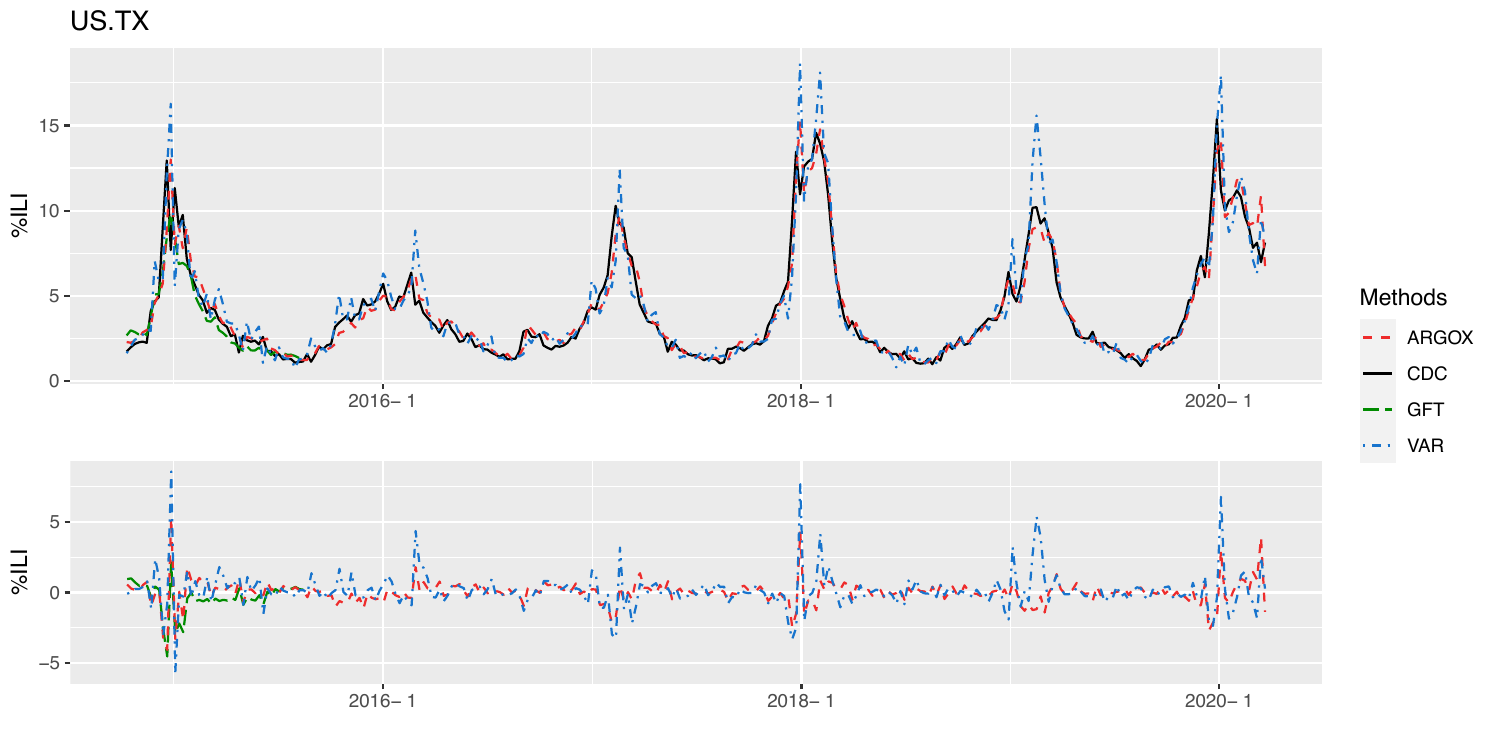} 
\caption{Plots of the \%ILI estimates (top) and the estimation errors (bottom) for Texas (TX).}
\end{figure}
\newpage      
\begin{table}[ht]
\centering
\begin{tabular}{crrrrrrrr}
  \hline
  & Whole period '14-'20 & Overall '14-'17 & '14-'15 & '15-'16 & '16-'17 & '17-'18 & '18-'19 & '19-'20 \\ 
  \hline  \multicolumn{1}{l}{MSE}\\ARGOX & \textbf{0.202} & \textbf{0.125} & \textbf{0.092} & \textbf{0.129} & 0.221 & \textbf{0.142} & \textbf{0.393} & \textbf{0.773} \\ 
  VAR & 0.875 & 0.742 & 0.960 & 1.257 & 0.363 & 0.232 & 1.888 & 2.574 \\ 
  GFT & -- & -- & 1.059 & -- & -- & -- & -- & -- \\ 
  Lu et al. (2019) & -- & 0.185 & 0.235 & 0.221 & 0.217 & -- & -- & -- \\ 
  naive & 0.255 & 0.182 & 0.238 & 0.184 & \textbf{0.193} & 0.197 & 0.433 & 0.938 \\ 
   \hline  \multicolumn{1}{l}{MAE}\\ARGOX & \textbf{0.305} & \textbf{0.253} & \textbf{0.224} & \textbf{0.281} & 0.321 & \textbf{0.286} & \textbf{0.476} & \textbf{0.672} \\ 
  VAR & 0.633 & 0.594 & 0.698 & 0.881 & 0.392 & 0.356 & 1.086 & 1.313 \\ 
  GFT & -- & -- & 0.930 & -- & -- & -- & -- & -- \\ 
  Lu et al. (2019) & -- & -- & -- & -- & -- & -- & -- & -- \\ 
  naive & 0.334 & 0.293 & 0.334 & 0.329 & \textbf{0.289} & 0.298 & 0.515 & 0.730 \\ 
   \hline  \multicolumn{1}{l}{Correlation}\\ARGOX & \textbf{0.961} & \textbf{0.946} & \textbf{0.972} & \textbf{0.917} & 0.883 & \textbf{0.900} & \textbf{0.934} & \textbf{0.915} \\ 
  VAR & 0.840 & 0.818 & 0.795 & 0.698 & 0.858 & 0.847 & 0.704 & 0.720 \\ 
  GFT & -- & -- & 0.949 & -- & -- & -- & -- & -- \\ 
  Lu et al. (2019) & -- & 0.921 & 0.930 & 0.849 & 0.880 & -- & -- & -- \\ 
  naive & 0.951 & 0.923 & 0.923 & 0.885 & \textbf{0.889} & 0.868 & 0.921 & 0.900 \\ 
   \hline
\end{tabular}
\caption{Comparison of different methods for state-level \%ILI estimation in Utah (UT).  The MSE, MAE, and correlation are reported. The method with the best performance is highlighted in boldface for each metric in each period. \label{tab_state44}} 
\end{table}

\begin{figure}[!h] 
  \centering 
\includegraphics[width=\linewidth, page=44]{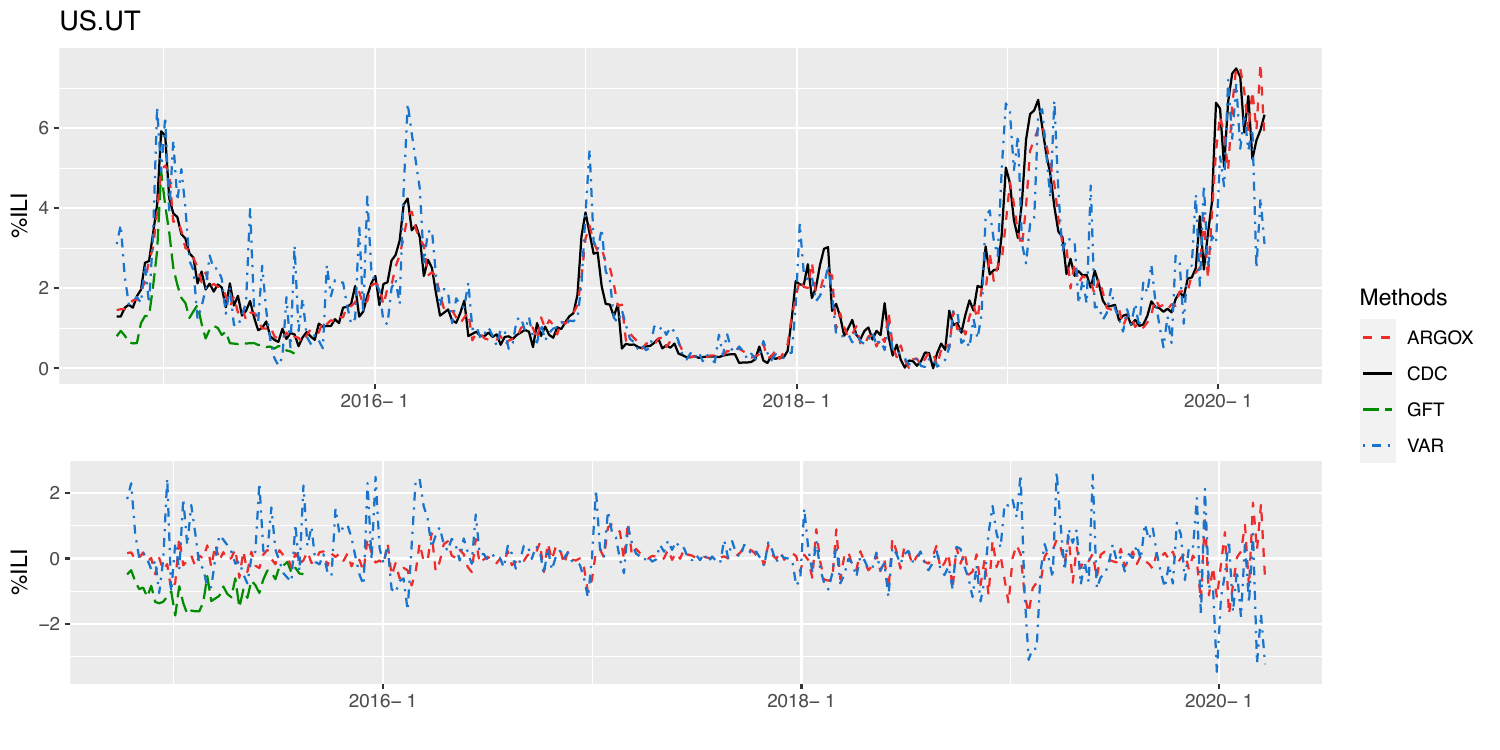} 
\caption{Plots of the \%ILI estimates (top) and the estimation errors (bottom) for Utah (UT).}
\end{figure}
\newpage      
\begin{table}[ht]
\centering
\begin{tabular}{crrrrrrrr}
  \hline
  & Whole period '14-'20 & Overall '14-'17 & '14-'15 & '15-'16 & '16-'17 & '17-'18 & '18-'19 & '19-'20 \\ 
  \hline  \multicolumn{1}{l}{MSE}\\ARGOX & \textbf{0.241} & \textbf{0.265} & \textbf{0.513} & 0.181 & 0.101 & \textbf{0.284} & \textbf{0.287} & \textbf{0.261} \\ 
  VAR & 0.579 & 0.597 & 0.655 & 0.692 & 0.597 & 0.801 & 0.967 & 0.355 \\ 
  GFT & -- & -- & 1.330 & -- & -- & -- & -- & -- \\ 
  Lu et al. (2019) & -- & 0.317 & 0.845 & \textbf{0.176} & \textbf{0.101} & -- & -- & -- \\ 
  naive & 0.286 & 0.315 & 0.606 & 0.220 & 0.127 & 0.333 & 0.348 & 0.323 \\ 
   \hline  \multicolumn{1}{l}{MAE}\\ARGOX & \textbf{0.349} & \textbf{0.355} & \textbf{0.492} & \textbf{0.325} & \textbf{0.252} & \textbf{0.407} & \textbf{0.409} & \textbf{0.342} \\ 
  VAR & 0.558 & 0.580 & 0.615 & 0.661 & 0.510 & 0.670 & 0.727 & 0.471 \\ 
  GFT & -- & -- & 0.761 & -- & -- & -- & -- & -- \\ 
  Lu et al. (2019) & -- & -- & -- & -- & -- & -- & -- & -- \\ 
  naive & 0.380 & 0.384 & 0.521 & 0.360 & 0.264 & 0.451 & 0.447 & 0.400 \\ 
   \hline  \multicolumn{1}{l}{Correlation}\\ARGOX & \textbf{0.918} & \textbf{0.902} & \textbf{0.914} & 0.789 & \textbf{0.799} & \textbf{0.920} & \textbf{0.880} & \textbf{0.946} \\ 
  VAR & 0.824 & 0.790 & 0.883 & 0.609 & 0.478 & 0.840 & 0.683 & 0.928 \\ 
  GFT & -- & -- & 0.851 & -- & -- & -- & -- & -- \\ 
  Lu et al. (2019) & -- & 0.889 & 0.845 & \textbf{0.833} & 0.782 & -- & -- & -- \\ 
  naive & 0.905 & 0.886 & 0.896 & 0.753 & 0.754 & 0.909 & 0.859 & 0.939 \\ 
   \hline
\end{tabular}
\caption{Comparison of different methods for state-level \%ILI estimation in Vermont (VT).  The MSE, MAE, and correlation are reported. The method with the best performance is highlighted in boldface for each metric in each period. \label{tab_state45}} 
\end{table}

\begin{figure}[!h] 
  \centering 
\includegraphics[width=\linewidth, page=45]{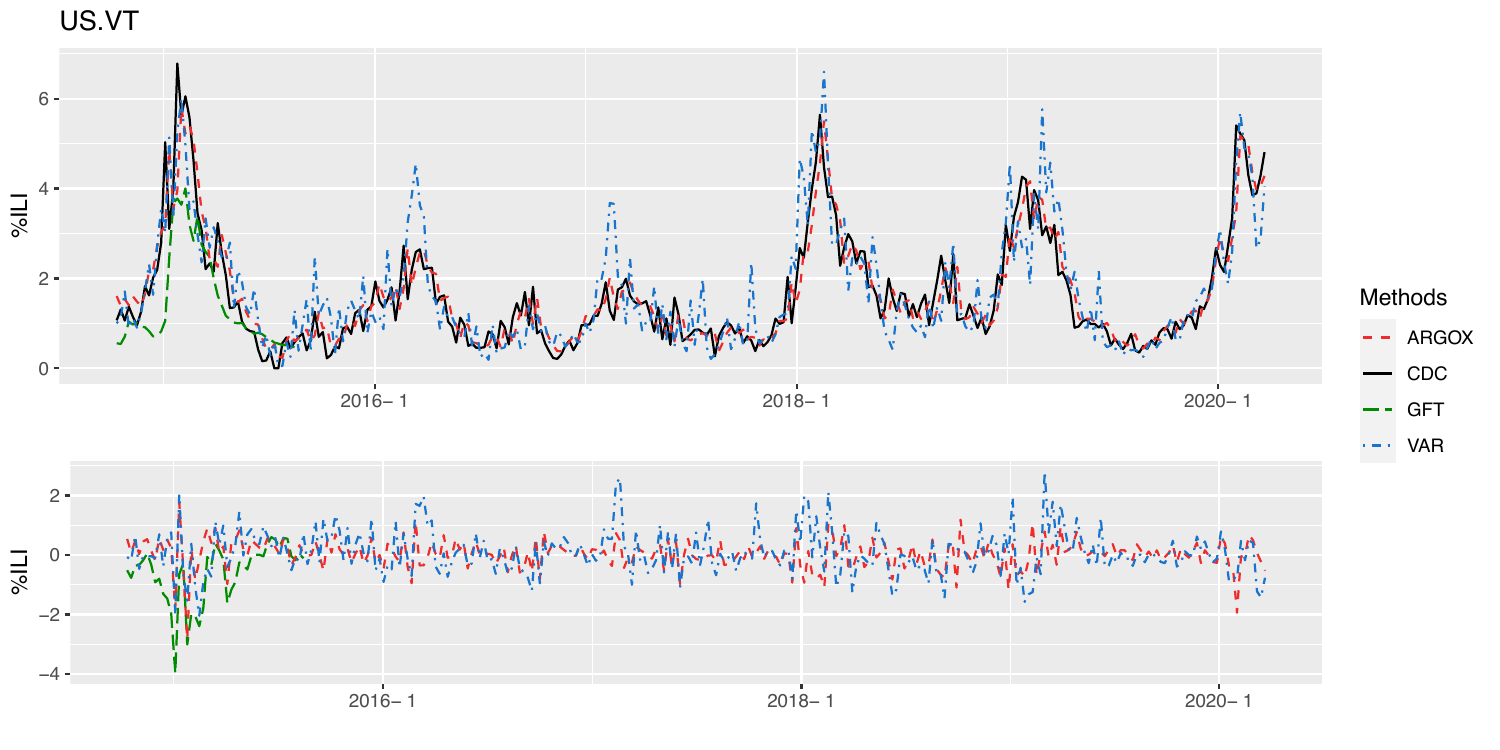} 
\caption{Plots of the \%ILI estimates (top) and the estimation errors (bottom) for Vermont (VT).}
\end{figure}
\newpage      
\begin{table}[ht]
\centering
\begin{tabular}{crrrrrrrr}
  \hline
  & Whole period '14-'20 & Overall '14-'17 & '14-'15 & '15-'16 & '16-'17 & '17-'18 & '18-'19 & '19-'20 \\ 
  \hline  \multicolumn{1}{l}{MSE}\\ARGOX & \textbf{0.169} & 0.178 & 0.329 & 0.058 & 0.214 & \textbf{0.376} & \textbf{0.203} & \textbf{0.163} \\ 
  VAR & 0.325 & 0.263 & 0.438 & 0.048 & 0.408 & 0.651 & 0.297 & 0.975 \\ 
  GFT & -- & -- & 0.984 & -- & -- & -- & -- & -- \\ 
  Lu et al. (2019) & -- & \textbf{0.108} & \textbf{0.211} & \textbf{0.024} & \textbf{0.183} & -- & -- & -- \\ 
  naive & 0.396 & 0.389 & 0.798 & 0.084 & 0.417 & 0.952 & 0.314 & 0.695 \\ 
   \hline  \multicolumn{1}{l}{MAE}\\ARGOX & \textbf{0.238} & \textbf{0.232} & \textbf{0.287} & 0.181 & \textbf{0.307} & \textbf{0.400} & \textbf{0.316} & \textbf{0.310} \\ 
  VAR & 0.327 & 0.299 & 0.373 & \textbf{0.159} & 0.455 & 0.518 & 0.387 & 0.609 \\ 
  GFT & -- & -- & 0.497 & -- & -- & -- & -- & -- \\ 
  Lu et al. (2019) & -- & -- & -- & -- & -- & -- & -- & -- \\ 
  naive & 0.335 & 0.310 & 0.418 & 0.240 & 0.381 & 0.595 & 0.411 & 0.643 \\ 
   \hline  \multicolumn{1}{l}{Correlation}\\ARGOX & \textbf{0.982} & 0.959 & 0.960 & 0.923 & 0.943 & \textbf{0.983} & \textbf{0.972} & \textbf{0.987} \\ 
  VAR & 0.965 & 0.946 & 0.956 & 0.937 & 0.892 & 0.960 & 0.959 & 0.929 \\ 
  GFT & -- & -- & 0.977 & -- & -- & -- & -- & -- \\ 
  Lu et al. (2019) & -- & \textbf{0.975} & \textbf{0.983} & \textbf{0.970} & \textbf{0.950} & -- & -- & -- \\ 
  naive & 0.955 & 0.907 & 0.895 & 0.889 & 0.885 & 0.941 & 0.956 & 0.948 \\ 
   \hline
\end{tabular}
\caption{Comparison of different methods for state-level \%ILI estimation in Virginia (VA).  The MSE, MAE, and correlation are reported. The method with the best performance is highlighted in boldface for each metric in each period. \label{tab_state46}} 
\end{table}

\begin{figure}[!h] 
  \centering 
\includegraphics[width=\linewidth, page=46]{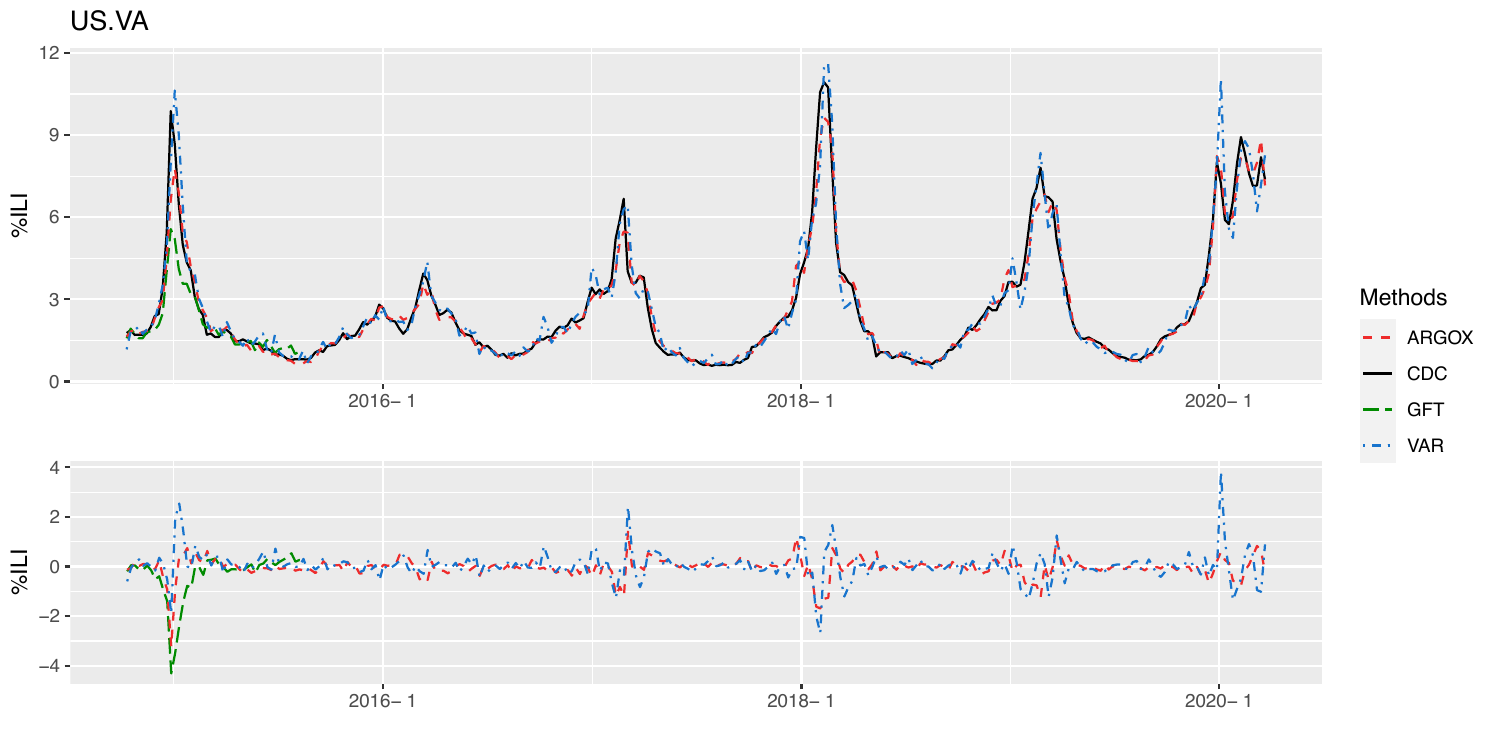} 
\caption{Plots of the \%ILI estimates (top) and the estimation errors (bottom) for Virginia (VA).}
\end{figure}
\newpage      
\begin{table}[ht]
\centering
\begin{tabular}{crrrrrrrr}
  \hline
  & Whole period '14-'20 & Overall '14-'17 & '14-'15 & '15-'16 & '16-'17 & '17-'18 & '18-'19 & '19-'20 \\ 
  \hline  \multicolumn{1}{l}{MSE}\\ARGOX & \textbf{0.192} & 0.141 & 0.122 & \textbf{0.092} & 0.312 & \textbf{0.411} & \textbf{0.211} & \textbf{0.552} \\ 
  VAR & 0.561 & 0.257 & 0.305 & 0.213 & 0.401 & 0.558 & 0.322 & 3.864 \\ 
  GFT & -- & -- & 0.562 & -- & -- & -- & -- & -- \\ 
  Lu et al. (2019) & -- & \textbf{0.115} & \textbf{0.076} & 0.115 & \textbf{0.269} & -- & -- & -- \\ 
  naive & 0.263 & 0.157 & 0.145 & 0.114 & 0.322 & 0.551 & 0.302 & 0.966 \\ 
   \hline  \multicolumn{1}{l}{MAE}\\ARGOX & \textbf{0.261} & \textbf{0.242} & \textbf{0.237} & \textbf{0.252} & 0.352 & \textbf{0.325} & \textbf{0.289} & \textbf{0.585} \\ 
  VAR & 0.406 & 0.337 & 0.396 & 0.364 & 0.383 & 0.450 & 0.329 & 1.431 \\ 
  GFT & -- & -- & 0.654 & -- & -- & -- & -- & -- \\ 
  Lu et al. (2019) & -- & -- & -- & -- & -- & -- & -- & -- \\ 
  naive & 0.304 & 0.262 & 0.277 & 0.283 & \textbf{0.352} & 0.415 & 0.358 & 0.723 \\ 
   \hline  \multicolumn{1}{l}{Correlation}\\ARGOX & \textbf{0.954} & 0.901 & 0.939 & \textbf{0.842} & 0.825 & \textbf{0.883} & \textbf{0.914} & \textbf{0.951} \\ 
  VAR & 0.861 & 0.836 & 0.867 & 0.604 & 0.796 & 0.848 & 0.882 & 0.698 \\ 
  GFT & -- & -- & 0.963 & -- & -- & -- & -- & -- \\ 
  Lu et al. (2019) & -- & \textbf{0.913} & \textbf{0.967} & 0.798 & 0.823 & -- & -- & -- \\ 
  naive & 0.937 & 0.894 & 0.928 & 0.809 & \textbf{0.834} & 0.850 & 0.876 & 0.902 \\ 
   \hline
\end{tabular}
\caption{Comparison of different methods for state-level \%ILI estimation in Washington (WA).  The MSE, MAE, and correlation are reported. The method with the best performance is highlighted in boldface for each metric in each period. \label{tab_state47}} 
\end{table}

\begin{figure}[!h] 
  \centering 
\includegraphics[width=\linewidth, page=47]{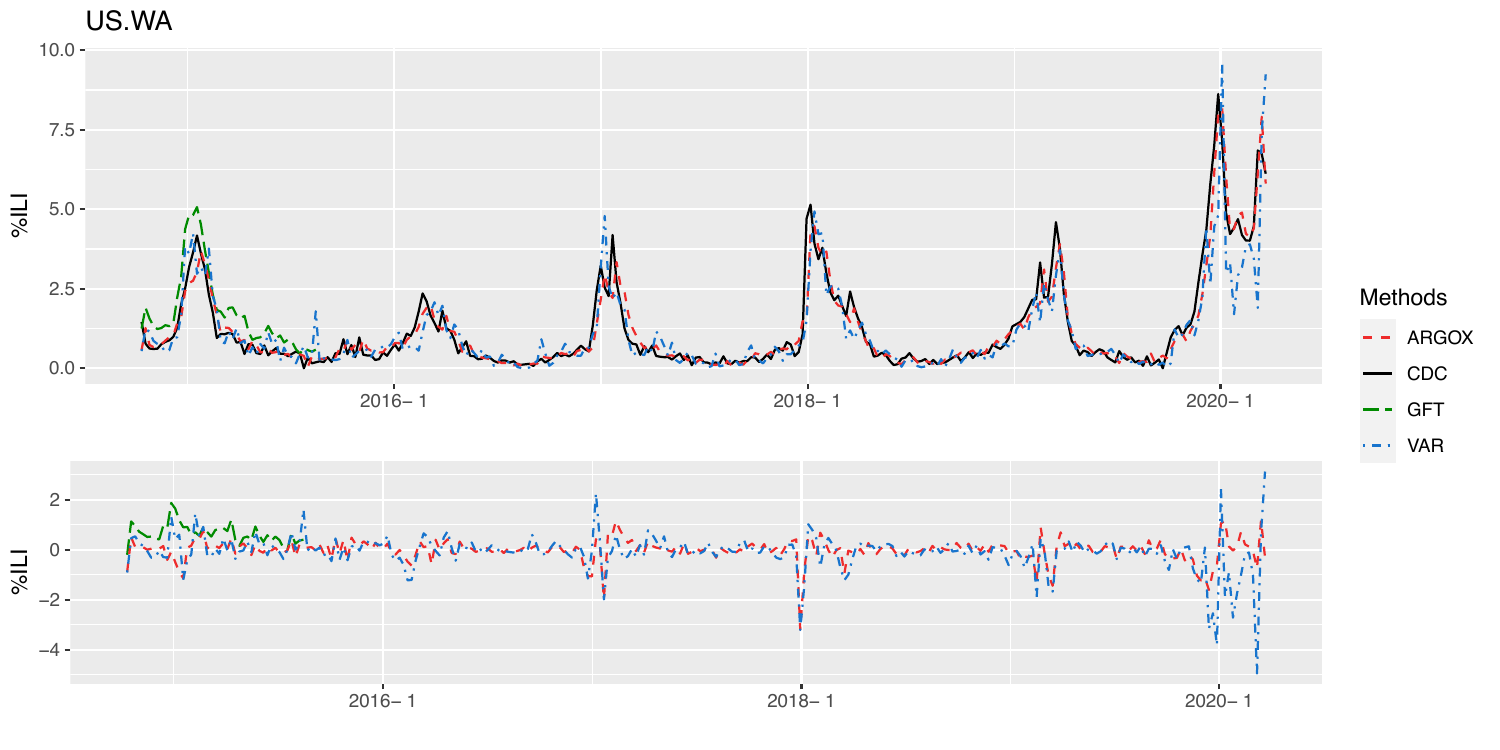} 
\caption{Plots of the \%ILI estimates (top) and the estimation errors (bottom) for Washington (WA).}
\end{figure}
\newpage      
\begin{table}[ht]
\centering
\begin{tabular}{crrrrrrrr}
  \hline
  & Whole period '14-'20 & Overall '14-'17 & '14-'15 & '15-'16 & '16-'17 & '17-'18 & '18-'19 & '19-'20 \\ 
  \hline  \multicolumn{1}{l}{MSE}\\ARGOX & \textbf{0.264} & \textbf{0.171} & 0.303 & \textbf{0.119} & \textbf{0.140} & \textbf{0.702} & \textbf{0.383} & \textbf{0.605} \\ 
  VAR & 0.600 & 0.307 & 0.341 & 0.222 & 0.523 & 1.791 & 0.489 & 2.093 \\ 
  GFT & -- & -- & 0.974 & -- & -- & -- & -- & -- \\ 
  Lu et al. (2019) & -- & 0.179 & \textbf{0.264} & 0.185 & -- & -- & -- & -- \\ 
  naive & 0.413 & 0.261 & 0.520 & 0.153 & 0.181 & 1.212 & 0.478 & 1.000 \\ 
   \hline  \multicolumn{1}{l}{MAE}\\ARGOX & \textbf{0.314} & \textbf{0.285} & \textbf{0.345} & \textbf{0.274} & 0.302 & \textbf{0.536} & \textbf{0.417} & \textbf{0.518} \\ 
  VAR & 0.458 & 0.385 & 0.405 & 0.355 & 0.529 & 0.778 & 0.522 & 1.101 \\ 
  GFT & -- & -- & 0.909 & -- & -- & -- & -- & -- \\ 
  Lu et al. (2019) & -- & -- & -- & -- & -- & -- & -- & -- \\ 
  naive & 0.355 & 0.317 & 0.442 & 0.281 & \textbf{0.298} & 0.649 & 0.437 & 0.645 \\ 
   \hline  \multicolumn{1}{l}{Correlation}\\ARGOX & \textbf{0.957} & 0.958 & 0.967 & \textbf{0.826} & \textbf{0.939} & \textbf{0.929} & \textbf{0.925} & \textbf{0.943} \\ 
  VAR & 0.920 & 0.939 & 0.964 & 0.787 & 0.896 & 0.894 & 0.909 & 0.835 \\ 
  GFT & -- & -- & 0.969 & -- & -- & -- & -- & -- \\ 
  Lu et al. (2019) & -- & \textbf{0.965} & \textbf{0.979} & 0.793 & -- & -- & -- & -- \\ 
  naive & 0.933 & 0.933 & 0.936 & 0.798 & 0.921 & 0.876 & 0.907 & 0.912 \\ 
   \hline
\end{tabular}
\caption{Comparison of different methods for state-level \%ILI estimation in West Virginia (WV).  The MSE, MAE, and correlation are reported. The method with the best performance is highlighted in boldface for each metric in each period. \label{tab_state48}} 
\end{table}

\begin{figure}[!h] 
  \centering 
\includegraphics[width=\linewidth, page=48]{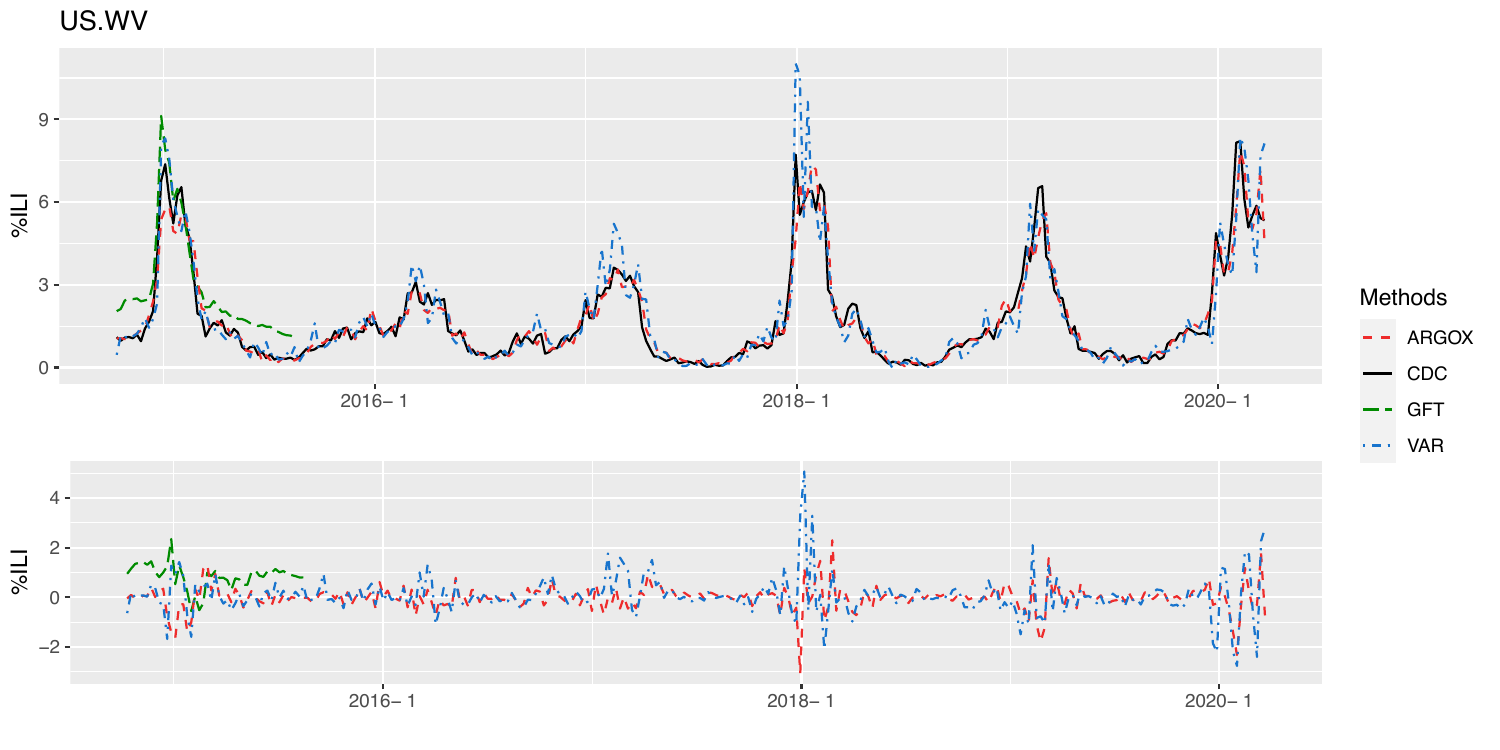} 
\caption{Plots of the \%ILI estimates (top) and the estimation errors (bottom) for West Virginia (WV).}
\end{figure}
\newpage      
\begin{table}[ht]
\centering
\begin{tabular}{crrrrrrrr}
  \hline
  & Whole period '14-'20 & Overall '14-'17 & '14-'15 & '15-'16 & '16-'17 & '17-'18 & '18-'19 & '19-'20 \\ 
  \hline  \multicolumn{1}{l}{MSE}\\ARGOX & \textbf{0.142} & \textbf{0.184} & \textbf{0.263} & \textbf{0.111} & \textbf{0.144} & \textbf{0.138} & \textbf{0.077} & \textbf{0.182} \\ 
  VAR & 0.870 & 0.802 & 1.446 & 0.474 & 0.681 & 1.958 & 0.343 & 0.788 \\ 
  GFT & -- & -- & 0.562 & -- & -- & -- & -- & -- \\ 
  Lu et al. (2019) & -- & 0.271 & 0.630 & 0.144 & 0.162 & -- & -- & -- \\ 
  naive & 0.203 & 0.274 & 0.458 & 0.120 & 0.180 & 0.204 & 0.086 & 0.243 \\ 
   \hline  \multicolumn{1}{l}{MAE}\\ARGOX & \textbf{0.276} & \textbf{0.320} & \textbf{0.380} & \textbf{0.260} & \textbf{0.311} & \textbf{0.300} & \textbf{0.216} & \textbf{0.328} \\ 
  VAR & 0.583 & 0.618 & 0.881 & 0.557 & 0.530 & 0.891 & 0.432 & 0.666 \\ 
  GFT & -- & -- & 0.604 & -- & -- & -- & -- & -- \\ 
  Lu et al. (2019) & -- & -- & -- & -- & -- & -- & -- & -- \\ 
  naive & 0.308 & 0.360 & 0.418 & 0.298 & 0.344 & 0.344 & 0.224 & 0.381 \\ 
   \hline  \multicolumn{1}{l}{Correlation}\\ARGOX & \textbf{0.956} & \textbf{0.928} & \textbf{0.945} & \textbf{0.752} & \textbf{0.895} & \textbf{0.951} & \textbf{0.888} & \textbf{0.974} \\ 
  VAR & 0.779 & 0.738 & 0.695 & 0.543 & 0.757 & 0.749 & 0.475 & 0.860 \\ 
  GFT & -- & -- & 0.897 & -- & -- & -- & -- & -- \\ 
  Lu et al. (2019) & -- & 0.898 & 0.868 & 0.696 & 0.895 & -- & -- & -- \\ 
  naive & 0.937 & 0.896 & 0.902 & 0.743 & 0.869 & 0.922 & 0.885 & 0.962 \\ 
   \hline
\end{tabular}
\caption{Comparison of different methods for state-level \%ILI estimation in Wisconsin (WI).  The MSE, MAE, and correlation are reported. The method with the best performance is highlighted in boldface for each metric in each period. \label{tab_state49}} 
\end{table}

\begin{figure}[!h] 
  \centering 
\includegraphics[width=\linewidth, page=49]{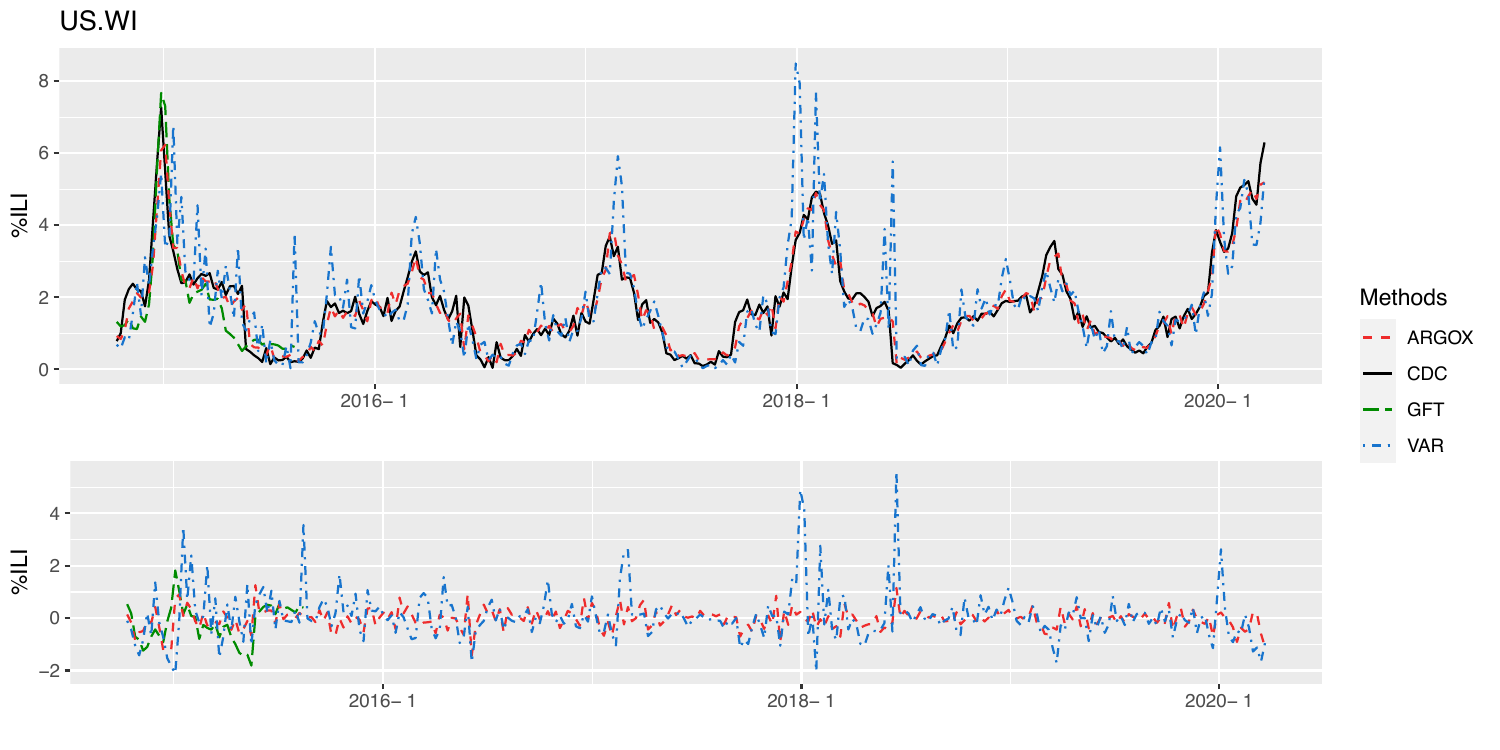} 
\caption{Plots of the \%ILI estimates (top) and the estimation errors (bottom) for Wisconsin (WI).}
\end{figure}
\newpage      
\begin{table}[ht]
\centering
\begin{tabular}{crrrrrrrr}
  \hline
  & Whole period '14-'20 & Overall '14-'17 & '14-'15 & '15-'16 & '16-'17 & '17-'18 & '18-'19 & '19-'20 \\ 
  \hline  \multicolumn{1}{l}{MSE}\\ARGOX & \textbf{0.276} & \textbf{0.143} & 0.130 & \textbf{0.128} & \textbf{0.278} & \textbf{0.468} & \textbf{0.556} & \textbf{0.912} \\ 
  VAR & 0.739 & 0.473 & 0.489 & 0.290 & 0.946 & 1.503 & 1.603 & 1.554 \\ 
  GFT & -- & -- & 0.318 & -- & -- & -- & -- & -- \\ 
  Lu et al. (2019) & -- & -- & -- & -- & -- & -- & -- & -- \\ 
  naive & 0.325 & 0.156 & \textbf{0.115} & 0.139 & 0.339 & 0.636 & 0.681 & 0.992 \\ 
   \hline  \multicolumn{1}{l}{MAE}\\ARGOX & \textbf{0.341} & \textbf{0.260} & 0.277 & \textbf{0.241} & \textbf{0.394} & \textbf{0.472} & \textbf{0.558} & 0.711 \\ 
  VAR & 0.503 & 0.402 & 0.393 & 0.390 & 0.615 & 0.919 & 0.730 & 0.887 \\ 
  GFT & -- & -- & 0.480 & -- & -- & -- & -- & -- \\ 
  Lu et al. (2019) & -- & -- & -- & -- & -- & -- & -- & -- \\ 
  naive & 0.362 & 0.271 & \textbf{0.248} & 0.265 & 0.448 & 0.585 & 0.593 & \textbf{0.690} \\ 
   \hline  \multicolumn{1}{l}{Correlation}\\ARGOX & \textbf{0.947} & \textbf{0.921} & 0.938 & \textbf{0.884} & \textbf{0.866} & \textbf{0.961} & \textbf{0.908} & \textbf{0.864} \\ 
  VAR & 0.880 & 0.846 & 0.930 & 0.728 & 0.740 & 0.848 & 0.862 & 0.803 \\ 
  GFT & -- & -- & 0.891 & -- & -- & -- & -- & -- \\ 
  Lu et al. (2019) & -- & -- & -- & -- & -- & -- & -- & -- \\ 
  naive & 0.939 & 0.918 & \textbf{0.946} & 0.881 & 0.842 & 0.940 & 0.890 & 0.858 \\ 
   \hline
\end{tabular}
\caption{Comparison of different methods for state-level \%ILI estimation in Wyoming (WY).  The MSE, MAE, and correlation are reported. The method with the best performance is highlighted in boldface for each metric in each period. \label{tab_state50}} 
\end{table}

\begin{figure}[!h] 
  \centering 
\includegraphics[width=\linewidth, page=50]{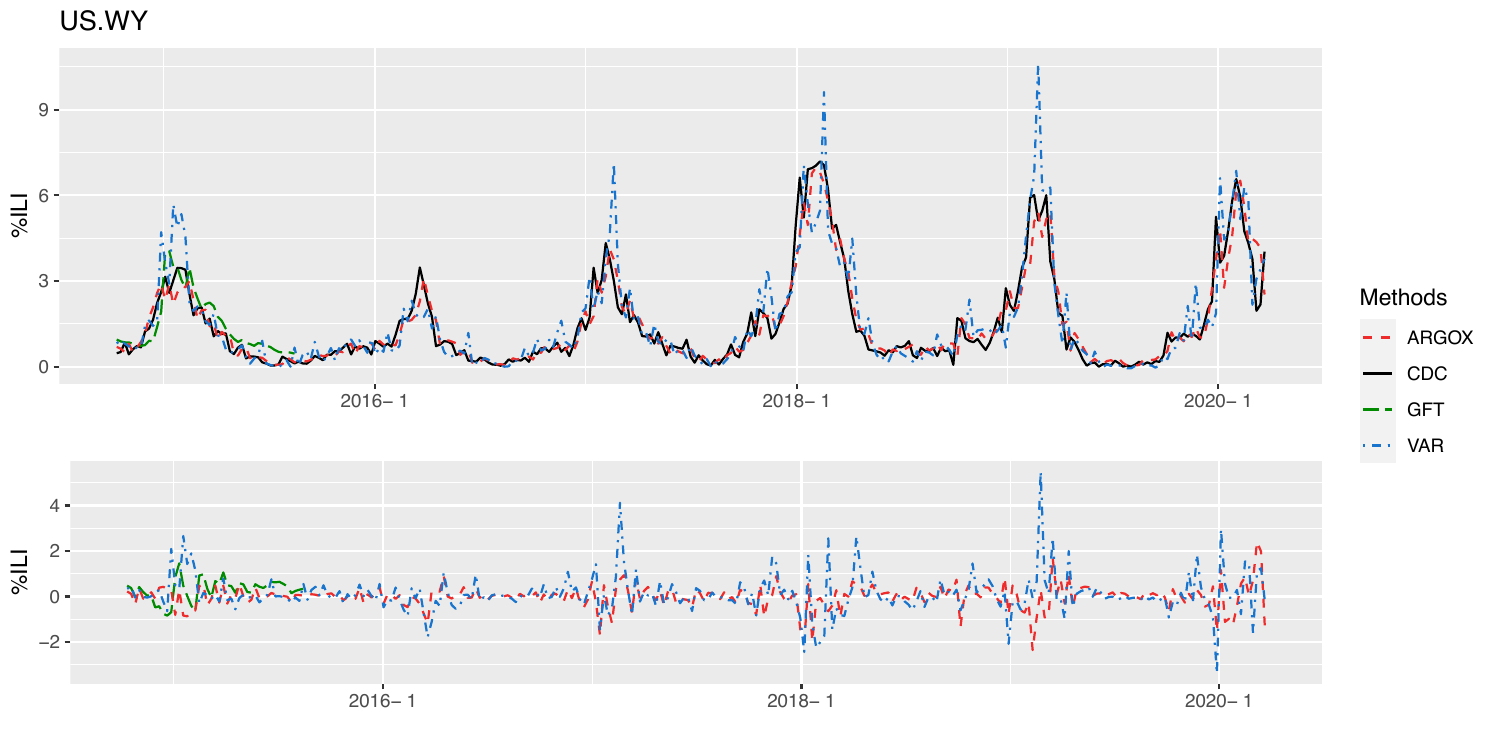} 
\caption{Plots of the \%ILI estimates (top) and the estimation errors (bottom) for Wyoming (WY).}
\end{figure}
\newpage      
\begin{table}[ht]
\centering
\begin{tabular}{crrrrrrrr}
  \hline
  & Whole period '14-'20 & Overall '14-'17 & '14-'15 & '15-'16 & '16-'17 & '17-'18 & '18-'19 & '19-'20 \\ 
  \hline  \multicolumn{1}{l}{MSE}\\ARGOX & \textbf{0.204} & \textbf{0.063} & \textbf{0.041} & \textbf{0.041} & \textbf{0.154} & \textbf{0.258} & \textbf{0.045} & 1.599 \\ 
  VAR & 0.242 & 0.105 & 0.080 & 0.097 & 0.187 & 0.339 & 0.187 & \textbf{1.399} \\ 
  GFT & -- & -- & 0.257 & -- & -- & -- & -- & -- \\ 
  Lu et al. (2019) & -- & -- & -- & -- & -- & -- & -- & -- \\ 
  naive & 0.269 & 0.072 & 0.048 & 0.069 & 0.154 & 0.581 & 0.105 & 1.768 \\ 
   \hline  \multicolumn{1}{l}{MAE}\\ARGOX & \textbf{0.220} & \textbf{0.178} & \textbf{0.153} & \textbf{0.172} & 0.285 & \textbf{0.347} & \textbf{0.167} & \textbf{0.632} \\ 
  VAR & 0.301 & 0.238 & 0.223 & 0.250 & 0.299 & 0.386 & 0.336 & 0.745 \\ 
  GFT & -- & -- & 0.436 & -- & -- & -- & -- & -- \\ 
  Lu et al. (2019) & -- & -- & -- & -- & -- & -- & -- & -- \\ 
  naive & 0.267 & 0.188 & 0.168 & 0.207 & \textbf{0.273} & 0.454 & 0.246 & 0.849 \\ 
   \hline  \multicolumn{1}{l}{Correlation}\\ARGOX & \textbf{0.959} & \textbf{0.956} & \textbf{0.962} & \textbf{0.956} & 0.907 & \textbf{0.972} & \textbf{0.971} & 0.887 \\ 
  VAR & 0.949 & 0.934 & 0.938 & 0.907 & \textbf{0.909} & 0.960 & 0.882 & \textbf{0.898} \\ 
  GFT & -- & -- & 0.949 & -- & -- & -- & -- & -- \\ 
  Lu et al. (2019) & -- & -- & -- & -- & -- & -- & -- & -- \\ 
  naive & 0.944 & 0.949 & 0.950 & 0.929 & 0.902 & 0.928 & 0.929 & 0.878 \\ 
   \hline
\end{tabular}
\caption{Comparison of different methods for state-level \%ILI estimation in New York City (NYC).  The MSE, MAE, and correlation are reported. The method with the best performance is highlighted in boldface for each metric in each period. \label{tab_state51}} 
\end{table}

\begin{figure}[!h] 
  \centering 
\includegraphics[width=\linewidth, page=51]{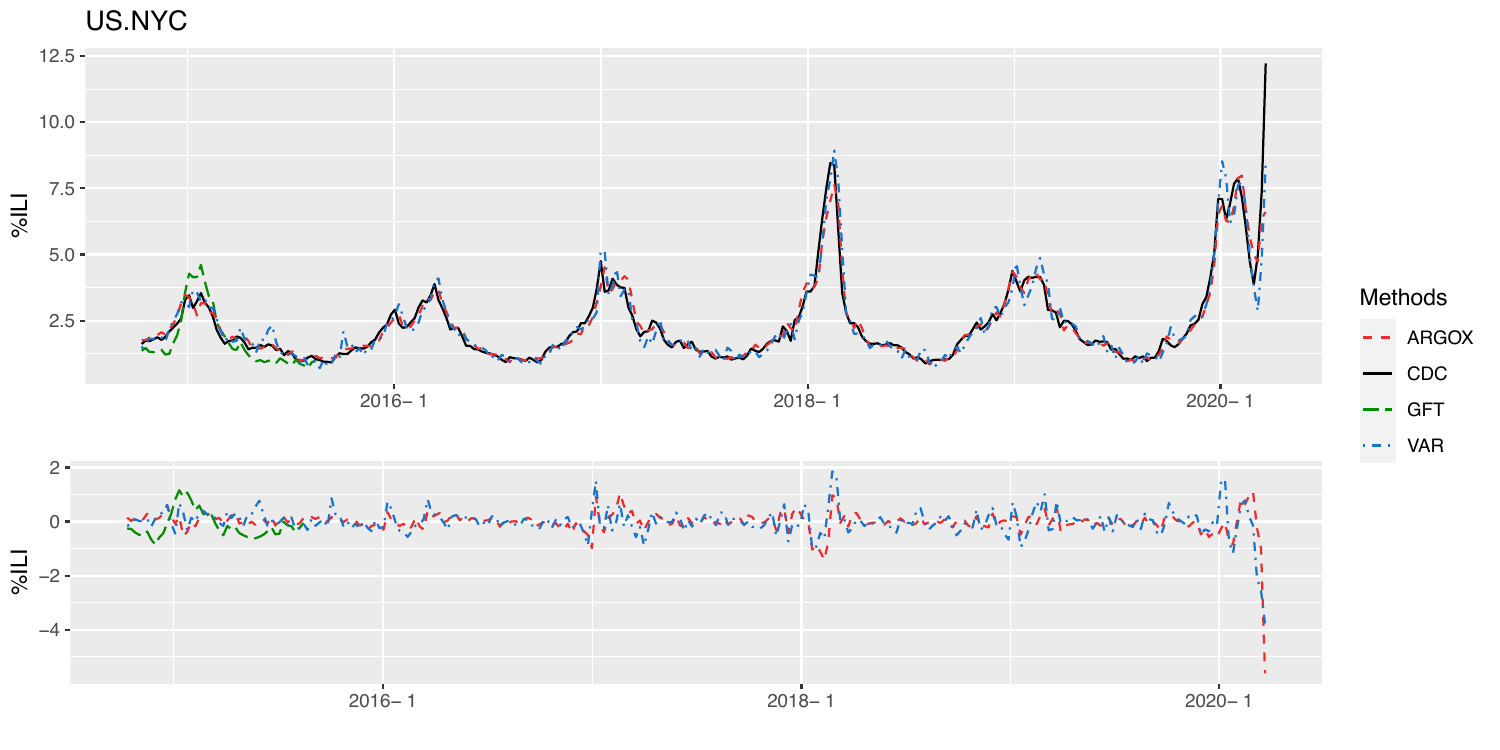} 
\caption{Plots of the \%ILI estimates (top) and the estimation errors (bottom) for New York City (NYC).}
\end{figure}
\newpage

\end{appendices}


\begin{thebibliography}{10}

\bibitem{fluburden}
{US Centers for Disease Control and Prevention (CDC)} (2020) Past seasons
  estimated influenza disease burden
  (\url{https://www.cdc.gov/flu/about/burden/past-seasons.html}).
\newblock Accessed: 2020-05-07.

\bibitem{Ginsberg_2009}
Ginsberg J, et~al. (2009) Detecting influenza epidemics using search engine
  query data.
\newblock {\em Nature} 457:1012--1014.

\bibitem{yang2017advances}
Yang S, et~al. (2017) Advances in using internet searches to track dengue.
\newblock {\em PLoS computational biology} 13(7):e1005607.

\bibitem{scott2014predicting}
Scott SL, Varian HR (2014) Predicting the present with {B}ayesian structural
  time series.
\newblock {\em International Journal of Mathematical Modelling and Numerical
  Optimisation} 5(1-2):4--23.

\bibitem{scott2015bayesian}
Scott SL, Varian HR (2015) Bayesian variable selection for nowcasting economic
  time series in {\em Economic Analysis of the Digital Economy}, eds.{}
  Goldfarb A, Greenstein SM, Tucker CE.
\newblock (University of Chicago Press), pp. 119--135.

\bibitem{wu2014}
Wu L, Brynjolfsson E (2015) The future of prediction: how {G}oogle searches
  foreshadow housing prices and sales in {\em Economic Analysis of the Digital
  Economy}, eds.{} Avi~Goldfarb SG, Tucker C.
\newblock (University of Chicago Press), pp. 89--118.

\bibitem{Shaman_2012}
Shaman J, Karspeck A (2012) Forecasting seasonal outbreaks of influenza.
\newblock {\em Proceedings of the National Academy of Sciences}
  109(50):20425--20430.

\bibitem{mcneil_2020}
McNeil DG (2020) Can smart thermometers track the spread of the coronavirus?
  (\url{https://www.nytimes.com/2020/03/18/health/coronavirus-fever-thermometers.html}).
\newblock Accessed: 2020-04-12.

\bibitem{yang2015accurate}
Yang S, Santillana M, Kou SC (2015) Accurate estimation of influenza epidemics
  using google search data via argo.
\newblock {\em Proceedings of the National Academy of Sciences}
  112(47):14473--14478.

\bibitem{Yang2017athgt}
Yang S, et~al. (2017) Using electronic health records and internet search
  information for accurate influenza forecasting.
\newblock {\em BMC Infectious Diseases} 17(1):332.

\bibitem{Yang03032015}
Yang W, Lipsitch M, Shaman J (2015) Inference of seasonal and pandemic
  influenza transmission dynamics.
\newblock {\em Proceedings of the National Academy of Sciences}
  112(9):2723--2728.

\bibitem{shaman2013real}
Shaman J, Karspeck A, Yang W, Tamerius J, Lipsitch M (2013) Real-time influenza
  forecasts during the 2012--2013 season.
\newblock {\em Nature Communications} 4(2837):2837.

\bibitem{yang2014comparison}
Yang W, Karspeck A, Shaman J (2014) Comparison of filtering methods for the
  modeling and retrospective forecasting of influenza epidemics.
\newblock {\em PLoS Comput Biol} 10(4):e1003583.

\bibitem{shaman2015improved}
Shaman J, Kandula S (2015) Improved discrimination of influenza forecast
  accuracy using consecutive predictions.
\newblock {\em PLoS currents outbreaks}.
\newblock doi:10.1371/currents.outbreaks.8a6a3df285af7ca973fab4b22e10911e.

\bibitem{cdcflusight}
(2020) Flusight: Flu forecasting | {CDC}
  (\url{https://www.cdc.gov/flu/weekly/flusight/index.html}).
\newblock Accessed: 2020-04-12.

\bibitem{brooks2015flexible}
Brooks LC, Farrow DC, Hyun S, Tibshirani RJ, Rosenfeld R (2015) Flexible
  modeling of epidemics with an empirical {B}ayes framework.
\newblock {\em PLoS Comput Biol} 11(8):e1004382.

\bibitem{Farrow2017-dj}
Farrow DC, et~al. (2017) A human judgment approach to epidemiological
  forecasting.
\newblock {\em PLoS Comput. Biol.} 13(3):e1005248.

\bibitem{yang2016forecasting}
Yang W, Olson DR, Shaman J (2016) Forecasting influenza outbreaks in boroughs
  and neighborhoods of {N}ew {Y}ork {C}ity.
\newblock {\em PLoS Computational Biology} 12(11):e1005201.

\bibitem{Davidson2015}
Davidson MW, Haim DA, Radin JM (2015) Using networks to combine ``big data''
  and traditional surveillance to improve influenza predictions.
\newblock {\em Scientific Reports} 5:8154.

\bibitem{zou2018multi}
Zou B, Lampos V, Cox I (2018) Multi-task learning improves disease models from
  web search in {\em Proceedings of the 2018 World Wide Web Conference}.
\newblock pp. 87--96.

\bibitem{lu2019improved}
Lu FS, Hattab MW, Clemente CL, Biggerstaff M, Santillana M (2019) Improved
  state-level influenza nowcasting in the united states leveraging
  internet-based data and network approaches.
\newblock {\em Nature communications} 10(1):1--10.

\bibitem{ning2019accurate}
Ning S, Yang S, Kou S (2019) Accurate regional influenza epidemics tracking
  using internet search data.
\newblock {\em Scientific reports} 9(1):5238.

\bibitem{reich2019accuracy}
Reich NG, et~al. (2019) Accuracy of real-time multi-model ensemble forecasts
  for seasonal influenza in the us.
\newblock {\em PLoS computational biology} 15(11):e1007486.

\bibitem{Burkom_etal_2007}
Burkom HS, Murphy SP, Shmueli G (2007) Automated time series forecasting for
  biosurveillance.
\newblock {\em Statistics in Medicine} 26(22):4202--4218.

\bibitem{biggerstaff2016results}
Biggerstaff M, et~al. (2016) Results from the {C}enters for {D}isease {C}ontrol
  and {P}revention's predict the 2013--2014 influenza season challenge.
\newblock {\em BMC Infectious Diseases} 16(1):1--10.

\bibitem{Santillana2015_ensemble}
Santillana M, et~al. (2015) Combining search, social media, and traditional
  data sources to improve influenza surveillance.
\newblock {\em PLoS Computational Biology} 11(10):e1004513.

\bibitem{Lazer_etal_14}
Lazer D, Kennedy R, King G, Vespignani A (2014) The parable of {G}oogle flu:
  traps in big data analysis.
\newblock {\em Science} 343(6176):1203--1205.

\bibitem{butler2013}
Butler D (2013) When {G}oogle got flu wrong.
\newblock {\em Nature} 494(7436):155--156.

\bibitem{lampos2020tracking}
Lampos V, et~al. (2020) Tracking covid-19 using online search.
\newblock {\em arXiv preprint arXiv:2003.08086}.

\bibitem{lipsitch_2011}
Lipsitch M, et~al. (2011) Improving the evidence base for decision making
  during a pandemic: the example of 2009 influenza {A/H1N1}.
\newblock {\em Biosecurity and bioterrorism: biodefense strategy, practice, and
  science} 9(2):89--115.

\bibitem{nsoesie2014systematic}
Nsoesie EO, Brownstein JS, Ramakrishnan N, Marathe MV (2014) A systematic
  review of studies on forecasting the dynamics of influenza outbreaks.
\newblock {\em Influenza and other respiratory viruses} 8(3):309--316.

\bibitem{chretien2014influenza}
Chretien JP, George D, Shaman J, Chitale RA, McKenzie FE (2014) Influenza
  forecasting in human populations: a scoping review.
\newblock {\em PloS One} 9(4):e94130.

\bibitem{stephens-davidowitz_2020}
Stephens-davidowitz S (2020) Google searches can help us find emerging covid-19
  outbreaks
  (\url{https://www.nytimes.com/2020/04/05/opinion/coronavirus-google-searches.html}).
\newblock Accessed: 2020-05-07.

\bibitem{hoerl1970ridge}
Hoerl AE, Kennard RW (1970) Ridge regression: Biased estimation for
  nonorthogonal problems.
\newblock {\em Technometrics} 12(1):55--67.

\bibitem{rcore}
{R Core Team} (2016) {\em R: A Language and Environment for Statistical
  Computing} (R Foundation for Statistical Computing, Vienna, Austria).

\bibitem{glmnet}
Friedman J, Hastie T, Tibshirani R (2010) Regularization paths for generalized
  linear models via coordinate descent.
\newblock {\em Journal of Statistical Software} 33(1):1--22.

\end{thebibliography}
\end{document}